\documentclass[journal]{IEEEtran}

\usepackage{xcolor,soul,framed} 
\usepackage[justification=centering]{caption}

\colorlet{shadecolor}{yellow}
\usepackage[pdftex]{graphicx}
\graphicspath{{../pdf/}{../jpeg/}}
\DeclareGraphicsExtensions{.pdf,.jpeg,.png}

\usepackage{units}
\usepackage[cmex10]{amsmath}
\usepackage{array}
\usepackage{mdwmath}
\usepackage{mdwtab}
\usepackage{eqparbox}
\usepackage{url}
\usepackage{mathtools}
\usepackage{geometry}
\usepackage{hyperref}
\usepackage{xcolor}
\usepackage{cite}
\usepackage{multirow}
\usepackage{multicol}
\usepackage{booktabs}
\usepackage{subfig}
\usepackage[T1]{fontenc}
\usepackage{lipsum}
\usepackage[utf8]{inputenc}
\usepackage{array}
\usepackage{lettrine}
\usepackage{authblk}
\usepackage{graphicx}
\usepackage{stfloats}
\usepackage{amsmath}
\usepackage{epsfig}
\usepackage{amssymb}
\usepackage{caption}
\usepackage{comment}
\usepackage[utf8]{inputenc}
\usepackage[mathletters]{ucs}
\usepackage{tablefootnote}
\usepackage{algcompatible}
\usepackage{algorithm}
\usepackage{algpseudocode}
\usepackage{setspace}
\usepackage{bbm}
\usepackage[english]{babel}
\usepackage{amsthm}
\usepackage{multirow}
\usepackage{nicematrix}
\usepackage[nopostdot,toc,acronym,nomain,nonumberlist]{glossaries}
\makeglossaries
\setacronymstyle{long-short}
\newacronym[type=\glsdefaulttype]{mmWave}{mmWave}{millimeter-wave}

\newacronym[type=\glsdefaulttype]{THz}{THz}{terahertz}

\newacronym[type=\glsdefaulttype]{LoS}{LoS}{line-of-sight}

\newacronym[type=\glsdefaulttype]{NLoS}{NLoS}{non-line-of-sight}

\newacronym[type=\glsdefaulttype]{RIS}{RIS}{Reconfigurable intelligent surfaces}

\newacronym[type=\glsdefaulttype]{IRS}{IRS}{intelligent reflecting surfaces}

\newacronym[type=\glsdefaulttype]{MUI}{MUI}{multi-user interference}

\newacronym[type=\glsdefaulttype]{NOMA}{NOMA}{non-orthogonal multiple access}

\newacronym[type=\glsdefaulttype]{SWIPT}{SWIPT}{simultaneous wireless information and power transfer}

\newacronym[type=\glsdefaulttype]{PLS}{PLS}{physical layer security}

\newacronym[type=\glsdefaulttype]{UAV}{UAV}{unmanned aerial vehicles}

\newacronym[type=\glsdefaulttype]{SNR}{SNR}{signal-to-noise ratio}

\newacronym[type=\glsdefaulttype]{STAR-RIS}{STAR-RIS}{Simultaneous transmitting and reflecting RISs}

\newacronym[type=\glsdefaulttype]{BD-RIS}{BD-RIS}{Beyond-diagonal RISs}

\newacronym[type=\glsdefaulttype]{BS}{BS}{base-stations}

\newacronym[type=\glsdefaulttype]{MSE}{MSE}{mean squared error}

\newacronym[type=\glsdefaulttype]{CRB}{CRB}{Cramér-Rao bound}

\newacronym[type=\glsdefaulttype]{FIM}{FIM}{Fisher information matrix}

\newacronym[type=\glsdefaulttype]{TP}{TP}{true-positive}

\newacronym[type=\glsdefaulttype]{FP}{FP}{false-positive}

\newacronym[type=\glsdefaulttype]{FN}{FN}{false-negative}

\newacronym[type=\glsdefaulttype]{TN}{TN}{true-negative}

\newacronym[type=\glsdefaulttype]{AoD}{AoD}{angles of departure}

\newacronym[type=\glsdefaulttype]{AoA}{AoA}{angles of arrival}

\newacronym[type=\glsdefaulttype]{MIMO}{MIMO}{Multiple-input multiple-output}

\newacronym[type=\glsdefaulttype]{RF}{RF}{radio-frequency}

\newacronym[type=\glsdefaulttype]{ISAC}{ISAC}{integrated sensing and communications}

\newacronym[type=\glsdefaulttype]{RCC}{RCC}{Radio-communications co-existence}

\newacronym[type=\glsdefaulttype]{DL}{DL}{downlink}

\newacronym[type=\glsdefaulttype]{DFRC}{DFRC}{Dual-function radar-communications}

\newacronym[type=\glsdefaulttype]{SCD}{SCD}{Sensing-centric design}

\newacronym[type=\glsdefaulttype]{CCD}{CCD}{Communication-centric design}

\newacronym[type=\glsdefaulttype]{AO}{AO}{alternating optimization}

\newacronym[type=\glsdefaulttype]{SINR}{SINR}{signal-to-interference-plus-noise ratio}

\newacronym[type=\glsdefaulttype]{SDR}{SDR}{Semi-definite Relaxation}

\newacronym[type=\glsdefaulttype]{QCQP}{QCQP}{Quadratically constrained quadratic programming}

\newacronym[type=\glsdefaulttype]{CCA}{CCA}{Convex sequential approximation}

\newacronym[type=\glsdefaulttype]{PDD}{PDD}{Penalty dual decomposition}

\newacronym[type=\glsdefaulttype]{FCC}{FCC}{Federal Communications Commission}

\newacronym[type=\glsdefaulttype]{QoS}{QoS}{quality-of-service}

\newacronym[type=\glsdefaulttype]{CSI}{CSI}{Channel State Information}

\newacronym[type=\glsdefaulttype]{QAM}{QAM}{quadrature amplitude modulation}

\newacronym[type=\glsdefaulttype]{AWGN}{AWGN}{additive white Gaussian noise}

\newacronym[type=\glsdefaulttype]{IID}{IID}{independent and identically distributed}

\newacronym[type=\glsdefaulttype]{CM}{CM}{constant modulus}

\newacronym[type=\glsdefaulttype]{SCA}{SCA}{Successive convex approximation}

\newacronym[type=\glsdefaulttype]{ADMM}{ADMM}{Alternative direction method of multipliers}

\newacronym[type=\glsdefaulttype]{ZF}{ZF}{Zero-forcing}

\newacronym[type=\glsdefaulttype]{KKT}{KKT}{Karush-Kuhn-Tucker}

\newacronym[type=\glsdefaulttype]{OFDM}{OFDM}{orthogonal frequency-division multiplexing}

\newacronym[type=\glsdefaulttype]{TDM}{TDM}{time-division multiplexing}

\newacronym[type=\glsdefaulttype]{FDM}{FDM}{frequency-division multiplexing}

\newacronym[type=\glsdefaulttype]{IoT}{IoT}{internet-of-things}

\newacronym[type=\glsdefaulttype]{6G}{6G}{sixth-generation}

\newacronym[type=\glsdefaulttype]{Gbps}{Gbps}{gigabits per second}

\newacronym[type=\glsdefaulttype]{Tb}{Tb}{terabits}

\newacronym[type=\glsdefaulttype]{Hz}{Hz}{hertz}

\newacronym[type=\glsdefaulttype]{SIC}{SICS}{successive interference cancellation}

\newacronym[type=\glsdefaulttype]{RSMA}{RSMA}{Rate-splitting multiple access}

\newacronym[type=\glsdefaulttype]{SDMA}{SDMA}{space-division multiple access}

\newacronym[type=\glsdefaulttype]{MM}{MM}{majorization minimization}

\newacronym[type=\glsdefaulttype]{UL}{UL}{uplink}

\newacronym[type=\glsdefaulttype]{CDMA}{CDMA}{code-division multiple access}

\newacronym[type=\glsdefaulttype]{V2V}{V2V}{vehicle-to-vehicle}

\newacronym[type=\glsdefaulttype]{V2I}{V2I}{vehicle-to-infrastructure}

\newacronym[type=\glsdefaulttype]{V2X}{V2X}{vehicle-to-everything}

\hypersetup{%
  colorlinks=true,
  linkcolor=black,
  citecolor=black,
  urlcolor=black,
  linktoc=all,
}
\usepackage{supertabular,booktabs}

\usepackage{longtable}

\newcommand\like[1]{\begin{picture}(1,1)
\ifnum0=#1\put(.5,.35){\circle{1}}\else
\ifnum10=#1\put(.5,.35){\circle*{1}}\else
\put(.5,.35){\circle{1}}\put(.5,.35){\circle*{.#1}}
\fi\fi\end{picture}}

\usepackage[skip=2pt,font=footnotesize]{caption}

\begin{document}
   \title{A Survey on Integrated Sensing and Communication with Intelligent Metasurfaces: Trends, Challenges, and Opportunities}
    	\newgeometry {top=25.4mm,left=19.1mm, right= 19.1mm,bottom =19.1mm}%
\author{Ahmed Magbool,~\IEEEmembership{Graduate Student Member,~IEEE,} Vaibhav Kumar,~\IEEEmembership{Member,~IEEE,} \\ \vspace{-0.3cm} Qingqing Wu,~\IEEEmembership{Senior Member,~IEEE,} 
Marco Di Renzo,~\IEEEmembership{Fellow,~IEEE,}  \\  and Mark F. Flanagan,~\IEEEmembership{Senior Member,~IEEE}\thanks{Ahmed Magbool and Mark F. Flanagan are with the School of Electrical and Electronic Engineering, University College Dublin, Belfield, Dublin 4, Ireland. Email: ahmed.magbool@ucdconnect.ie,  mark.flanagan@ieee.org. \par
Vaibhav Kumar is with the Engineering Division, New York University (NYU), Abu Dhabi, UAE. Email: vaibhav.kumar@ieee.org. \par
Qingqing Wu is with the Department of Electronic Engineering, Shanghai Jiao Tong University, Shanghai 200240, China. Email: qingqing-wu@sjtu.edu.cn. \par
Marco Di Renzo is with Universit\'e Paris-Saclay, CNRS, CentraleSupe\'elec, Laboratoire des Signaux et Syst\'emes, 3 Rue Joliot-Curie, 91192 Gif-sur- Yvette, France. Email: marco.di-renzo@universite-paris-saclay.fr. \par
The work of A. Magbool and M. Flanagan was supported by Science Foundation Ireland under Grant Number 13/RC/2077\_P2. \par
The work of M. Di Renzo was supported in part by the European Commission through the Horizon Europe project titled COVER under grant agreement number 101086228, the Horizon Europe project titled UNITE under grant agreement number 101129618, and the Horizon Europe project titled INSTINCT under grant agreement number 101139161, as well as by the Agence Nationale de la Recherche (ANR) through the France 2030 project titled ANR-PEPR Networks of the Future under grant agreement NF-YACARI 22-PEFT-0005, and the ANR-CHISTERA project titled PASSIONATE under grant agreements CHIST-ERA-22-WAI-04 and
ANR-23-CHR4-0003-01.
}
}
\maketitle

\begin{abstract}
The emergence of various technologies demanding both high data rates and precise sensing performance, such as autonomous vehicles and internet of things devices, has propelled an increasing popularity of integrated sensing and communication (ISAC) in recent years. ISAC offers an efficient framework for communication and sensing, where both functionalities are carried out simultaneously or in a coordinated manner. There are two levels of integration in ISAC. The first is radio-communications coexistence (RCC), where communication and radar systems use distinct hardware, waveforms and signal processing, but share the spectrum and coordinate its use. The second level of integration is dual-function radar-communications (DFRC), where communication and sensing utilize the same hardware, waveform and signal processing. At the architectural level, intelligent metasurfaces have been identified as a key enabler for the upcoming sixth-generation (6G) of wireless communication due to their ability to control the propagation environment in an energy-efficient manner. Due to the potential of metasurfaces to enhance both communication and sensing performance, numerous papers have explored the performance gains of using metasurfaces to improve ISAC. In addition, certain ISAC frameworks can address some limitations associated with using RIS for communications. Therefore, integrating ISAC with metasurfaces can mutually enhance both technologies. This survey reviews the existing literature on metasurface-assisted ISAC, detailing the associated challenges and opportunities. To provide a comprehensive overview, we commence by offering relevant background information on the fundamentals of ISAC and metasurfaces. The core part of the paper then summarizes the state-of-the-art studies on metasurface-assisted ISAC with metasurfaces employed as separate entities placed between the transmitter and receiver, also known as reconfigurable intelligent surfaces, with an emphasis on its two levels of integration: RCC and DFRC. We also review the current works in the area of holographic ISAC where metasurfaces are used to form part of ISAC transmitter and receiver. Within each category, the lessons learned, challenges, opportunities and future research directions are also highlighted.
\end{abstract}

\IEEEpeerreviewmaketitle

\section{Introduction}
The proliferation of personal communication devices, internet-of-things (IoT) devices, smart sensors, radars, and similar technologies continues to escalate dramatically. In this context, the upcoming sixth-generation (6G) of cellular wireless communication aims to connect more than 50 billion devices in environmentally friendly and sustainable ways~\cite{2015_Qualcomm}. The key performance indicators for 6G networks include providing extreme data rates, enhanced spectral efficiency and coverage, extensive bandwidth, improved energy efficiency, ultra-low latency, and extremely high reliability~\cite{2020_Nandana,2024_Magbool}.

Various 6G applications, such as autonomous driving, smart city solutions, and IoT, demand both precise position information and high data rates simultaneously. In addition, the exploration of additional spectrum in high-frequency bands such as millimeter-wave (mmWave) and terahertz (THz) bands holds great potential for sensing applications, resulting in millimeter-range accuracy positioning~\cite{2023_Behravan,2021_Wymeersch}. Motivated by these advancements, integrated sensing and communication (ISAC) has garnered increased attention in recent years. Traditionally, communication and sensing functions operated independently, necessitating distinct communication transmitters/receivers and radar transmitters/receivers, additional spectrum, and separate beamformers and waveforms. Only recently have researchers recognized the effectiveness of sharing these resources (or some of them) for communication and radar sensing, leading to more efficient utilization of these resources~\cite{2022_Zhang3}.

At the architectural level, metasurfaces represent another promising technology for 6G networks. An electromagnetic metasurface, also referred to as a reconfigurable holographic surface (RHS), is a surface constructed from electromagnetic materials with capabilities beyond those found in naturally occurring materials~\cite{2019_Renzo}. Comprising extremely small sub-wavelength scattering elements, metasurfaces are electrically large in transverse size~\cite{2015_Achouri}. RHSs can function as transmitters, receivers, or reflectors, enabling the creation of reconfigurable wireless environments~\cite{2020_Huang}.

The most common types of RHSs are reconfigurable intelligent surfaces (RISs), which refer to using RHSs as separate entities positioned between the transmitter and the receiver. An RIS is composed of a planar array containing a large number of low-cost, nearly-passive meta-elements, often referred to as elements, ports, atoms, or tiles. Notably, these elements can dynamically alter the phase, and sometimes the amplitude, of incident signals~\cite{2021_Liu_Comm_Sur}. Consequently, RISs can offer additional degrees of freedom for beamforming design, resulting in notable beamforming gains~\cite{2020_DiRenzo_Jour_sel}. To provide the reader with an incentive to understand the advantages of metasurface-assisted ISAC, we introduce several use cases in the following subsection.

\subsection{Motivation and Applications of Metasurface-Assisted ISAC}
\begin{figure*}
  \centering
  \begin{tabular}{c c c }
    \includegraphics[width=0.95\columnwidth]{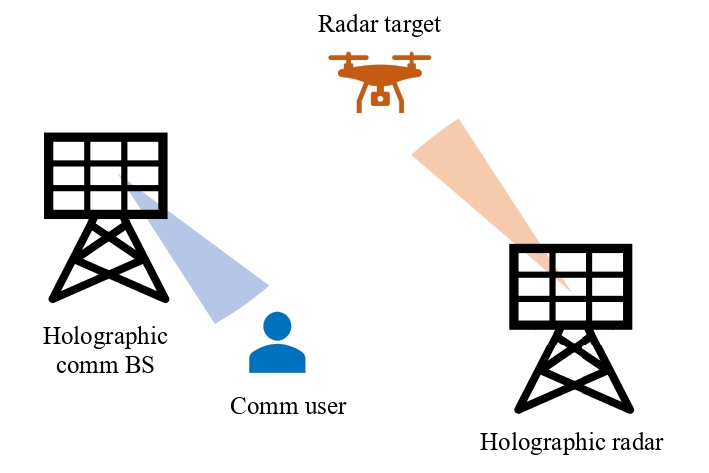} &
      \includegraphics[width=0.95\columnwidth]{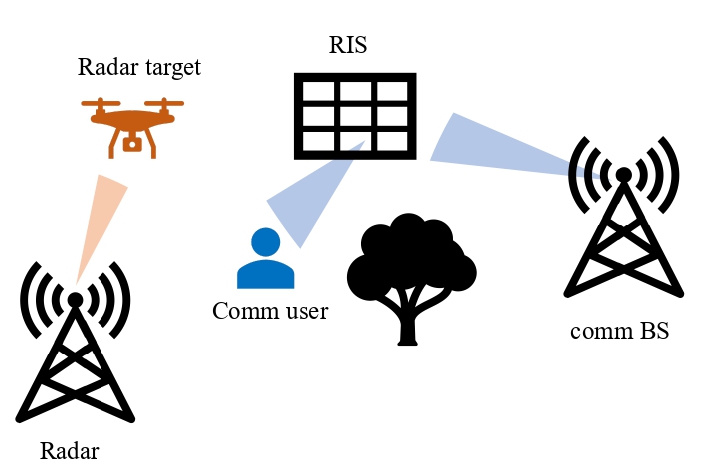}  \\
      \scriptsize (a) Sharp beamforming in RCC using RHSs.   &
      \scriptsize (b) NLoS communication in RCC using RISs.  &\\
    \includegraphics[width=0.95\columnwidth]{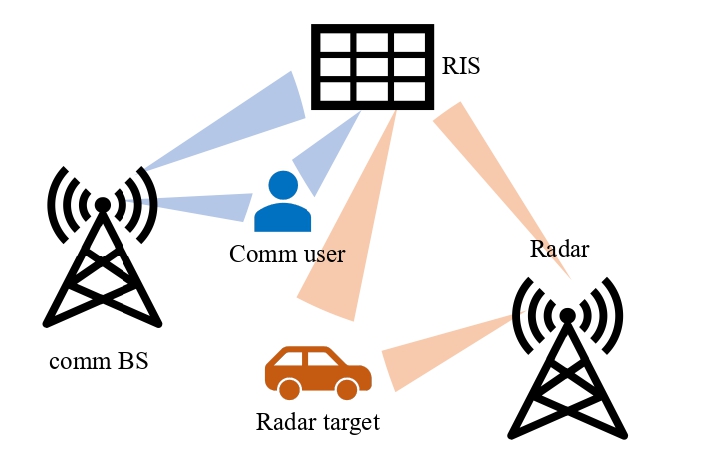} &
      \includegraphics[width=0.95\columnwidth]{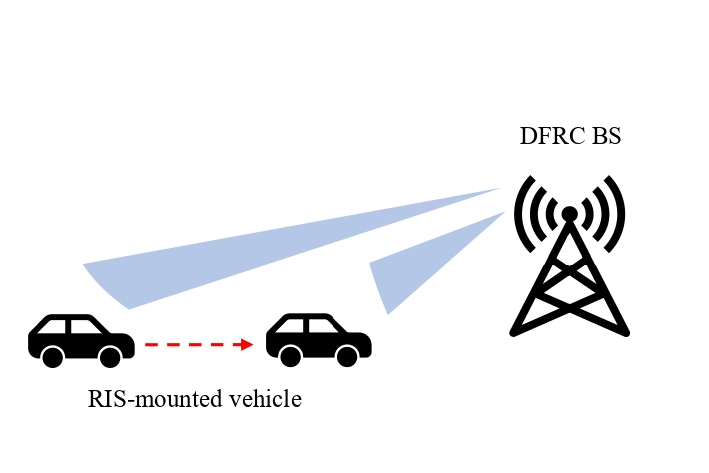}  \\
      \scriptsize (c) Interference management in RCC using RISs.  &
      \scriptsize (d) Predictive beamforming for RIS-mounted vehicles.  \\
      \includegraphics[width=0.95\columnwidth]{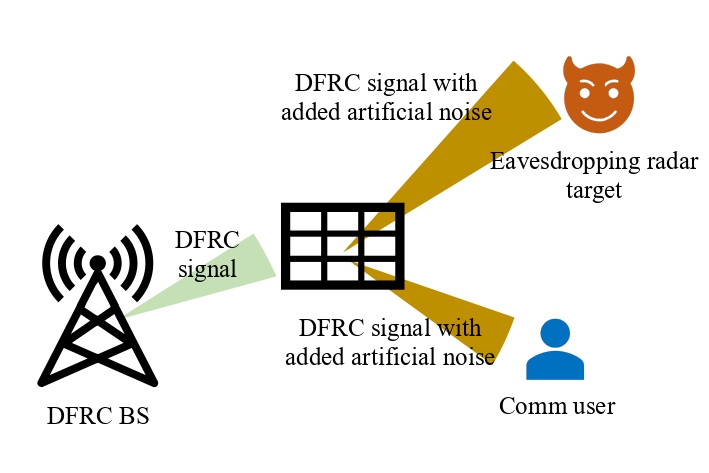} & \includegraphics[width=0.95\columnwidth]{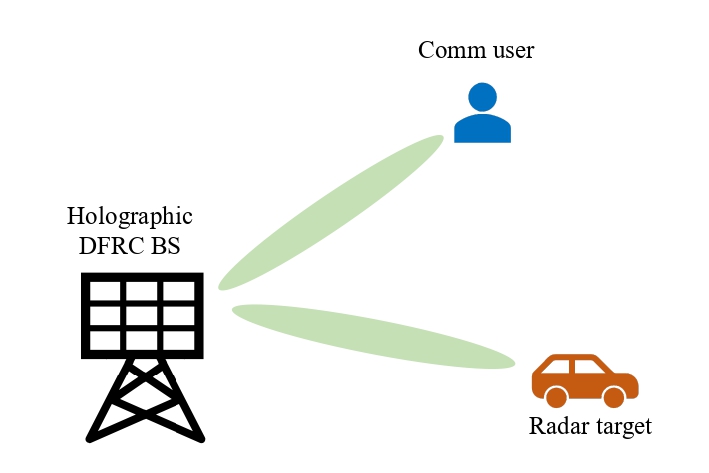}\\
        \scriptsize (e) PLS in DFRC using RIS.   &
      \scriptsize  (f) Near-field beam-focusing in DFRC using RHSs.  \\
  \end{tabular}
    \medskip
  \caption{Applications of metasurface-assisted ISAC.}
  \label{mot_app}
\end{figure*}

The integration of ISAC and metasurfaces has attracted increasing interest recently, driven by their several applications~\cite{2023_Haobo,2022_Elbir,2023_Chepuri}. Due to their ability to manipulate the propagation environment, metasurfaces can offer enhancements to both communication and sensing functionalities in ISAC systems. Also, metasurfaces can contribute to improving the performance tradeoffs between communication and sensing arising from the shared use of hardware, signal processing resources, and radio spectrum. 

Not only can RIS enhance the performance of ISAC, but ISAC can also address some of the limitations of using RIS for communication. For instance, channel estimation is a significant challenge in RIS-assisted communication systems, especially because a large-dimensional RIS is needed for communication in practical scenarios. In such a system, an ISAC framework can be used to transmit data to users while simultaneously exploiting the received echo signals to estimate their angles of departure (AoD) and angles of arrival (AoA), thereby constructing the channel matrices based on this information~\cite{2022_Albanese}.

In this subsection, we list some prominent use cases of metasurface-assisted ISAC. While the later sections of this survey provide more details and applications, we provide these use cases to motivate the use of metasurfaces to enhance ISAC systems.

\textit{1) Sharp beamforming in radio-communications co-existence (RCC) using RHSs:} RCC refers to the co-existence of radar and communication transmission within the same spectrum, but with separate hardware and waveforms. As a result, advanced signal processing techniques should be utilized to manage the interference between the two systems. Fig.~\ref{mot_app}(a) shows an example of an RCC system with RHSs employed at the communication base station (BS) and the radar. Since RHSs comprise a large number of antennas, they can form sharp beams toward the communication user and the radar target, mitigating the interference between the radar and communication signals.

\textit{2) Non-line-of-sight (NLoS) communication and/or sensing in RCC:} In RCC systems where an obstacle obstructs the line-of-sight (LoS) link between the communication BS and the user, RISs can be employed by the communication system to establish an intelligent NLoS link, as illustrated in Fig.~\ref{mot_app}(b). The same rationale can be applied to systems with a blocked LoS radar link or blocked LoS links for both radar and communication.

\textit{3) Interference management in RCC:} RISs have the capability to manipulate the propagation environment, steering signals toward specific physical directions. Hence, in RCC systems, an RIS can be employed to guide the radar signal to the designated target and the communication signal to the intended user as shown in Fig.~\ref{mot_app}(c). This aids in minimizing the interference between the communication and radar systems.

\begin{table*}
\footnotesize
\centering
\caption{Comparison between the relevant papers on metasurface-assisted ISAC and this survey article.}
\begin{tabular} {|m{1.2cm} | m{1cm}| m{1.8cm} |m{3cm}| m{2.2cm}| m{2.4cm} | m{2.2cm} |} 
 \hline 
 Reference & Year & Type & Background on metasurfaces, sensing and ISAC   & RIS-assisted RCC & RIS-assisted DFRC & Holographic ISAC  \\  [0.5ex] 
 \hline 
 \hline 
    \cite{2022_Elbir} & 2022 & Magazine & \like{4}  & \like{4} & \like{4} & \like{0}  \\
    \hline 
    \cite{2023_Chepuri} & 2023 & Tutorial-style magazine & \like{4}  & \like{0} & \like{4} & \like{0}  \\
   \hline 
    \cite{2023_Liu2} & 2023 & Magazine & \like{4}  & \like{4} & \like{4} & \like{0}  \\
   \hline 
   \cite{2023_Meng4} & 2023 & Magazine & \like{4}  & \like{4} & \like{4} & \like{0}  \\
   \hline 
   \cite{2023_Asif2} & 2023 & Mini-review & \like{4}  & \like{4} & \like{4} & \like{0}  \\
   \hline
   \cite{2023_Haobo} & 2023 & Magazine & \like{4}  & \like{0} & \like{0} & \like{10}  \\
   \hline
   This paper & 2024 & Survey & \like{10}  & \like{10} & \like{10} & \like{10}  \\
   \hline
\end{tabular}
\caption*{\like{0} : The category has not been discussed in the paper. \\ 
\like{4} : The category has been discussed in the paper but not in great detail. \\ 
\like{10} : The category has been discussed in the paper in great detail.}
\label{tab:comp}
\end{table*}

\textit{4) Predictive beamforming for RIS-mounted vehicles:} Predictive beamforming is the process of predicting the direction of the beam based on received echo measurements. This process can alleviate the burden of pilot overhead and the computational complexity associated with channel estimation, establishing a low-complexity beamforming design framework. In this context, the BS transmits a communication signal intended for users inside a vehicle while concurrently collecting the echo signals reflected from the vehicle's body to predict the channel. Installing RISs on the top of vehicles can enhance the quality of received echo signals, leading to an improved channel prediction as shown in Fig.~\ref{mot_app}(d).

\textit{5) Physical layer security (PLS) in dual-function radar-communications (DFRC) using RIS:} In DFRC, both communication and sensing utilize the same hardware, beamforming design, and radio spectrum. Thus, DFRC allows a higher degree of integration between communication and sensing compared to RCC. The transmission scheme in DFRC involves a DFRC BS transmitting a combined communication and radar signal, referred to as DFRC signal. While the communication part of the DFRC signal is intended to be decoded by the communication users, the radar targets can also decode the information-bearing communication signals using sophisticated receivers, increasing the risk of data leakage to potential eavesdropping targets, as illustrated in Fig.~\ref{mot_app}(e). In such scenarios, RIS can be employed to introduce an artificial noise with specific signatures known to the communication user, thereby enhancing PLS. Alternatively, RIS can be employed to steer the sensing signals in the target's direction, which improves the signal-to-interference-plus-noise ratio (SINR) at the communication user, and reduces the SINR at the target to decode the communication signal. This in turn improves the PLS of the DFRC system. 

\textit{6) Near-field beam-focusing in DFRC using RHSs:} As RHSs possess relatively large surface areas, their near-field regions are notably large, particularly at high carrier frequencies, such as in mmWave and THz bands. In the near-field, RHSs have the capability to dynamically concentrate electromagnetic waves toward specific locations as shown in Fig.~\ref{mot_app}(f), in contrast to the angle-dependent beams observed in the far-field. This dynamic focusing allows for precise beamforming, which is essential for accurately directing signals within confined spaces. This precision not only improves the accuracy of the radar sensing but also enhances the data rate of the communication users.

\subsection{Contributions and Relevant Works}
This paper constitutes a comprehensive survey of the state-of-the-art research on metasurface-assisted ISAC. To the best of the authors' knowledge, this survey stands as the first of its kind, offering a detailed overview of metasurface-assisted ISAC.

\begin{figure*} 
         \centering
         \includegraphics[width=1.7\columnwidth]{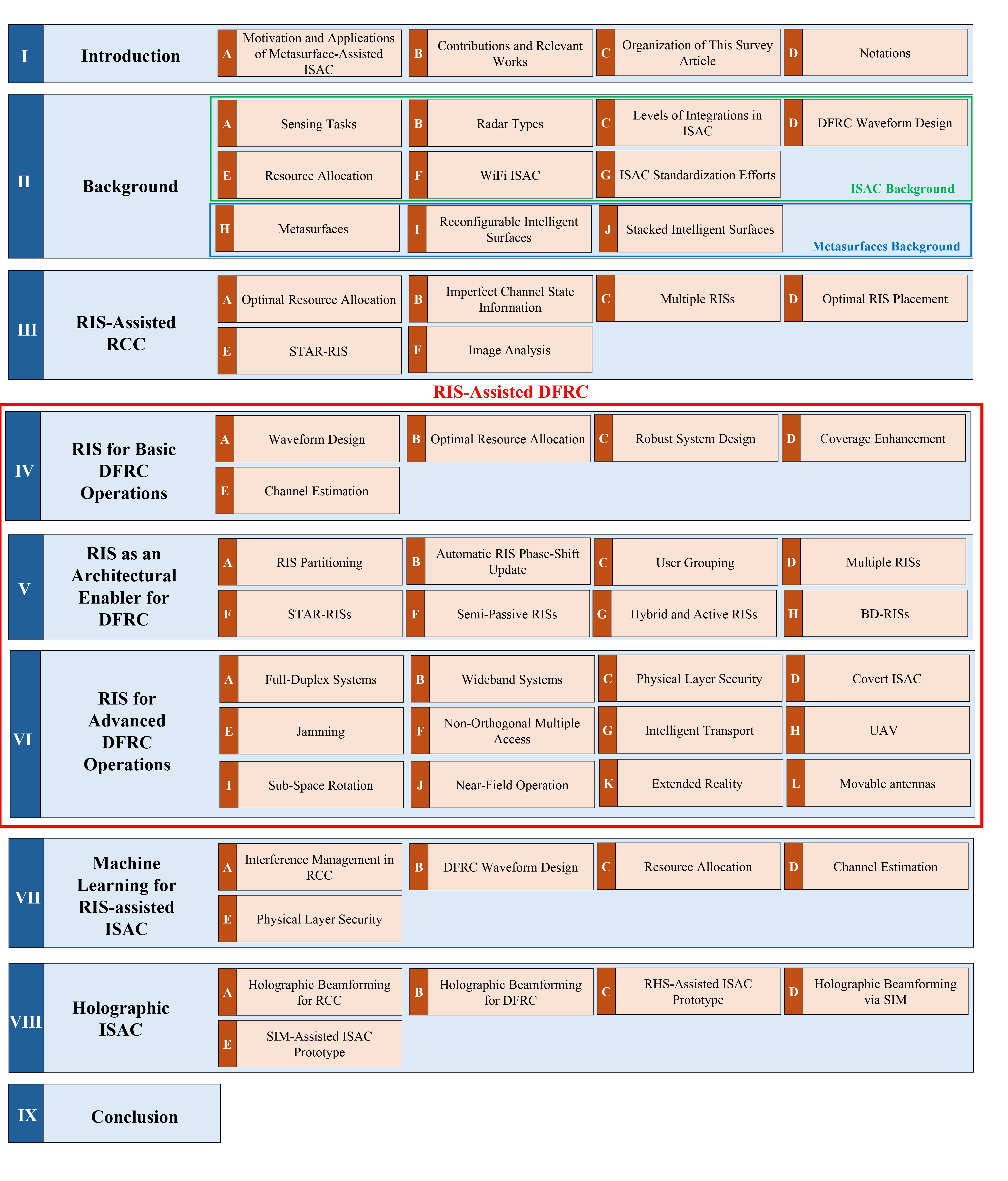}

        \caption{The organization of this paper.}
        \label{fig:outline}
\end{figure*}

While there are several surveys addressing ISAC in general, such as~\cite{2022_Liu,2022_Zhang3,2021_DeLima,2021_Cui} and~\cite{2019_Zheng}, these review articles cover general aspects of ISAC and lack a specific focus on metasurface-assisted ISAC systems. Some emphasis on  metasurface-assisted ISAC is present in~\cite{2022_Elbir,2023_Liu2,2023_Asif2,2023_Chepuri,2023_Meng4} and~\cite{2023_Haobo}. However,~\cite{2022_Elbir,2023_Liu2} and~\cite{2023_Meng4} concentrates on providing a brief background on RIS-assisted ISAC, without delving into many details and without discussing holographic ISAC. Similarly,~\cite{2023_Asif2} is an eight-page mini-review article offering a concise discussion of certain aspects of RIS-assisted ISAC without comprehensive coverage of the area of metasurface-assisted ISAC.~\cite{2023_Chepuri} is a tutorial-style magazine paper providing a high-level overview, mainly focusing on offering a brief background on RIS-assisted ISAC and exploiting RIS to improve the performance of ISAC by reducing the levels of coupling between communications and sensing. Finally,~\cite{2023_Haobo} is a magazine-style paper intended to provide an overview of holographic ISAC. In particular,~\cite{2023_Haobo} outlines the general concept of RHSs and their use for enhancing the performance of ISAC. It also presents a prototype of holographic ISAC with some experimental results. Table~\ref{tab:comp} provides a comparison between the relevant papers and this survey article.

\begin{table}
  \caption{List of Acronyms. }
  \centering
  \begin{tabular}{| c | c |}
    \hline
    1D & One-Dimensional \\
    2D & Two-Dimensional \\
     3D & Three-Dimensional \\
     3GPP & Third Generation Partnership Project \\
     5G & Fifth Generation \\
     6G & Sixth Generation \\
     ADC & Analog-to-Digital Converter \\
     ADMM & Alternative Direction Method of Multipliers \\
     AO & Alternating Optimization \\
     AoA & Angle of Arrival \\
     AoD & Angle of Departure \\
     AWGN & Additive White Gaussian Noise\\
     BD & Beyond-Diagonal \\
     BS & Base Station \\
     CCD & Communication-Centric Design \\
     CDMA & Code-Division Multiple Access \\
     CM & Constant Modulus \\
     C-RAN & Cloud Radio Access Network \\
     CRB & Cramér-Rao Bound \\
     CSI & Channel State Information \\
     DAC & Digital-to-Analog Converter \\
     DFRC & Dual-Function Radar-Communications \\
     DL & Downlink \\
     ETSI & European Telecommunication Standard \\
     & Institution\\
     FCC & Federal Communications Commission\\
     FDM & Frequency-Division Multiplexing \\
     FMCW & Frequency-Modulated Continuous-Wave \\
     FPGA & Field Programmable Gate Array \\
     HDMA & Holographic-Pattern Division \\
     & Multiple Access \\
     i.i.d & Independent and Identically Distributed \\
     IoT & Internet-of-Things \\
     IRS & Intelligent Reflecting Surface \\
     ISAC & Integrated Sensing and Communication  \\
     JCAS & Joint Communication and Radar Sensing \\
     JCR & Joint Communication Radar \\
     JRC & Joint Radar Communication \\
     KKT & Karush-Kuhn-Tucker \\
     LoS & Line-of-Sight \\
     LSTM & Long Short-Term Memory \\
     MAC & Media Access Control \\
     MIMO & Multiple-Input Multiple-Output \\
     MM & Majorization-Minimization  \\
     mmWave & Millimeter-Wave \\
     MSE & Mean Squared error \\
     MUI & Multi-User Interference \\
     MUSIC & Multiple Signal Classification \\
     NLoS & Non-Line-of-Sight \\
     NOMA & Non-Orthogonal Multiple Access \\
     OFDM & Orthogonal Frequency-Division Multiplexing \\
     PPD & Penalty Dual Decomposition \\
     PLS & Physical Layer Security \\
     QAM & Quadrature Amplitude Modulation \\
    \hline
  \end{tabular}
  \label{acr}
\end{table}

\setcounter{table}{1}
\begin{table}
  \caption{List of Acronyms (continued). }
  \centering
  \begin{tabular}{| c | c |}
    \hline
     QCQP & Quadratically Constrained Quadratic \\
     & Programming \\
     QoS & Quality-of-Service \\
     RCC & Radio-Communications Coexistence \\
     RF & Radio-Frequency \\
     RHS & Reconfigurable Holographic Surface \\
     RIS & Reconfigurable Intelligent Surface \\
     RSMA & Rate-Splitting Multiple Access \\
     SCA & Successive Convex Approximation \\
     SCD & Sensing-Centric Design \\
     SDMA & Space-Division Multiple Access \\
     SDR & Semi-Definite Relaxation \\
     SIC & Successive Interference Cancellation \\
     SIM & Stacked Intelligent Metasurface \\
     SINR & Signal-to-Interference-Plus-Noise Ratio \\
     SNR & Signal-to-Noise Ratio \\
     SOCP & Second-Order Cone Programming \\
     STAR & Simultaneous Transmitting and Reflecting \\
     SV & Saleh-Valenzuela \\
     TDM & Time-Division Multiplexing \\
     THz & Terahertz \\
     UAV & Unmanned Aerial Vehicle \\
     UL & Uplink \\
     V2I & Vehicle-to-Infrastructure \\
     V2V & Vehicle-to-Vehicle \\
     V2X & Vehicle-to-Everything \\
     WiFi & Wireless Fidelity \\
     XR & Extended Reality \\
     ZF & Zero-Forcing\\
    \hline
  \end{tabular}
\end{table}

\subsection{Organization of This Survey Article}
This survey article provides a comprehensive overview of existing research, challenges, and opportunities in the area of metasurface-aided ISAC. The outline of the paper is illustrated in Fig.~\ref{fig:outline}. For clarity, we begin with a brief background on ISAC and metasurfaces in Section~\ref{sec:BG}. Following this, we summarize existing research on RIS-assisted ISAC, where metasurfaces are employed as separate entities between the transmitters and the receivers at the two integration levels in ISAC: RCC in Section~\ref{sec:RIS_RCC}, and DFRC when RIS is utilized for basic DFRC operations, as an architectural enabler for DFRC, and for advanced DFRC operations in Sections~\ref{sec:RIS_DFRC1}, ~\ref{sec:RIS_DFRC2}, and~\ref{sec:RIS_DFRC3}, respectively. Section~\ref{sec:ML_RIS_ISAC} offers a disussion on current works on machine learning for RIS-assisted ISAC. Section~\ref{sec:holo} summarizes the state of the art in the area of holographic ISAC, which includes utilizing metasurfaces at the transmitter and/or the receiver side. Finally, the paper concludes in Section~\ref{sec:conc}. A list of the acronyms used in the paper is given in Table~\ref{acr}.

\subsection{Notations} 
Bold lowercase and uppercase letters denote vectors and matrices, respectively. $||\mathbf{\cdot} ||_2$ and $||\mathbf{\cdot} ||_\mathsf{F}$ represent the Euclidean vector norm and the Frobenius matrix norm, respectively. $\mathbf{(\cdot)}^\mathsf{T}$, $\mathbf{(\cdot)}^\mathsf{H}$ denote the matrix transpose and conjugate transpose, respectively. $\mathbb{C}$ indicates the set of complex numbers, and $j = \sqrt{-1}$ is the imaginary unit. $\mathbb{E}\{ \cdot \}$ represents the expectation operator.

\section{Background}  \label{sec:BG}
This section presents essential background information on the key concepts related to this survey paper, divided into two parts. The first part covers various important aspects of ISAC, including sensing tasks, radar types, ISAC integration levels, DFRC waveform design, resource allocation, Wireless Fidelity (WiFi) based ISAC, and efforts toward ISAC standardization. The second part focuses on relevant concepts related to metasurfaces, including RISs and stacked intelligent metasurfaces (SIMs).

\subsection{Sensing Tasks}
Radar sensing refers to the process of gathering relevant information about targets by transmitting a probing signal and processing the reflected echo. Unlike active localization (sometimes referred to as positioning), the targets in sensing are passive, meaning they do not actively transmit or receive signals. Localization, on the other hand, involves objects with active antennas that transmit, receive, and process signals to participate in the localization process.

While the definition of sensing is broad, it encompasses three main tasks. These tasks are summarized in Table~\ref{tab:sens_tasks} and explained with their relevant performance metrics below.

\textit{1) Estimation:} Parameter estimation refers to the extraction of useful parameters that describe the state of the sensed target, such as position, velocity, or temperature, from the received echo signals~\cite{2022_Zhang3}. Parameter estimation is the most common task in sensing, to the extent that \textit{radar sensing} and \textit{parameter estimation} are often used interchangeably in many papers. A widely used performance metric for quantifying parameter estimation is the \textit{mean squared error (MSE)}. The MSE is defined as the expected value of the squared difference between the true and estimated parameters~\cite{2022_Liu5}. In many situations, however, obtaining the exact MSE is challenging. In such cases, it is useful to consider the \textit{Cram\'er-Rao bound (CRB)}, which is defined as a lower bound on the variance of the error for an unbiased estimator over the estimated variables~\cite{2022_Liu5,2022_Liu}. 

\textit{2) Detection:} The second task of sensing is state detection, which involves making decisions about the sensed target based on observed echo signals. These decisions can be binary, such as determining the presence or absence of the target, or multi-class, such as classifying an object's weight into categories like light, medium, or heavy~\cite{2022_Liu5}. To measure the performance of the detection methods, a widely-used metric is the \textit{detection probability}. The detection probability is the probability that the state detected is the true state. Another important performance metric is the \textit{false-alarm probability}, which is the probability that the state detected is not the true state.

\textit{3) Recognition:} The third task of sensing is recognition, which involves obtaining useful information about the sensed target from received echo signals. This definition includes two categories. The first is object recognition. An example of this is an autonomous car determining whether the surrounding object is another car or a pedestrian. The second category is event recognition, such as identifying an alphabet from hand air-writing. The \textit{confusion matrix} can provide valuable information about the performance of the recognition task. In a two-class recognition task, the confusion matrix includes four entries: true-positive rate, false-positive rate, false-negative rate, and true-negative rate. 

\begin{table*}
\footnotesize
\centering
\caption{Sensing tasks and their most common performance metrics.}
\begin{tabular} {|m{1.5cm}| m{5cm} | m{5cm} |m{4cm}|} 
 \hline 
 Task & Definition & Examples  & Most common performance metrics  \\  [0.5ex] 
 \hline 
 Estimation & Extracting useful parameters that describe the state of the sensed target. & Position - Velocity - Temperature - Angle & MSE - CRB \\
 \hline 
 Detection & Making decisions about the sensed target. & Target presence - Motion detection - Weight classification & Detection probability - False-alarm probability \\
 \hline 
  Recognition & Obtaining useful information about the sensed target. & Pedestrian detection in autonomous vehicles - Hand air-writing & True-positive rate - False-positive rate - False-negative rate - True-negative rate \\
 \hline 
\end{tabular}
\label{tab:sens_tasks}
\end{table*}

\subsection{Radar Types}
\begin{figure*} 
         \centering
         \includegraphics[width=2\columnwidth]{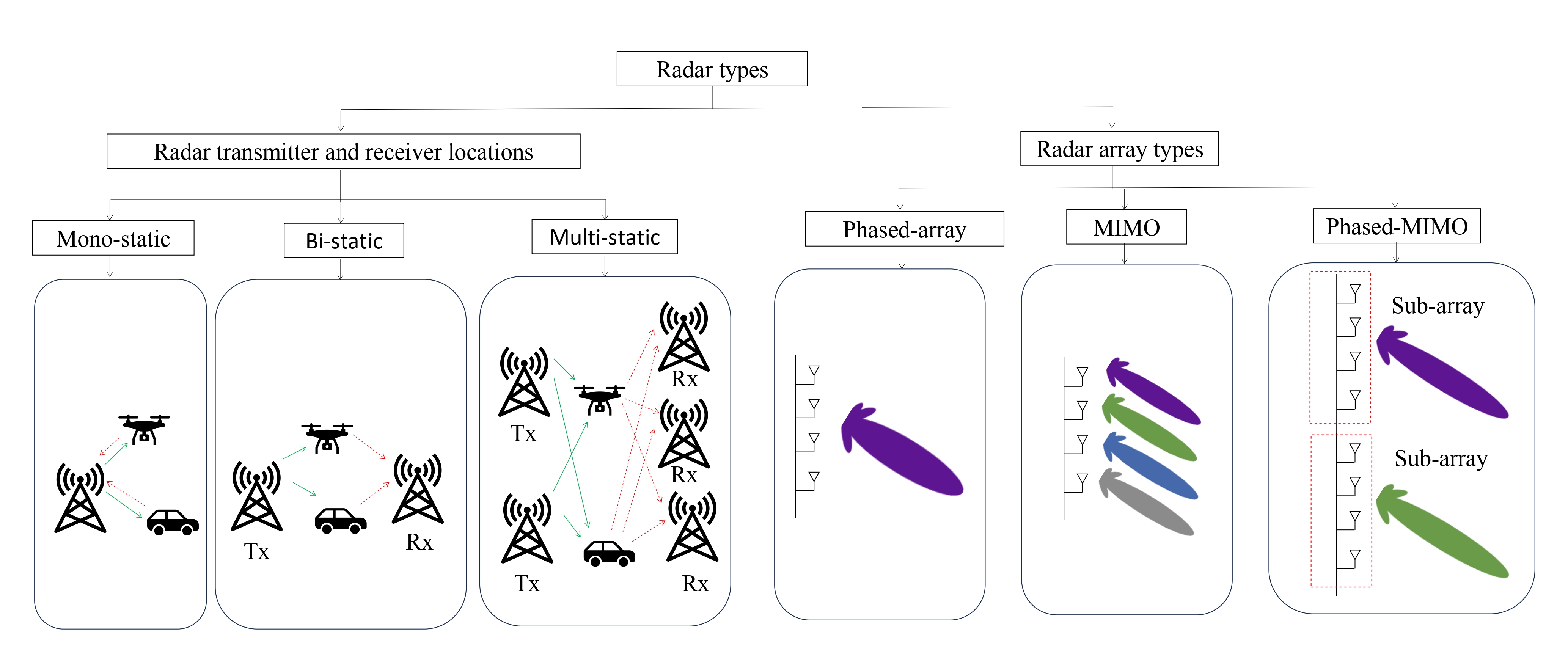}
         \vspace{0.05cm}
        \caption{Radar types.}
        \label{fig:radar_types}
\end{figure*}

Radars can be classified into multiple categories based on the locations of the transmitter and the receiver or the type of array, as shown in Fig.~\ref{fig:radar_types}.

Based on the \textbf{locations of the radar transmitter and receiver}, radars can be classified into three groups:

\textit{1) Mono-static radars:} Mono-static radars are radars that use the same antenna array for both the transmission of the probing signals and the reception of the echo signals. The definition of mono-static radars also includes radars that use one antenna array for transmission and another separate antenna array for reception, but both are co-located within the same radar unit~\cite{2022_Liu}. These radars usually have fewer parameters to estimate, for example, the AoD information can be obtained from the AoA using geometric relationships. However, a drawback of mono-static radars is the interference between the transmit radar signal and the received echo signal. This interference can be mitigated by using separate time slots for transmission and reception at the cost of lower time exploitation efficiency.

\textit{2) Bi-static radars:} Bi-static radars use separate transmit and receive arrays which are located in different geographic locations. While more parameters need to be estimated by bi-static radars compared to their mono-static counterparts, they experience lower levels of interference.

\textit{3) Multi-static radars:} Multi-static radars operate by utilizing multiple radar units that are spatially separated and work together. An example of a multi-static radar system involves one or more transmitters emit radar signals while multiple receivers, positioned at different locations, receive the reflected signals from the targets. The diverse spatial arrangement of these components enhances detection capabilities, as it allows for better angular resolution and improved tracking of moving objects. In general, multi-static radars have the ability to cover larger areas without requiring a single, powerful transmitter. However, the complexity of coordinating multiple radar units and data fusion from various sources are some of the challenges of multi-static radars.

Based on the \textbf{array type}, radars can be classified into three classes:

\textit{1) Phased-array radars:} Phased arrays employ a transmit beamformer to steer a single baseband signal to the desired direction, enhancing the beamforming gain~\cite{2014_Davis}. Phased-array radars are similar to communication transmitters that use an analog beamformer to transmit a single data stream.

\textit{2) Multiple-input multiple-output (MIMO) radars:} Unlike phased-array radars, MIMO radars transmit an independent signal from each transmit antenna~\cite{2004_Fishler}. Since each signal undergoes independent fading, MIMO radars experience almost constant receive signal-to-noise ratio (SNR), resulting in a consistent and predictable system performance. MIMO radars are analogous to communication systems that use a digital beamformer with an independent radio-frequency (RF) chain connected to each individual antenna. However, MIMO radars suffer from a high hardware complexity. Moreover, the computational complexity of processing the independent signals in MIMO radars is high compared to that of phased-array radars.

\textit{3) Phased-MIMO radars:} Phased-MIMO radars strike a balance between radar performance and complexity. In this architecture, the antenna array is divided into a number of sub-arrays. Phased-array transmission is employed within each sub-array, and different sub-arrays transmit independent signals~\cite{2010_Hassanien}. In the realm of communications, a similar system architecture is referred to as \textit{array-of-sub-arrays} architecture.

\subsection{Levels of Integration in ISAC}
\begin{figure*}
  \centering
  \begin{tabular}{c c}
    \includegraphics[width=0.95\columnwidth]{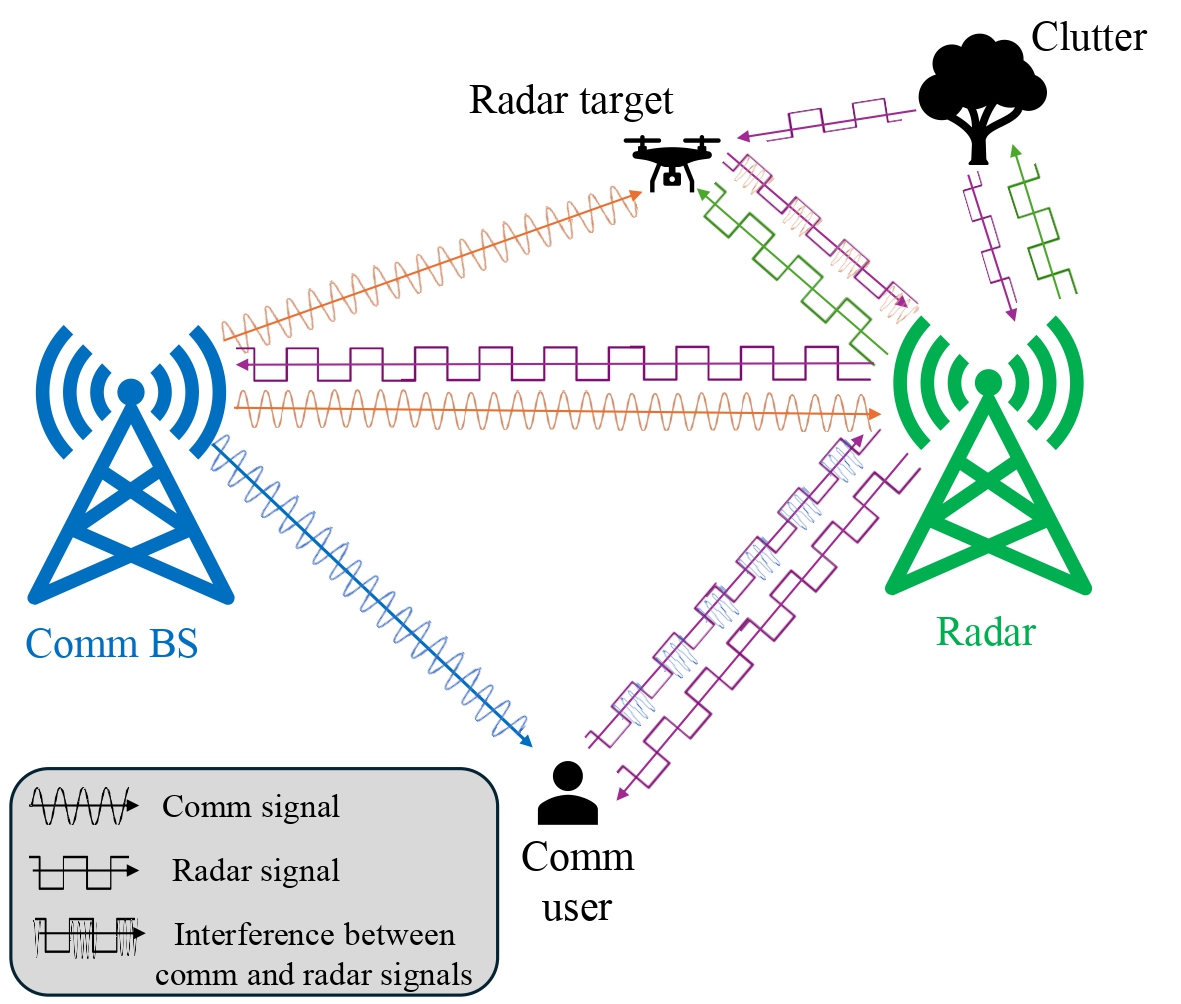} &
      \includegraphics[width=0.95\columnwidth]{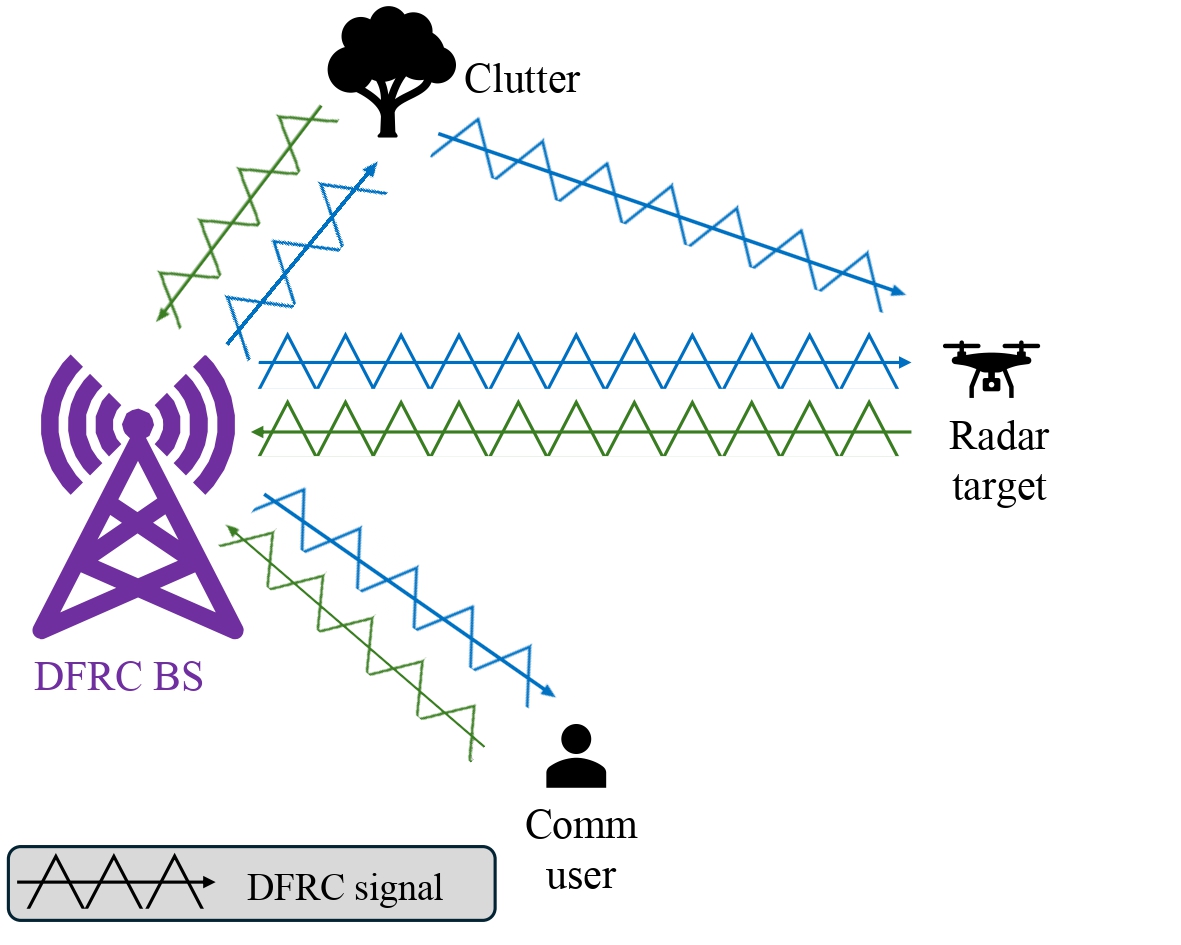}  \\
      \scriptsize (a) & \scriptsize (b) 
  \end{tabular}
    \medskip
  \caption{Two levels of integration between radar sensing and communications: (a) an RCC system with a BS communicating with a user and a mono-static radar sensing a target in a presence of a clutter, and (b) a DFRC system with a DFRC BS transmitting a DFRC signal to communicate with a user and simultaneously sense a target in a presence of a clutter.}
  \label{fig:int_lvl}
\end{figure*}

Integration between communication and sensing can occur at two levels in ISAC as shown in Fig.~\ref{fig:int_lvl}.

\textit{1) Radio-communications co-existence:} In RCC systems, both the communication BS and radar transmit their respective signals in a coordinated way. There are two types of co-existence in RCC, co-existence by spectral overlap and co-existence by cognition~\cite{2018_Liu}. 

Spectral overlap co-existence implies that both communication and radar sensing use the same spectrum, simultaneously. Consider, for example, the system shown in Fig.~\ref{fig:int_lvl}(a), where a communication BS transmits a downlink (DL) signal and a radar uses the same spectrum to send a sensing signal for target sensing. In this shared transmission scheme, both communications and sensing functionalities experience interference. For instance, the radar receiver experiences two types of interference. The first is the conventional radar interference of signals being reflected by clutter. The second type of interference includes the interfering echo signals generated by the radar signal bouncing back from the communication user and the BS, as well as the interfering communication signal from the BS. In such a system, the primary challenge lies in mitigating interference resulting from the coexistence of radar and communication transmissions as well as conventional radar challenges through resource allocation and power control methods.

Coexistence by cognition, on the other hand, aims to prevent interference caused by spectral overlap by assigning different transmission phases to the communication and radar systems~\cite{2019_Zheng}. An example of co-existence by cognition involves sharing channel information between the radar and communication systems. This sharing can occur through channel estimation by the communication system, which is then shared with the radar system, or by having the radar system sense the environment and share the relevant information with the communication system to aid in channel estimation. 

\textit{2) Dual-function radar-communications:} A higher level of integration between communications and sensing involves utilizing the same hardware, waveform and radio spectrum to perform both functions~\cite{2022_Liu5,2016_Hassanien}. The fundamental concept of DFRC is to transmit a combined radar-communications signal to the intended communication users and radar targets as shown in Fig.~\ref{fig:int_lvl}(b). In this setup, communication users extract their relevant information by treating the received sensing signal as additional noise or by using interference cancellation, while received echoes from the targets serve the radar sensing purpose. While this full integration allows mutual assistance between the two systems, it introduces several challenges. First, since communication and sensing traditionally utilize different waveforms, each tailored to its specific purpose, designing a unified waveform is a challenging task. In addition, communication and sensing are evaluated using different performance metrics, making optimal resource allocation a complex endeavor.

It is worth noting that a number of terms have been used in the literature to describe related research topics. These terms include joint radar communication (JRC), joint communication radar (JCR)~\cite{2019_Mishra}, and joint communication and radar sensing (JCAS)~\cite{2021_Zhang5}. The key distinction between the \textit{joint} aspect in these terms and the \textit{integration} aspect in ISAC is that JRC, JCR and JCAS prioritize the integration of sensing functionalities into cellular network infrastructure. On the other hand, ISAC, while somewhat interchangeable with the other terms, places a greater emphasis on the shift in network architecture and its impact on electronics, involved objects, or user equipment. Additionally, ISAC underscores a focus on resource efficiency, an aspect usually ignored by the others~\cite{2021_Cui}.

\subsection{DFRC Waveform Design}
Since the transmit waveforms in communication and sensing systems are fundamentally different, finding a common DFRC waveform presents a significant challenge. In essence, there are three primary waveform design methods for DFRC~\cite{2022_Liu5}.

\textit{1) Sensing-centric design (SCD):} A pure radar sensing waveform is not readily suitable for communication purposes since it lacks signaling information. The core concept of SCD involves incorporating some useful data into the sensing waveform without causing significant degradation in sensing performance. Thus, the primary objective here is to preserve the effectiveness of the sensing performance while adding some communication capabilities.

An example of SCD for DFRC is the use of frequency modulated continuous waveform (FMCW) to convey information along with its traditional use for radar sensing as demonstrated in~\cite{2021_Ma1}. FMCW is commonly used in automotive radar sensing due to its low-complexity signal processing and hardware design. To enable FMCW for communication, the authors of~\cite{2021_Ma1} proposed utilizing an array sparsification approach at the transmitter to create a virtual MIMO array, which combines a reduced number of radio frequency modules with narrowband FMCW signals.

\textit{2) Communication-centric design (CCD):} In theory, any communication waveform can be employed for mono-static sensing as the transmitter has complete knowledge of the waveform. However, the unpredictability introduced by communication data can substantially reduce the system's sensing performance. To address this, CCD incorporates the sensing functionality into an existing communication waveform. The primary focus in such a design is ensuring the reliability of communication while including some sensing capabilities. 

For instance, in~\cite{2021_Keskin1}, the authors explored the use of an orthogonal frequency-division multiplexing (OFDM) waveform for DFRC. In this study, target parameters were estimated from the backscattered signals. The authors sought to achieve a balanced performance tradeoff between radar sensing and communications by optimizing subcarrier powers within a specific time-frequency region. This optimization involved considering a side-lobe power level constraint in the delay-Doppler ambiguity domain to ensure accurate radar sensing.

\textit{3) Joint design:} The SCD and CCD still prioritize one DFRC functionality over the other, and as a result, do not provide a high level of integration between communications and sensing. To address this, joint waveform design has been considered more appealing. Joint design aims at creating a new DFRC waveform from the ground up rather than relying on existing sensing and/or communication waveforms. This approach provides additional degrees of freedom and flexibility, resulting in simultaneous improvements in both sensing and communication performance.

An example of joint waveform design for DFRC was presented in~\cite{2018_Liu2}, where the authors designed a DFRC waveform to minimize the multiuser interference. In this work, both omnidirectional and directional beampattern design problems were addressed to achieve globally optimal solutions. Building on this, the authors formulated an optimization problem with objective function equal to the weighted sum of the radar and communication multiuser interference energy, enabling a flexible tradeoff between radar and communications performance.

\subsection{Resource Allocation} \label{sec:ISAC_RA}
The task of resource allocation in ISAC aims at determining optimal values for available resources, such as the transmit beamformer, receive beamformer, power allocation, etc., to optimize a system performance metric. Resource allocation in both RCC and DFRC systems poses challenges due to the absence of a unified performance metric that can effectively quantify the quality of ISAC systems. Since communication and sensing systems are assessed using different evaluation criteria, resource allocation in ISAC can be treated as a multi-objective optimization problem.

Consider, for example, an ISAC system with a communication performance metric expressed as $\mathcal{C}(\mathcal{R})$ and a sensing performance metric of $\mathcal{S}(\mathcal{R})$, with $\mathcal{R}$ representing the set of available resources. If the objective is to maximize both $\mathcal{C}(\mathcal{R})$ and $\mathcal{S}(\mathcal{R})$, the following multi-objective optimization problem is formulated:
\begin{subequations}
\begin{align}
           \max_{  \mathcal{R}} \ \ & [\mathcal{C}(\mathcal{R}), \mathcal{S}(\mathcal{R}) ] \label{optm_MO_obj} \\
        \text{s.t.} \ \ \ & \mathcal{T}(\mathcal{R}),
\end{align} 
\label{eq:MOO}
\end{subequations}
\hspace{-0.2cm} where $\mathcal{T}(\mathcal{R})$ denotes the constraint set, e.g., power budget constraint. From a mathematical point of view, the optimization problem~\eqref{eq:MOO} is not well-defined, as the two objective functions can have different trends, preventing the existence of a simultaneously global optimal solution for both of them. A field known as \textit{multi-objective optimization}~\cite{2016_Deb,2018_Gunantara} employs various methods to address such optimization problems.

There are two common ways to handle multiple objective functions in ISAC resource allocation optimization problems. The first approach is to optimize one of the objective functions while setting a threshold in the other. For example, if we choose to maximize the sensing performance metric while setting a threshold for the communication metric, we can reformulate~\eqref{eq:MOO} as
\begin{subequations}
\begin{align}
           \max_{  \mathcal{R}} \ \ & \mathcal{S}(\mathcal{R})  \\
        \text{s.t.} \ \ \ & \mathcal{C}(\mathcal{R}) \geq \mathcal{C}_\text{th}, \\
        &\mathcal{T}(\mathcal{R}),
\end{align} 
\label{eq:MOO_1}
\end{subequations}
\hspace{-0.2cm} where $\mathcal{C}_\text{th}$ is the threshold value on the communication metric.

\begin{figure} 
         \centering
    \includegraphics[width=1\columnwidth]{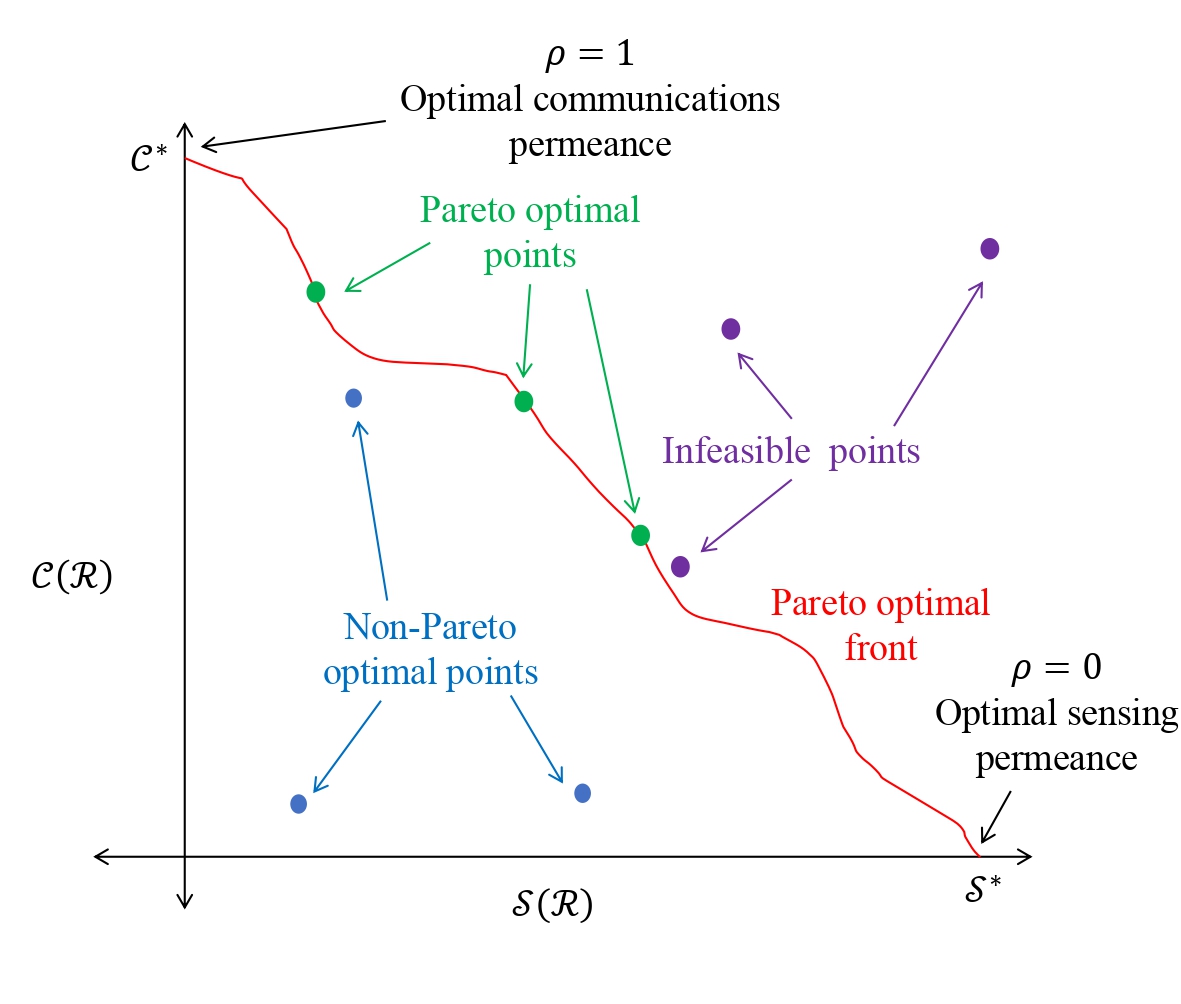}
        \caption{The Pareto optimal front produced by solving~\eqref{eq:MOO_2} when varying $\rho$ between zero and one.}
        \label{fig_part_opt}
\end{figure}

The second option is to create a new objective function that is function of both $\mathcal{C}(\mathcal{R})$ and $\mathcal{S}(\mathcal{R})$. A commonly used function is the weighted sum, which is expressed as:
\begin{equation}
    \mathcal{G}(\mathcal{R}) = \rho \mathcal{C}(\mathcal{R}) + (1-\rho) \mathcal{S}(\mathcal{R}),
    \label{eq:wsum}
\end{equation}
where $\rho \in [0,1]$ is a design parameter included to prioritize $\mathcal{C}(\mathcal{R})$ or $\mathcal{S}(\mathcal{R})$. At one extreme, selecting $\rho=1$ implies that the communication performance metric is solely optimized without considering the sensing performance. At the other extreme, setting $\rho=0$ implies that the optimization problem focuses only on the sensing aspect of the system. Varying $\rho$ and solving the following optimization problem:
\begin{subequations}
\begin{align}
           \max_{  \mathcal{R}} \ \ &  \mathcal{G}(\mathcal{R}) \\
        \text{s.t.} \ \ \
        &\mathcal{T}(\mathcal{R}),
\end{align} 
\label{eq:MOO_2}
\end{subequations}
\hspace{-0.27cm} produces what is known as the \textit{Pareto optimal front}~\cite{2018_Gunantara}. The points on the Pareto optimal front represent the optimal communication (or sensing) performance for a given value of the sensing (or communication) performance, as shown in Fig.~\ref{fig_part_opt}.

\subsection{WiFi ISAC}
The use of WiFi for sensing has garnered growing interest recently, particularly in human-centric applications such as sleep monitoring, fall detection, and health monitoring, alongside traditional uses like proximity detection and target counting. Detailed discussions on WiFi sensing, including fundamental concepts, principles, challenges, and applications, are presented in~\cite{2022_Chenshu}.

Notably, the IEEE formed a new task group, 802.11bf, tasked with developing an amendment that specifies the required physical layer and media access control (MAC) protocols to facilitate WiFi sensing across all spectrum bands. This includes the sub-7 GHz bands (2.4 GHz, 5 GHz, and 6 GHz) as well as the 60 GHz mmWave band~\cite{2023_Cheng}.

Researchers have advanced one step further by investigating the use of WiFi for ISAC. Specifically, target parameters such as AoA and AoD can be derived from channel state information (CSI), acquired through dedicated WiFi packets. However, this process poses challenges for existing WiFi systems, as they require high-quality CSI measurements. To overcome this issue,~\cite{2024_Yinghui} introduced an ISAC framework that integrates sensing functionality effectively. Their approach refines channel estimation using general WiFi packets, enabling CSI calibration across diverse WiFi communication modes and ensuring precise CSI measurements for upper-layer sensing applications. The authors also implemented a prototype and conducted extensive experiments involving 15 participants to demonstrate the effectiveness of ISAC deployment over WiFi.

\subsection{ISAC Standardization Efforts}
Standardization plays a crucial role in the progression of new technologies by establishing a cohesive framework that fosters collaborative efforts among developers, academics, industry. This subsection delves into the advancements in ISAC standardization achieved by the Third Generation Partnership Project (3GPP) and the European Telecommunication Standard Institution (ETSI).

\textit{1) 3GPP:} While fifth-generation (5G) cellular wireless networks primarily focused on delivering communication services, ISAC is poised to play a pivotal role in facilitating the evolution of inventive communication-assisted services. The initiation of ISAC standards involved an inaugural 3GPP study, which explored three scenarios: object identification and tracking, environment monitoring, and motion monitoring~\cite{2023_Kaushik}.

The second phase of the 3GPP (Release 19) study aims to conduct a more comprehensive identification of the fundamental requirements of ISAC. While current channel models mainly define various aspects of multi-path propagation, it is crucial, from a sensing perspective, to consider the reflectors of nearby objects and structures, as well as the impact of a particular cluster on both sensing and communications. Consequently, the initial step in the physical layer design of ISAC must involve an analysis of the channel model\footnote{More details about 3GPP's ISAC standardization efforts can be found at \url{https://portal.3gpp.org/desktopmodules/Specifications/SpecificationDetails.aspx?specificationId=4044}}~\cite{2023_Kaushik,2023_Haojin,2023_Ruiqi}.

\textit{2) ETSI:} The role of standardization in resolving interoperability issues, ensuring data integrity, and mitigating potential risks associated with ISAC deployment is examined within the context of European standardization. Specifically, ETSI has been at the forefront of standardization efforts related to networking and communication technologies. ETSI's standardization initiatives for ISAC prioritize four key areas. The first area concentrates on the design aspects of ISAC, encompassing waveform design, modulation, and beamforming. The second area focuses on standardizing the architectural aspects of ISAC, including MIMO and RIS. The third area addresses the capabilities of passive sensing using pure communication waveforms. Finally, the fourth area explores the utilization of ISAC at high-frequency bands, such as mmWave and THz bands~\cite{2023_Kaushik}.

Nevertheless, achieving harmony in standards proves to be challenging due to the extensive scope of ISAC, particularly considering its diverse application domains. Moreover, the rapid pace of technological advancement necessitates adaptable and forward-looking approaches to standardization. This approach would seamlessly integrate with other ongoing ETSI groups focused on emerging technologies, such as ETSI THz for THz communications and ETSI RIS for RIS-assisted communications\footnote{More details about ETSI's ISAC standardization efforts can be found at \url{https://portal.etsi.org/tb.aspx?tbid=912\#/}}~\cite{2023_Kaushik,2022_Jian2}.

\subsection{Metasurfaces}
Massive MIMO with large-scale phased arrays is considered one of the most promising technologies for 6G networks due to its ability to form highly directional beams and achieve spatial diversity. However, as massive MIMO evolves into ultra-massive MIMO, the cost and power consumption of phased arrays, reliant on high-resolution phase shifters, become prohibitive~\cite{2014_Larsson}. To overcome these limitations, a novel paradigm known as holographic radio has been proposed in recent years. This paradigm enables several integrated antenna elements at low hardware cost packed in a metasurface to achieve high directional gain.

An electromagnetic metasurface, also known as an RHS, is a surface composed of electromagnetic materials designed to exhibit capabilities not found in naturally occurring materials, as suggested by the Greek meaning of the word \textit{meta}, which means beyond~\cite{2019_Renzo}. Metasurfaces are made up of extremely small sub-wavelength scattering elements and are electrically large in transverse size~\cite{2015_Achouri}. Based on the holographic interference principle, the RHS antenna creates a holographic pattern on the surface using meta patches that record the interference between the desired object wave and the incident electromagnetic wave it generates, also known as the reference wave~\cite{2021_Deng}. The holographic pattern can modify the radiation characteristics of the reference wave as it propagates over the antenna surface, producing the intended radiation pattern. RHS can serve as a transmitter, receiver, or reflector, allowing for the creation of reconfigurable wireless environments~\cite{2020_Huang}.

 Based on hardware structure, RHSs can be classified into two categories~\cite{2020_Huang}.

\textit{1) Contiguous RHSs:} Contiguous RHS creates a spatially continuous transceiver aperture by integrating an extremely large number of elements onto a small surface area. For this, the concept of holography is utilized, which indicates the use of the interference principle of electromagnetic waves to record an electromagnetic field. The original field can then be reconstructed using the recorded electromagnetic field and the diffraction principle. A continuous aperture can achieve higher spatial resolution and allow the creation and the detection of electromagnetic waves with arbitrary spatial frequency components without undesirable side lobes. Contiguous RHSs can be seen as an asymptotic limit of massive MIMO by integrating an extremely large number of antennas.

\textit{2) Discrete RHSs:} A discrete RHS consists of several discrete unit cells constructed of software-tunable, low-power metamaterials. Various reconfigurable metamaterials, such as liquid crystals, micro-electromechanical systems, electromechanical switches, and off-the-shelf electronic components, can be used to electronically adjust the electromagnetic properties of the unit cells. Unlike contiguous RHSs, the aperture of a discrete RHS is not continuous. As a result, discrete RHSs differ from contiguous RHSs in terms of theory and implementation.

\subsection{Reconfigurable Intelligent Surfaces}
RISs are the most commonly used type of RHSs and represent the utilization of RHSs as separate entities between the transmitter and receiver. One motivation for the use of RISs comes from the nature of mmWave and THz propagation channels. The quasi-optical nature of signals in these bands leads to channels that are predominantly LoS, with only a few weak NLoS components surviving. This results in channel matrices that are ill-conditioned and highly correlated~\cite{2021_Sarieddeen}. Additionally, the severe path loss, especially when the LoS is obstructed, makes the NLoS components unreliable for communication over reasonable distances~\cite{2019_Xing}. To tackle these challenges, there has been a growing focus on the use of RISs, sometimes referred to as intelligent reflecting surfaces (IRSs). These surfaces consist of planar arrays comprising a large number of low-cost, nearly-passive meta-elements that can dynamically alter the phase, and sometimes the amplitude, of incident signals and provide additional passive beamforming gains\footnote{While the meta-atoms at RISs do not require power amplifiers, the tuning circuits that configure the RIS phase shifts consume a small amount of power. This makes RISs \textit{nearly-passive} rather than \textit{passive}.}. 

Various accurate models incorporating electromagnetic analysis have been proposed to characterize RIS-aided channels. These models consider factors such as RIS power loss~\cite{2022_Tang}, RIS atom coupling~\cite{2021_Gradoni}, RIS amplitude-phase coupling, frequency response of individual RIS elements~\cite{2022_Bjornson}, and near-field operation~\cite{2020_Emil}. However, a common simplified model is to represent the RIS matrix as a diagonal matrix. In this representation, independent diagonal elements are assumed, each with a unit modulus and an adjustable phase shift in the range $[ 0, 2 \pi )$.

In assessing the added value of RISs, it is worth discussing two important laws. 

\textit{1) The square law:} In the far-field region, the power received by the RIS from the transmitter is directly proportional to its surface area. Therefore, having a large number of RIS elements $N$ results in a larger surface area and, consequently, higher received power. The RIS then reflects the signal after adjusting the phases, leading to an additional array gain proportional to $N$~\cite{2020_Bjornson_Comm_Mag}. These two effects combined establish what is known as the \textit{square law}, which states that in a single-user scenario, the received SNR is proportional to $N^2$, where $N$ is  the number of RIS elements. For instance, doubling the number of RIS elements can result in a power gain of \unit[6]{dB}.

\textit{2) The double path-loss law:} In the far-field region, the power received at a user via the RIS channel is inversely proportional to the product of the path loss from the transmitter to the RIS and the path loss from the RIS to the receiving user~\cite{2021_Wu}. Consequently, in medium-to-long-distance communications over high frequencies, the reflected path remains significantly weaker than the direct path. In such situations, while the reflected path may not substantially enhance system performance, it can still be valuable when the LoS link is obscured.

RISs initially emerged as passive surfaces capable of adjusting the phases of incident signals. However, various other types of RIS have been proposed to address different scenarios. 

\begin{figure*}
  \centering
  \begin{tabular}{c c}
    \includegraphics[width=0.9\columnwidth]{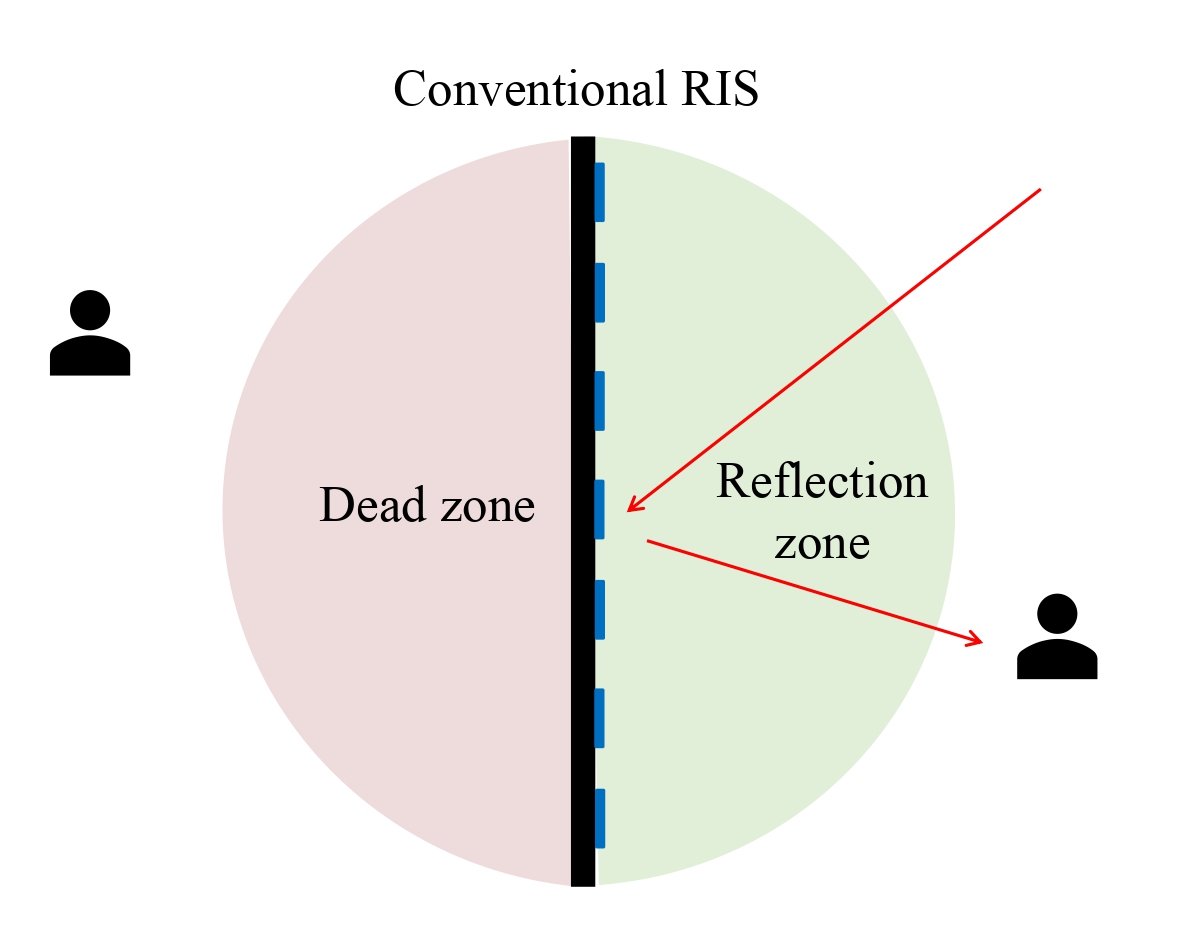} &
      \includegraphics[width=0.9\columnwidth]{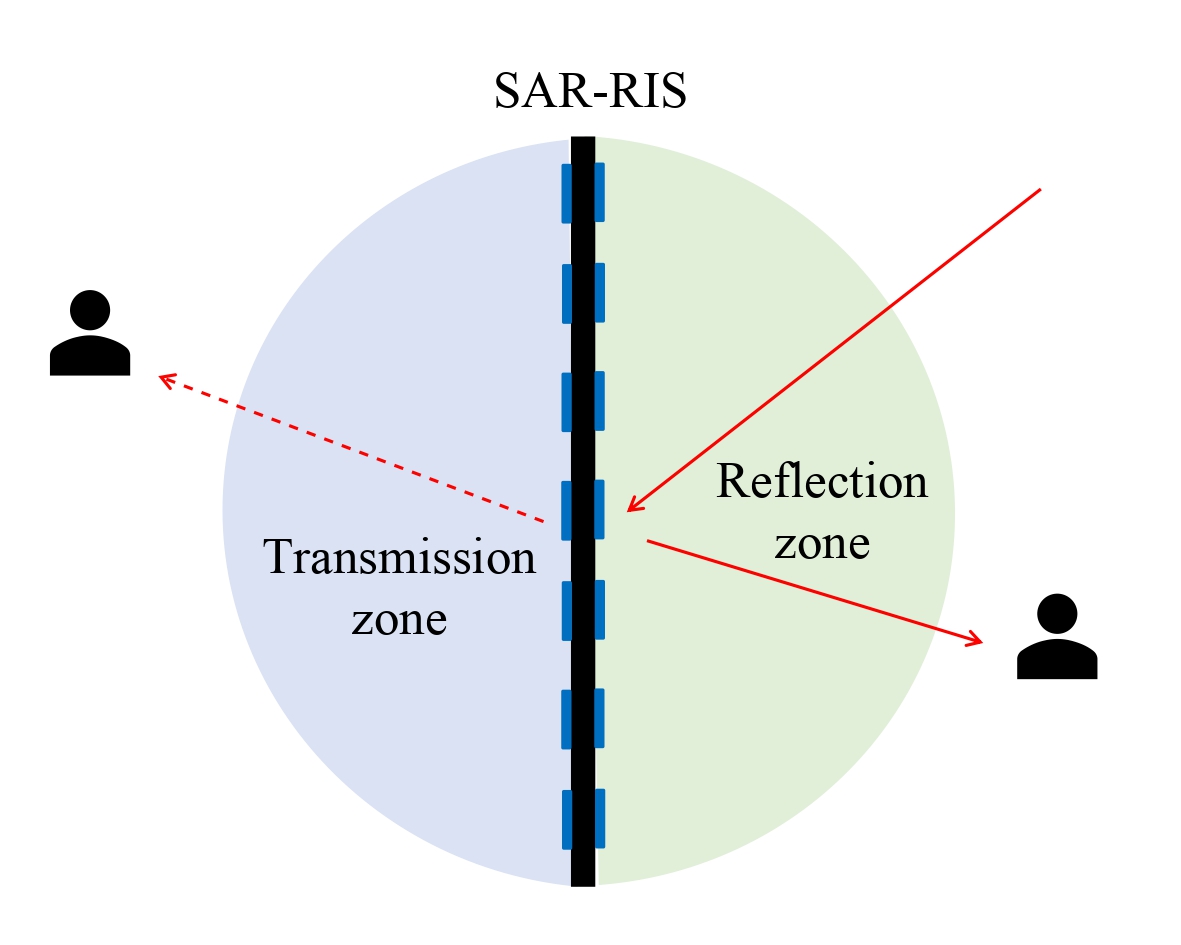}  \\
      \scriptsize (a)  &
      \scriptsize (b)\\
  \end{tabular}
    \medskip
  \caption{(a) The reflection zone and dead zone of a conventional RIS, and (b) the reflection and transmission zones of a STAR-RIS. STAR-RISs can extend the converge of conventional RISs from $180^\circ$ to $360^\circ$.}
  \label{fig:STAR_RIS}
\end{figure*}

\textit{1) Simultaneous transmitting and reflecting RISs (STAR-RIS):} The coverage zone of conventional RISs spans $180^\circ$ as shown in Fig.~\ref{fig:STAR_RIS}(a). This indicates that conventional RISs are blind to users located in their dead zones. In contrast, STAR-RISs, also referred to as omni-surfaces, can extend the coverage to span the entire space by deploying RIS elements on both sides of the surface~\cite{2021_Xu}. Each receiving element in a STAR-RIS is connected to two transmitting elements, one on each side of the surface as shown in Fig.~\ref{fig:STAR_RIS}(b). This creates two separate zones: the reflection zone and the transmission zone, each with its own RIS phase shift matrix.
    
    There are three main protocols that govern the operation of STAR-RISs. The first of these is the energy splitting protocol, which splits the signal between the reflection and transmission zones. The energy splitting protocol also requires a phase coupling model that characterizes the relationship between the angles in the reflection and transmission zones~\cite{2022_Yuanwei}. The second operational protocol is mode switching, which indicates that a subset of the RIS elements functions as fully reflective, while the rest operate as fully transmissive elements. The third operational protocol is the time splitting, which indicates that all RIS elements function as reflective elements for a specific fraction of the communication frame, and function as transmissive elements for the remainder of the frame.

\textit{2) Beyond-diagonal RISs (BD-RISs):} More recently, the authors of~\cite{2023_Li2} proposed eliminating the diagonal condition in the RIS matrix. From a hardware perspective, this is equivalent to allowing each receiving RIS element to connect with all of the transmit RIS elements. While this introduces additional hardware complexity, simulation results in~\cite{2023_Li5} show that this can lead to a notable improvement in performance. BD-RISs are also passive, and as a result they should adhere to the energy conservation condition. To address the hardware complexity challenge in BD-RISs, a group-connected architecture can be employed, where BD connection are employed within each group~\cite{2023_Li3}.

\textit{3) Active and absorptive RISs:} Some research works have explored the effect of adding amplification capabilities to the RIS. Thus, active RISs can address the double path-loss issue by deploying an amplifier with each RIS element~\cite{2023_Zhang3,2022_Zhi}. Due to the maximum power constraint imposed by the amplifiers, the amplification factor of each RIS elements should not exceed a certain threshold. However, signal models for active RISs differs from those of passive ones, primarily in two aspects. The first is the power budget constraint encompassing the overall system, including the transmitter, the receiver, and the RIS. The other difference is the non-negligible thermal noise, which is typically ignored in the case of a passive RIS as it has a negligible effect~\cite{2023_Zhang3}. 


Similar to active RISs, absorptive RISs can control the amplitude of the incident signals in a passive manner. This indicates that the amplitude amplification factors in absorptive RISs are confined to the range $[0, 1]$~\cite{2022_Albanese}. Hence, the noise model of absorptive RIS is similar to that of passive RIS, and no thermal noise needs to be accounted for~\cite{2023_wang4}. Absorptive RISs have been shown to be more effective at mitigating MUI compared with conventional passive RISs~\cite{2023_wang6}.

\subsection{Stacked Intelligent Metasurfaces}
A SIM consists of multiple layers of metasurfaces and is able to facilitate signal processing directly within the wave domain. SIMs can be utilized for two purposes: they can be integrated into transmitters and/or receivers for holographic beamforming and wave-domain signal processing, or they can function as independent components between the transmitter and receiver, offering improved channel reconfigurability compared to single-layer RISs.

\begin{figure*}
  \centering
  \begin{tabular}{c c}
    \includegraphics[width=0.9\columnwidth]{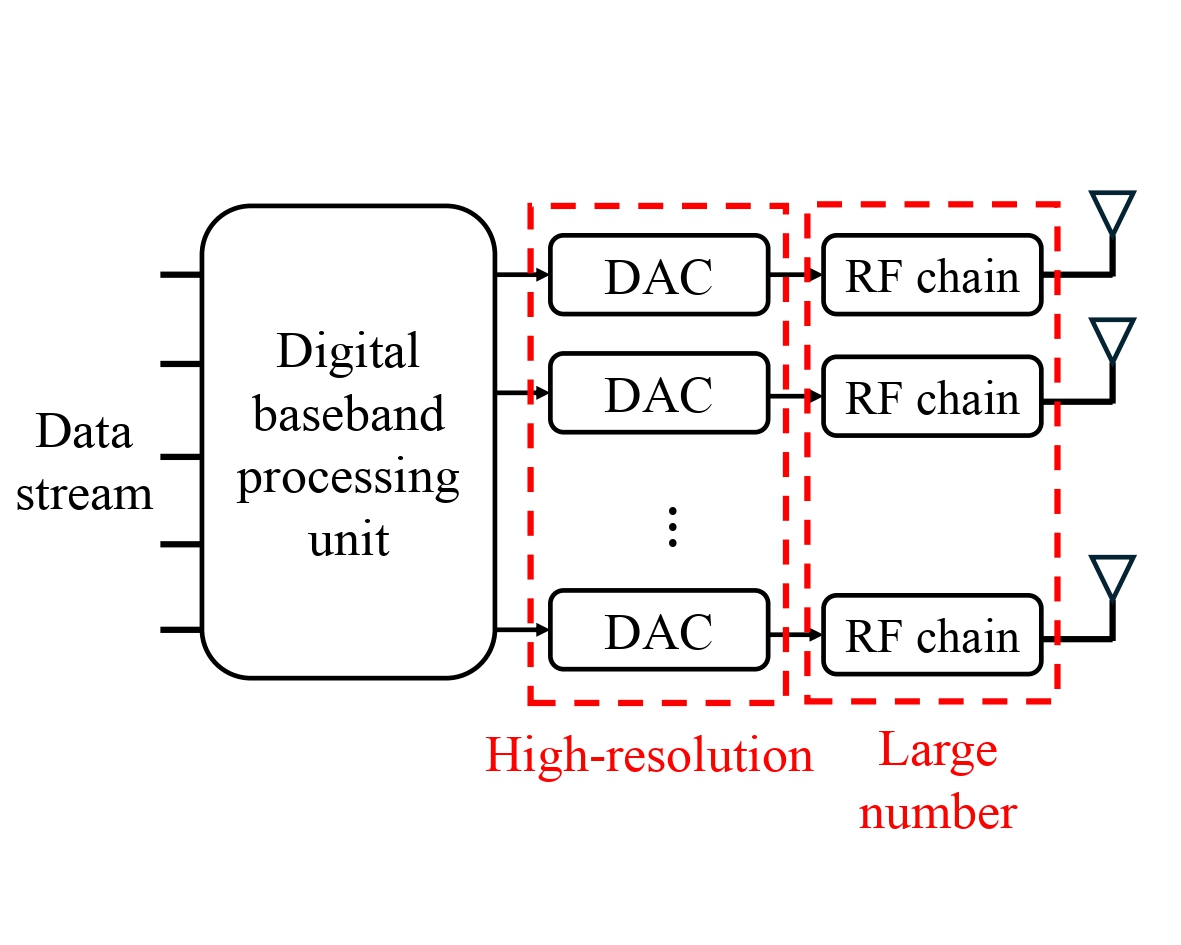} &
      \includegraphics[width=0.9\columnwidth]{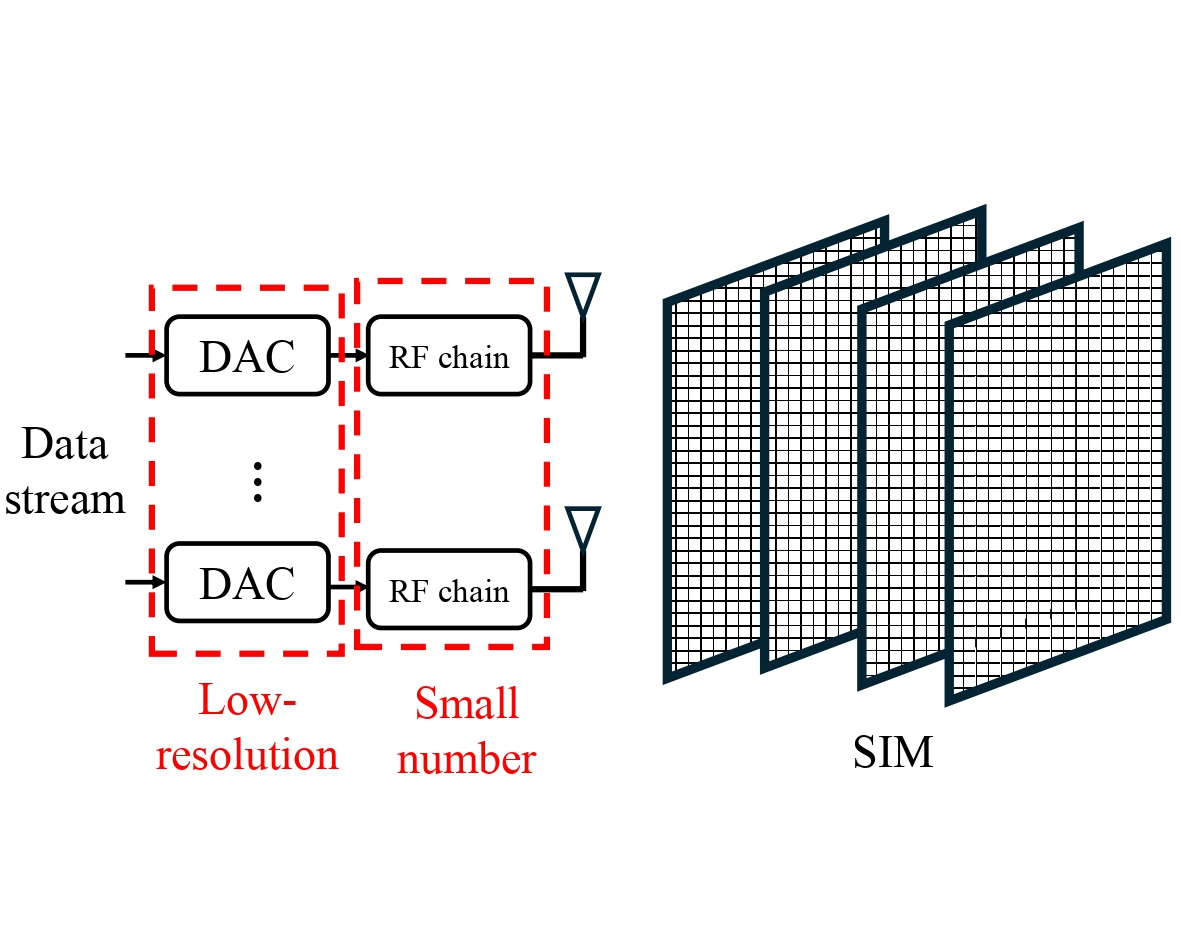}  \\
      \scriptsize (a)  &
      \scriptsize (b)\\
  \end{tabular}
    \medskip
  \caption{(a) A conventional MIMO transmitter, and (b) a SIM-based MIMO transmitter. }
  \label{fig:SIM}
\end{figure*}

\textit{1) SIMs as transmitters and/or receivers:} Conventional MIMO transceivers aim to transmit multiple data streams by first precoding them and then sending a superimposed signal to the transmit antennas. At the receiver side, a combiner is used to recover the different spatial streams. This process requires computationally demanding digital beamformers, high-resolution digital-to-analog (DACs) converters and analog-to-digital converters (ADCs), and a large number of RF chains~\cite{2024_Liu1} as shown in Fig.~\ref{fig:SIM}(a).

A SIM-based architecture was recently introduced as an alternative to reduce the hardware costs associated with conventional MIMO systems. The new architecture, depicted in Fig.~\ref{fig:SIM}(b), eliminates the need for a power-hungry digital baseband processing unit and replaces it with a SIM. It can also achieve similar performance to a MIMO transceiver using low-resolution DACs and fewer RF chains~\cite{2023_An1}. A key feature of SIM-based transceivers is wave-domain beamforming, which utilizes the parallel processing capabilities and ultra-low latency of the multi-layer diffractive structure of the SIM, making it a promising solution for next-generation wireless networks~\cite{2024_Liu1}.

\textit{2) SIMs as refracting devices:} Similar to RISs, SIMs can be positioned between the transmitter and receiver to add controllability to the communication channels. The first layer of a SIM may also partially reflect (and partially refract) the incident signals, thus operating as a reconfigurable device with controllable reflected and transmitted signals. Due to their multiple layers, SIMs are more effective at enhancing system performance compared to single-layer RISs~\cite{2023_An1}. Furthermore, by leveraging their programmable multi-layer architecture, SIMs can implement deep neural network functions as electromagnetic waves pass through them. However, the inference capability of a SIM is limited by the linearity of its transfer function~\cite{2024_Liu1}.

\section{RIS-Assisted RCC} \label{sec:RIS_RCC}
The vast majority of the research work in metasurface-assisted ISAC has focused on employing metasurfaces as separate entities positioned between the transmitter and the receiver, i.e., as RISs. Considering this, we first present the state of the art in the area of RIS-assisted RCC in this section. A summary of the research works presented in this section is provided in Table~\ref{tab:RIS_RCC}.

\subsection{Optimal Resource Allocation}
The underlying principle of optimal resource allocation, as elaborated in Section~\ref{sec:ISAC_RA}, involves the co-design of available resources, including RIS phase shifts, to optimize a designated performance metric for either communications or/and sensing.

Motivated by the announcement from the Federal Communications Commission (FCC) approving the use of the \unit[3.5]{GHz} band for communication purposes alongside its traditional application in radar sensing~\cite{2018_Rihan}, the authors in~\cite{2021_Wang2} pioneered the use of RIS for interference mitigation in RCC. Their primary objective was to enhance the radar detection probability in an RIS-assisted RCC system with multiple communication users and a single radar target, while adhering to quality-of-service (QoS) constraints on the communication SINRs. To tackle this optimization problem, the authors in~\cite{2021_Wang2} employed a semi-definite relaxation (SDR) approach based on quadratically constrained quadratic programming (QCQP). The simulation results presented in~\cite{2021_Wang2} demonstrate that the use of RIS can offer a substantial improvement in the detection probability, particularly when the SINR thresholds of the communication users are high. Consequently, both the communication and the sensing systems can achieve excellent performance. For instance, the detection probability can increase from 0.6 to 0.85 with the assistance of the RIS, which can further improve to 0.98 by increasing the number of RIS elements.

While this approach holds a very good potential for improving the performance of RCC systems in the sub-\unit[6]{GHz} band, its effectiveness at higher frequencies (e.g., mmWave and THz) was not established. In~\cite{2021_Shtaiwi}, the authors investigated a mmWave system with communication users and radar targets suffering from blocked LoS links. In such a scenario, the RIS serves a dual purpose. First, it becomes the sole means of communication and sensing, and second, it is used to eliminate mutual interference between the two systems. The objective in~\cite{2021_Shtaiwi} was to maximize the communication sum rate with a minimum disruption to the radar functionality. To achieve this, a constraint is imposed on the radar's power budget to keep interference power below a specified threshold. The paper presents two methods to obtain the optimal solution. The first is exhaustive search based, providing excellent performance but with high computational complexity, and the second is local search based, offering a sub-optimal solution with lower computational complexity. In an extended journal version~\cite{2023_Shtaiwi}, the authors introduced a more efficient manifold-based optimization approach to find the optimal solution.

\textbf{Lessons learned:} Research works have demonstrated that RISs can be effectively utilized to improve interference mitigation in RCC systems, which helps in boosting the performance of both radar and communication systems. Moreover, in systems with blocked direct links, RIS serves as both a communication and sensing enabler, as well as an interference mitigator. Nevertheless, the inclusion of an RIS generally requires more advanced signal processing and optimization techniques. Time complexity remains an important factor in this process, as an RIS with a large number of elements is usually required for reasonable performance improvement. This makes approaches such as exhaustive search unfavorable despite their excellent performance. Thus, a sub-optimal solution may offer a satisfactory balance between performance and complexity.

\textbf{Challenges, opportunities and open research directions:} Considering the intricate nature of optimization problems in RIS-assisted RCC systems and the interdependence of the optimization variables in the objective function and certain constraints, the adoption of alternating optimization (AO) is a conventional approach for establishing a solution framework for these optimization problems. It is worth noting, however, that AO algorithms can only converge to a sub-optimal solution~\cite{2003_Bezdek}. To achieve convergence to satisfactory sub-optimal solutions (or even the global solution), optimizing more than one variable jointly is essential. Another strategy for addressing the highly non-convex RCC problems involves employing multiple initializations and selecting the resulting resource allocation that leads to the best sub-optimal solution. Additionally, radar and communication systems can exchange valuable information, resulting in a lower interference level between the two systems. For instance, if the radar system shares its probing signal with the communication system before data transmission, the communication BS will have a greater capability of detecting the interfering radar signal and cancel its interference. Additionally, radar and communication systems can coordinate in using the RIS, enabling enhanced performance for each system during specific time slots.

\begin{table*}
\footnotesize
\centering
\caption{References on RIS-assisted RCC.}
\begin{tabular} {|m{2.2cm} | m{1cm}| m{0.5cm} |m{1cm}| m{1cm}| m{1.8cm} | m{1.7cm} | m{1.7cm} | m{3cm}|} 
 \hline 
 Main goal & Reference & Year  & Frequency band  & LoS availability & Communications metric & Sensing metric & Objective function & Methods employed  \\  [0.5ex] 
 \hline 
 \hline 
  Optimal resource allocation 
 & \cite{2021_Wang2} & 2020 & Sub-\unit[6]{GHz}  & Available & SINR & Detection probability & Detection probability & SDR - QCQP  \\   
 \cline{2-9}
  & \cite{2021_Shtaiwi} & 2021  & mmWave  & Not available & Sum rate & Sensing power & Sum rate  & Exhaustive search - Local search  \\ 
 \cline{2-9}
 & \cite{2023_Shtaiwi} & 2023 & mmWave  &  Not available & Sum rate &  Sensing power &  Sum rate & Lagrangian dual transformation - Penalty-based optimization - Manifold optimization - Riemannian conjugate gradient  \\ 
 \hline
 Optimal resource allocation with imperfect CSI & \cite{2023_Rihan} & 2023 & mmWave  &  Available & Communications rate - Outage probability & Receive radar SINR & Receive radar SINR & Exhaustive search - Local search - SCA - Relaxation and projection method \\ 
  \hline 
  Optimal resource allocation with multiple RISs & \cite{2022_He} & 2022  & sub-\unit[6]{GHz}  & Available & Average communication SINR & Receive radar SINR & Average communication SINR & PDD - Concave-convex procedure  \\   
  \cline{2-9}
 & \cite{2024_Mengyu} & 2024 & sub-\unit[6]{GHz}  &  Available & Achievable data rate & Radar SINR &  Achievable data rate & PDD - block coordinate descent - Lagrange dual method \\ 
 \hline 
  Optimal resource allocation with absorptive RISs & \cite{2023_wang4} & 2023  &  mmWave & Available & Radar interference mitigation & Communications interference mitigation & Interference suppression  & Fractional QCQP - SDR - Dinkelbach's algorithm \\   
 \cline{2-9}
   & \cite{2023_wang6} & 2023   &  mmWave & Available  & Radar interference mitigation  & Communications interference mitigation & Interference suppression & Fractional QCQP - SDR - Dinkelbach's algorithm  \\ 
 \hline 
  Optimal BS and RIS placement  & \cite{2022_Kafafy} & 2022 & sub-\unit[6]{GHz}  & Available & Coverage probability & Detection probability & Coverage probability & Genetic optimization - Selective Search  \\    
   \hline 
  Optimal resource allocation with STAR-RISs & \cite{2024_Papazafeiropoulos} & 2024  &  sub-\unit[6]{GHz} & Available & Spectral efficiency &  Radar SINR & Spectral efficiency  &AO - Projected gradient ascent \\   
     \hline 
  Optimal resource allocation for image analysis-focused RIS-assisted RCC & \cite{2024_Ning} & 2024  &  sub-\unit[6]{GHz} &Available for communication only & Image analysis accuracy &  Radar estimation information rate & Image analysis accuracy and radar estimation information rate  & AO - SCA \\  
 \hline 
\end{tabular}
\label{tab:RIS_RCC}
\end{table*}

\subsection{Imperfect CSI}
To evaluate the performance of an RIS-assisted RCC system in the presence of channel uncertainties,~\cite{2023_Rihan} investigated a scenario where full CSI is not available at the transmitter.

Recognizing that channel estimation errors usually occur in the RIS-user link due to user mobility, the method of~\cite{2023_Rihan} employs a statistical mean feedback model for the RIS-user link instead of relying on instantaneous channel information. The authors in~\cite{2023_Rihan} presented two cases. First, in the perfect CSI scenario, the objective is to optimize the receive radar SINR while maintaining a constraint on the communications rate. In the statistical CSI case, on the other hand, the goal is to maximize the average radar SINR while adhering to an outage probability constraint for the communication user. The optimization problems are addressed using a method that combines local search and successive convex approximation (SCA) along with relaxation and projection methods.

\textbf{Lessons learned:} Assuming perfect CSI for RIS-aided systems is impractical. Neglecting factors such as channel uncertainty can diminish the potential improvements offered by RIS, emphasizing the importance of holistic system design. One solution to account for channel uncertainties is to employ statistical models, which have been shown to be effective in realistic deployment scenarios where taking channel estimation error into account can restrict potential performance degradation. However, it remains crucial to accurately characterize the channel estimation error model using using appropriate statistical modeling techniques.

\textbf{Challenges, opportunities and open research directions:} Despite the critical significance of precise CSI imperfection models, research in this area remains relatively limited. Given that passive RISs lack the capability to transmit or process pilots for channel estimation, a predominant approach in the RIS-assisted channel estimation literature involves the estimation of the cascaded channel. This leads to channel estimation errors affecting both the transmitter-RIS channel and the RIS-receiver channel. Consequently, more advanced models must be employed to articulate the nature of these estimation errors.  In addition, various channel models, such as the Saleh-Valenzuela (SV) channel model~\cite{ch_model_2}, utilize AoA and AoD information to formulate far-field channels. Consequently, sharing the estimated AoA and AoD information between the radar and communication systems presents an opportunity to mitigate channel estimation errors.

\subsection{Multiple RISs}
\begin{figure} 
         \centering
    \includegraphics[width=0.9\columnwidth]{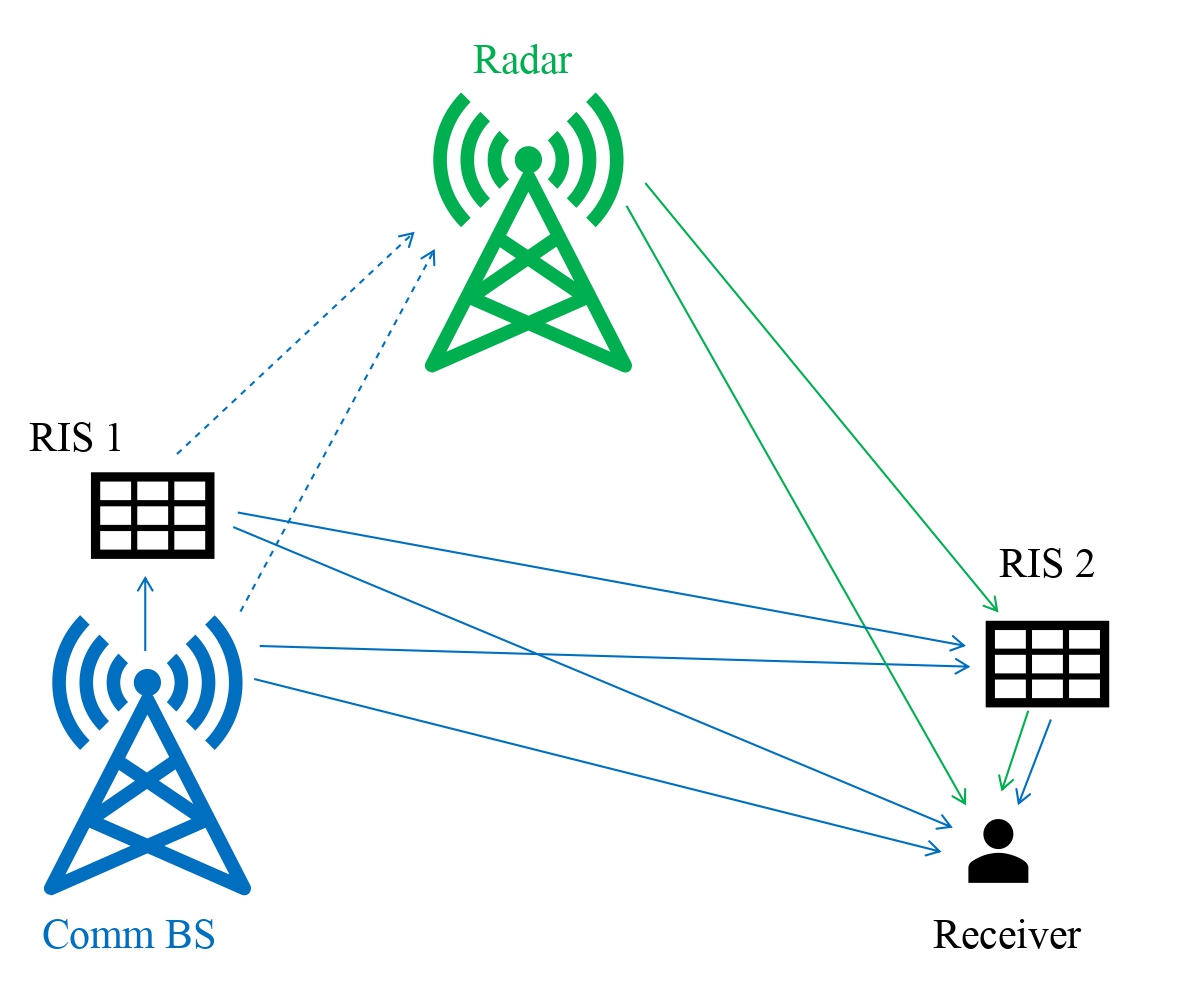}
        \caption{Multiple-RIS-assisted RCC system model proposed in~\cite{2022_He}.}
        \label{Fig:MRIS_RCC}
\end{figure}

\textit{1) Multiple passive RISs:} In RCC systems, both the radar and communication receivers encounter interference. To tackle this problem, the authors of~\cite{2022_He} introduced the use of two 
RISs as shown in Fig.~\ref{Fig:MRIS_RCC}. One RIS is placed in proximity to the communication transmitter, while the other is positioned near the communication receiver. The purpose of situating an RIS close to the transmitter is to alleviate interference from the communication transmitter to the radar receiver. Meanwhile, the RIS near the communication receiver is employed to suppress interference between the radar transmitter and the communication receiver. This setup proves beneficial when the communication transmitter and receiver either remain stationary or move within confined regions.

The optimization problem in~\cite{2022_He} is formulated to maximize the average communication SINR while adhering to a constraint on the receive radar SINR. The latter, while not a direct radar performance metric, is positively correlated with many radar performance metrics, including the detection probability. To address this optimization problem, the authors proposed a solution based on the penalty dual decomposition (PDD) method and concave-convex procedure. 

\textit{2) Multiple active RISs:} To address the double path-loss issues at the RIS link, the authors of~\cite{2024_Mengyu} presented a system model with two active RISs. The goal was to design the beamformer at the BS and the active RISs' amplification factors and reflection coefficients to maximize the achievable data rate of the communication system while adhering to a radar detection constraint. The authors utilized the PDD method to incorporate the challenging constraints into the objective function. The optimization problem was then decoupled into several smaller sub-optimization problems using an AO framework. It wa shsown in~\cite{2024_Mengyu} that, under the same power budget, deploying double active RISs in RCC systems can achieve 2.1 and 1.6 times higher data rates compared to single active RIS and double passive RISs, respectively.

\textbf{Lessons learned:} Deploying RISs strategically can significantly reduce interference in RCC systems. Placing an RIS near the communication transmitter helps to alleviate interference from the communication transmitter to the radar receiver. Similarly, positioning an RIS near the communication receiver suppresses interference from the radar transmitter to the communication receiver. This motivates the dual-RIS setup to enhance overall system performance by mitigating mutual interference, allowing both radar and communication systems to operate more effectively. This configuration is particularly beneficial in scenarios where the communication transmitter and receiver are stationary or move within confined regions. Using two active RISs in the system model can further address the double path-loss phenomenon encountered in RIS-assisted links. 

\textbf{Challenges, opportunities and open research directions:} While employing multiple RISs contributes to improved system performance, it introduces an additional complexity burden in terms of channel estimation, pilot overhead and resource allocation. Consequently, the feasibility of such systems, along with a thorough examination of their benefits and drawbacks, should be critically studied.

\subsection{Optimal RIS Placement}
While the previous papers primarily focused on the problem of resource allocation, the study in~\cite{2022_Kafafy} aimed at identifying the optimal positions for the communication BS and the RIS. The objective was to maximize the communication coverage probability for users located within a radar exclusion zone while ensuring a specific radar performance, which was modeled as a constraint on the radar detection probability.

The paper first analyzes the communication coverage probability and radar detection probability. It was observed that the presence of an RIS leads to a lower detection probability but a higher coverage probability, as the RIS is employed solely to enhance communication performance. To minimize the additional interference in the radar system resulting from the presence of the RIS, the distances between the radar and the communication BS, and that between the BS and the RIS, were optimized. The problem was formulated to maximize the coverage probability while maintaining a radar detection probability equivalent to that of a scenarios with no RIS. The authors utilized a selective search method based on a genetic approach to determine the optimal placements for the BS and RIS.

\textbf{Lessons learned:} Optimizing the placement of the RIS is important for achieving excellent communication and radar performance. Proper placement can help to achieve the desired balance between communication and radar performance, indicating that spatial configuration is a significant factor in system design. Optimal RIS placement can guide the deployment of communication infrastructure in environments where both communication and radar systems coexist, such as in urban areas near airports or military installations. This should also take into consideration the specific application scenarios in order to have an accurate system representation, such as the distribution of communication users and radar receivers, for accurate problem formulation.

\textbf{Challenges, opportunities and open research directions:} While various papers have explored the optimal placement of RISs for different purposes and system models, the complexity of this problem surpasses that of resource allocation optimization problems. This additional complexity arises from the fact that the objective function and the constraints, comprising the communication and radar performance metrics, can be easily expressed as functions of resources. Nevertheless, expressing these functions in terms of the location in the Cartesian plane requires additional derivations. Moreover, strategic RIS placement considerations generally extend beyond instantaneous decisions, necessitates the adoption of stochastic system modeling tools, such as stochastic geometry with accurate distributions describing the system model.

\subsection{STAR-RIS}
The authors of~\cite{2024_Papazafeiropoulos} proposed using STAR-RISs to extend the RIS coverage to $360^\circ$. Taking into account realistic conditions such as correlated fading and the presence of multiple users on both sides of the RIS, the authors derived closed-form expressions for the achievable rates of both the radar and the communication receivers using statistical CSI. Notably, one of the main findings of~\cite{2024_Papazafeiropoulos} is that spatial correlation can improve the STAR-RIS's capability to customize the wireless channel and manipulate propagation.

The authors then formulated a problem to maximize the spectral efficiency of the communication system while setting a threshold on the radar SINR. To address this non-convex optimization problem, an AO-based approach was employed to find the optimal STAR-RIS matrix and radar beamforming. For the STAR-RIS matrix, the optimal amplitudes and phase shifts were determined using the projected gradient ascent method for both energy splitting and mode switching protocols. It was shown that that the STAR-RIS architecture outperforms conventional RIS systems and the impact of various parameters on performance was highlighted. Specifically, in the moderate to high transmit SNR regime, the spectral efficiency of the STAR-RIS-assisted RCC system can be improved to 10.5 and 9.5 bit/sec/Hz, compared with 8 bit/sec/Hz for conventional RISs.

\textbf{Lessons learned:} Spatial correlation can enhance the STAR-RIS's ability to tailor the wireless channel and manipulate signal propagation, leading to better communication and radar performance under accurate system performance characterization. Moreover, STAR-RIS architectures outperform the conventional RIS systems as they allow for full space coverage and better propagation control.

\textbf{Challenges, opportunities and open research directions:} The complexity in optimizing STAR-RIS matrices and radar beamforming, particularly in non-convex problem settings, requires sophisticated algorithms and significant computational resources, especially with the presence of the constraints related to the operational protocols. Additionally, managing mutual interference and ensuring energy efficiency are significant concerns, alongside the technological demands and costs of advanced hardware capable of simultaneous transmission and reflection. Despite these challenges, STAR-RIS offers enhanced coverage and improved spectral efficiency, making it a promising solution for dense urban environments and high user density areas. Security, privacy concerns, and the integration of machine learning for dynamic optimization also represent important research directions to fully harness the potential of STAR-RIS in RCC systems.

\subsection{Image Analysis}
Unlike the RIS-assisted RCC works discussed earlier, an image analysis-focused RIS-assisted RCC system was proposed in~\cite{2024_Ning}. A wireless image sensor was designed to perform image sensing tasks, which involve capturing and transmitting images or visual data using wireless communication. Additionally, an edge server is connected to the BS to facilitate image analysis. The authors aimed to offload the captured image data to the edge server through uplink (UL) data transmission. This creates a more complex scenario involving co-channel interference between the radar echo signal and the UL data transmission. The authors analyzed the mutual interference between radar sensing and the offloading of image data, providing a quantitative assessment of the radar estimation information rate as a performance metric for radar sensing. They then formulated an optimization problem to maximize overall system performance by creating an objective function that incorporates both the radar estimation information rate and image analysis accuracy, which is used to design the RIS phase shift, image resolution, and transmit power of the image sensor. To determine the optimal solution for the formulated problem, the authors employed a two-tier AO-based SCA algorithm.

\begin{figure} 
         \centering
         \includegraphics[width=1\columnwidth]{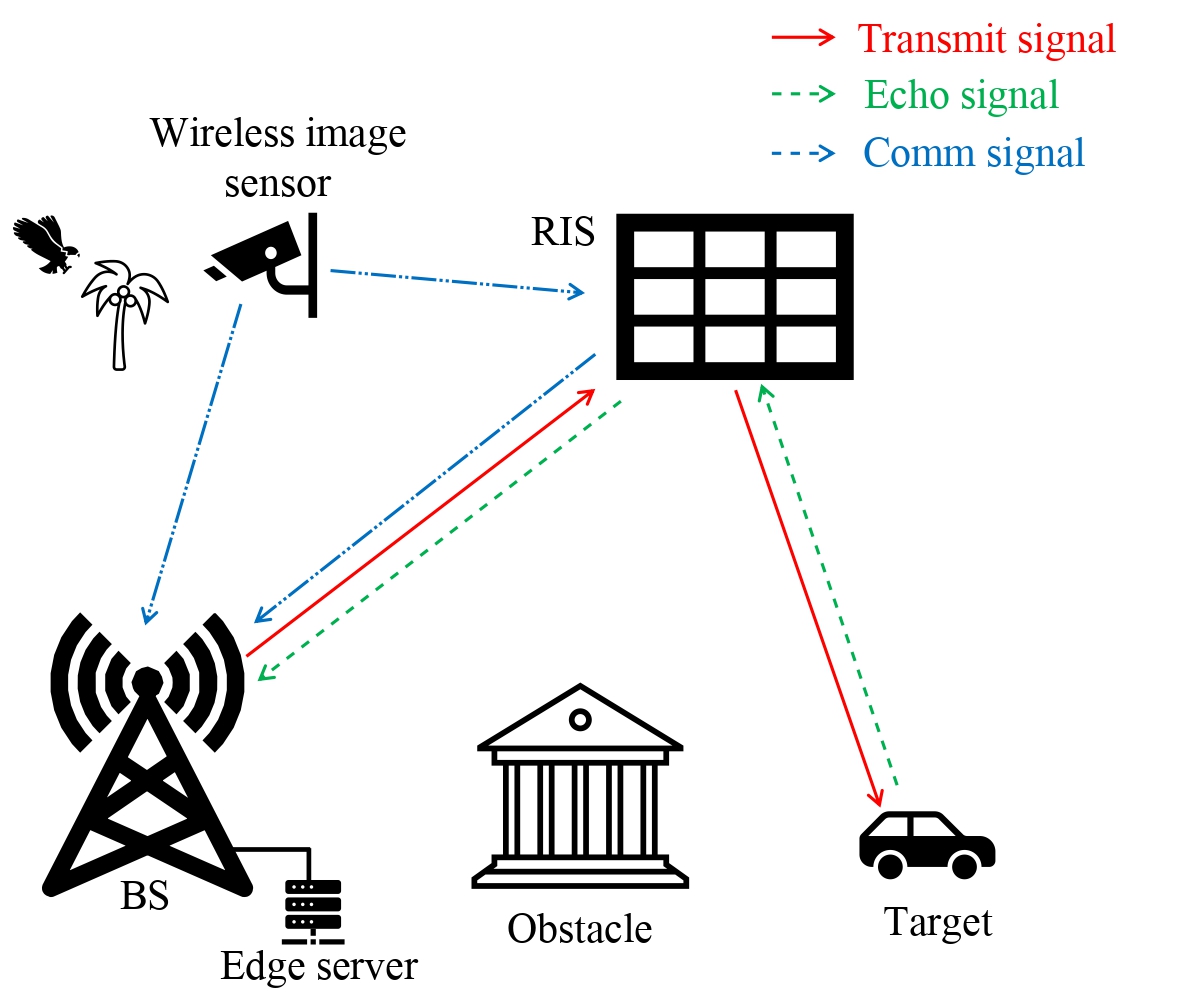}
         \vspace{0.05cm}
        \caption{Image analysis-focused RIS-assisted RCC system proposed in~\cite{2024_Ning}.}
        \label{fig:image}
\end{figure}

\textbf{Lessons learned:} The study of image analysis-focused RIS-assisted RCC systems underscores the effectiveness of using RISs to reduce the  mutual interference between radar sensing and image data offloading. However, this highlights the need for a effective optimization approaches that considers both radar estimation information rate and image analysis accuracy. Overall, this work exemplifies the promising synergy between radar and wireless image sensing technologies, paving the way for enhanced performance in various applications.

\textbf{Challenges, opportunities and open research directions:} RIS-assisted RCC systems with image analysis present several challenges, opportunities, and open research directions. One significant challenge is managing the co-channel interference between radar echo signals and UL image data transmission, which requires advanced optimization techniques to ensure effective performance. This situation creates opportunities for developing innovative algorithms that enhance the balance between radar sensing and image data offloading, particularly in dynamic environments. Additionally, there is potential for exploring the impact of various environmental factors on system performance and the effectiveness of different configurations of RIS and image sensors. Open research directions include investigating the scalability of these systems in large networks, improving energy efficiency, and addressing practical implementation challenges, such as hardware limitations and real-time processing requirements.

\section{RIS for Basic DFRC Operations} \label{sec:RIS_DFRC1}
DFRC allows for a greater degree of integration between communications and sensing compared to RCC at the hardware, spectrum, waveform, and beamforming design levels. In this section,we describe the exploitation of RISs for basic DFRC operations.

\begin{table*}
\footnotesize
\centering
\caption{References on waveform design and optimal resource allocation for RIS-assisted DFRC.}
\begin{tabular} {|m{2.2cm} | m{1cm}| m{0.5cm} |m{1cm}| m{1cm}| m{1.8cm} | m{1.7cm} | m{1.7cm} | m{3cm}|} 
 \hline 
 Waveform design & Reference & Year  & Frequency band  & LoS availability & Communications metric & Sensing metric & Objective function & Methods employed  \\  [0.5ex] 
 \hline 
 \hline 
  Beamforming-based joint design & \cite{2022_Song} & 2022  & sub-\unit[6]{GHz}  & Not available & Communications SNR & Minimum radar beampattern gain & Minimum radar beampattern gain & SDR  - Gaussian randomization - CVX toolbox \\    
  \cline{2-9}
  & \cite{2022_Zhang1} & 2022 &   sub-\unit[6]{GHz}  & Available & Sum rate & Radar mutual information & Weighted sum of communications sum rate and radar mutual information & SCA - the Lagrangian decomposition - Riemannian optimization  \\
  \cline{2-9}
  & \cite{2022_Zhengyu} & 2022 &  mmWave  & Available & Sum rate & Radar beampattern match & Sum rate & Quadratic transformation - MM  - Manifold optimization  \\
   \cline{2-9}
    & \cite{2022_Zhu} & 2022 &  mmWave  & Available & Sum rate & Radar beampattern match & Sum rate & Quadratic transformation - Eigenvalue decomposition - Lagrangian multiplier method - QCQP \\
    \cline{2-9}
   & \cite{2022_Yan} & 2022 &   sub-\unit[6]{GHz} & Available & Communications SNR & Radar SNR & Communications SNR & SCA - CVX toolbox \\
   \cline{2-9}
   & \cite{2022_Liu3} & 2022 &  sub-\unit[6]{GHz} & Available & Achievable sum rate & Radar SNR & Achievable sum rate & Fractional programming - MM - ADMM \\
   \cline{2-9}
   & \cite{2022_Jiang} & 2022  &  sub-\unit[6]{GHz} & Available & Communications SNR & Radar SNR & Radar SNR & Semi-definite programming - Bisection search - MM - QCQP - SDR \\
    \cline{2-9}
   & \cite{2022_wang5} & 2022  &  sub-\unit[6]{GHz} & Available & MUI energy & Ideal beampattern matching & Communication MUI energy & Singular value decomposition - SDR - Riemannian optimization \\
     \cline{2-9}
   & \cite{2022_Yu2} & 2022  &  sub-\unit[6]{GHz} & Available &  User's rate  & Radar SINR & Energy efficiency & Dinkelbach’s method - SDR - CVX toolbox \\
        \cline{2-9}
   & \cite{2022_Wang9} & 2022  &  mmWave & Available for communications only & User's SINR  & Radar beampattern mismatch &Radar beampattern mismatch & ADMM - Penalty-based method \\
   \cline{2-9}
   & \cite{2022_Wei2} & 2022  &  sub-\unit[6]{GHz} & Available & User's SINR  & Radar SINR & Radar SINR & ADMM - Dinkelbach’s algorithm - CVX toolbox \\
    \cline{2-9}
   & \cite{2022_Mai} & 2022  &  sub-\unit[6]{GHz} & Available & Radar interference energy & Beampattern error and cross correlation pattern & Weighted sum of radar interference energy, beampattern error and cross correlation pattern &  Riemannian optimization \\
   \cline{2-9}
   & \cite{2022_Zhang4} & 2022  &  sub-\unit[6]{GHz} & Available & Users' SINR & Beampattern matching & Radar beampattern matching &  SDR - Riemannian optimization - CVX toolbox \\
    \cline{2-9}
   & \cite{2022_Yikai} & 2022  &  sub-\unit[6]{GHz} & Available for communication only & Communications SNR & Radar SNR & Weighted sum of communications and radar SNRs &  Linear programming - Manifold optimization \\
   \cline{2-9}
   & \cite{2022_Luo} & 2022  &  sub-\unit[6]{GHz} & Not available & Achievable sum rate & Radar beampattern match & Achievable sum rate &  Lagrangian dual reformulation - MM - CVX toolbox - Riemannian optimization \\
   \cline{2-9}
   & \cite{2023_Sankar} & 2023 & mmWave  & Available & Worst-case SINR & Weighted sum of beampattern mismatch error and cross-correlation &  Weighted sum of beampattern mismatch error
 cross-correlation & Convex relaxation - SDR - Generalized Dinkelbach's method \\ 
 \cline{2-9}
  &\cite{2023_Luo} & 2023 &  sub-\unit[6]{GHz}  & Not available & Users' SNRs & Weighted radar sum SNR & Weighted radar sum SNR & 
 QCQP - Semi-definite programming - SDR - MM - Penalty-based algorithm \\  
   \hline 
\end{tabular}
\label{tab:RIS_DFRC_WFD_RA}
\end{table*}

\setcounter{table}{4}
\begin{table*}
\footnotesize
\centering
\caption{References on waveform design and optimal resource allocation for RIS-assisted DFRC (continued).}
\begin{tabular} {|m{2.2cm} | m{1cm}| m{0.5cm} |m{1cm}| m{1cm}| m{1.8cm} | m{1.7cm} | m{1.7cm} | m{3cm}|} 
 \hline 
 Waveform design & Reference & Year  & Frequency band  & LoS availability & Communications metric & Sensing metric & Objective function & Methods employed  \\  [0.5ex] 
 \hline 
 \hline   
 Beamforming-based joint design & \cite{2023_Xianxin2} & 2023 &  sub-\unit[6]{GHz}  & Available for communications only & Users' SINRs & CRB of AoA estimation & CRB of AoA estimation & 
 SDR- CVX toolbox  \\ 
 \cline{2-9}
 &\cite{2023_Xianxin} & 2023 &  sub-\unit[6]{GHz}  & Available for communications only & Users' SINRs & CRB of AoA estimation & CRB of AoA estimation & 
 SDR - SCA - CVX toolbox \\  
  \cline{2-9}
 &\cite{2023_Liao} & 2023 &  sub-\unit[6]{GHz}  & Available for communications only & Users' SINRs & Minimum sensing beampattern gain & Minimum sensing beampattern gain & 
 SDR - CVX toolbox \\ 
   \cline{2-9}
 & \cite{2023_Zhao2} & 2023 & sub-\unit[6]{GHz}/ mmWave & Available for communications only & Channel capacity and communications MSE & Detection probability & Detection probability & ADMM - Riemannian optimization - ZF \\
  \cline{2-9}
 & \cite{2023_Liu6} & 2023 & sub-\unit[6]{GHz} & Available & sum rate  & CRB and radar SNR & sum rate  & Fractional programming - MM - ADMM \\
   \cline{2-9}
 & \cite{2023_Esmaeilbeig} & 2023 & sub-\unit[6]{GHz} & Available for communication only & Communication SINR  & Radar SINR & Weighted sum of communication and radar  SINRs & Power method-like iteration method  \\
\cline{2-9}
 & \cite{2023_Tian}  & 2023 & mmWave & Available & Weighted sum rate & Radar beampattern matching & Weighted sum rate & Powell Hestenes Rockafellar method - Riemannian optimization - SCA \\
 \cline{2-9}
 & \cite{2023_Xu2}  & 2023 & Sub-\unit[6]{GHz} & Available & sum rate & Radar mutual information & sum rate & SDR - one-dimension iterative algorithm - Riemannian optimization - CVX toolbox \\
  \cline{2-9}
     & \cite{2023_Wu}  & 2023 & Sub-\unit[6]{GHz} & Available & Constructive interference & Target illumination power & Target illumination power & Relaxand-retract method - Differece-of-convex programming \\
   \cline{2-9}
 & \cite{2023_Li9}  & 2023 & Sub-\unit[6]{GHz} & Available for communications only & Communication SNR & Radar SNR  & Weighted sum of the Communication and radar SNRs  & MM - Riemannian manifold optimization - Branch-and-bound algorithm \\
  \cline{2-9}
 & \cite{2023_Li10}  & 2023 & Sub-\unit[6]{GHz}/ mmWave & Available for communications only & Communication SNR & Radar SNR  & Weighted sum of communication and radar SNRs  & MM - Riemannian manifold optimization - CVX toolbox \\
   \cline{2-9}
 & \cite{2023_Akshay}  & 2023 & Sub-\unit[6]{GHz} & Available & Communication SNR & Radar SNR  & Radar SNR  & Bisection search algorithm
 \\
 \cline{2-9}
 &\cite{2023_Xiao} & 2023 &  sub-\unit[6]{GHz}  & Not available & Users' SNR & Radar detection SNR & Radar detection SNR & Penalty-based method - SDR - Gaussian randomization - MM- CVX toolbox  \\
  \cline{2-9}
&\cite{2023_Prasobh} & 2023 &  mmWave  & Available for sensing only & Instantaneous SNR & Transmit beampattern error &  Transmit beampattern error & Semi-definite programming - SDR \\
\cline{2-9}
 &\cite{2023_Zhu2} & 2023 & sub-\unit[6]{GHz}  & Not available & communications SNR & Radiation pattern &  Weighted sum of communications SNR and radar radiation pattern & Back propagation - Factor graph - Adaptive sparse Bayesian learning \\
   \cline{2-9}
 &\cite{2023_Kumar} & 2023 & sub-\unit[6]{GHz}  & Available for communications only & Users' SNR & Beampattern gain &  Radar beampattern gain & SCA - Penalty-based method - QCQP \\
   \hline 
\end{tabular}
\end{table*}

\setcounter{table}{4}
\begin{table*}
\footnotesize
\centering
\caption{References on waveform design and optimal resource allocation for RIS-assisted DFRC (continued).}
\begin{tabular} {|m{2.2cm} | m{1cm}| m{0.5cm} |m{1cm}| m{1cm}| m{1.8cm} | m{1.7cm} | m{1.7cm} | m{3cm}|} 
 \hline 
 Waveform design & Reference & Year  & Frequency band  & LoS availability & Communications metric & Sensing metric & Objective function & Methods employed  \\  [0.5ex] 
 \hline 
 \hline 
  Beamforming-
based joint design & \cite{2024_Steven} & 2024 & sub-\unit[6]{GHz} & Available & {Communications SINR} & Sensing SNR & Sensing SNR & {SDR - CVX toolbox - MM} \\
  \cline{2-9}
     & \cite{2024_Yasheng} & {2024} & {sub-\unit[6]{GHz}} & {Available} & {Achievable sum rate} & {Sensing SNR} & {Achievable sum rate} & {Minimum mean-square error method - penalty-based method - MM - QCQP - SDR} \\
    \cline{2-9}
     & \cite{2024_Yongqing} & {2024} & {sub-\unit[6]{GHz}} & {Available} & {Communications rate} & {Sensing SINR} & {Sensing SINR} & {Block coordinate descent - Dinkelbach’s method - SCA } \\
         \cline{2-9}
     & \cite{2024_Shoushuo} & {2024} & {sub-\unit[6]{GHz}} & {Available} & {Communications SINR} & {Radar SNR} & {Radar SN} & {MM - ADMM } \\
  \hline 
    Optimal joint design & \cite{2022_Wang4} & 2022 & mmWave & Available & Communications MSE & CRB of AoA estimation & Communications MSE & Exact penalty method - Manifold optimization - SCA \\
     \cline{2-9}
       & \cite{2022_sun} & 2022 & sub-\unit[6]{GHz} & Available for Communications only & MUI energy & Radar output SNR & Weighted sum of communications MUI energy and radar output SNR & Dinkelbach’s method - Gradient projection method - Minimum variance distortionless response method \\
       \cline{2-9}
       & \cite{2022_Zhai} & 2022 & sub-\unit[6]{GHz} & Available  & MUI energy & Radar beampattern & Weighted sum of communications MUI energy and the radar SNR beampattern & Riemannian trust-region - Singular value decomposition \\
     \cline{2-9}
   & \cite{2022_Liu4} & 2022  &  sub-\unit[6]{GHz} & Available & MUI energy & Radar output SINR & Radar output SINR & ADMM - MM - CVX toolbox \\
        \cline{2-9}
   & \cite{2022_Tian} & 2022  &  mmWave & Available & MUI energy  & Radar beampattern & Weighted sum of the communications  MUI energy and Radar beampattern energy & Sum-path-gain maximization - KKT conditions \\
    \cline{2-9}
    & \cite{2023_Zhong} & 2023 &  sub-\unit[6]{GHz} & Available & MUI energy & Radar SINR & Weighted sum of communications MUI and Radar SINR &  Dinkelbach’s method - Quadratic fractional programming \\
        \cline{2-9}
    & \cite{2023_An} & 2023 &  sub-\unit[6]{GHz} & Available for communications only & MUI energy & Radar SINR & Weighted sum of communications MUI energy and radar SINR & Riemannian optimization - Dinkelbatch's method \\
           \cline{2-9}
    & \cite{2024_Zhong} & {2024} &  {sub-\unit[6]{GHz}} & {Available} & {MUI energy} & {beampattern matching} & {Weighted sum of communications MUI energy and radar beampattern matching} & {Parallel product complex circle manifold algorithm - unconstrained
coupling quartic programming - parallel conjugate gradient method } \\
   \hline 
\end{tabular}
\end{table*}

\subsection{Waveform Design}
\label{sec:wfd}

To enhance the integration between communications and sensing in DFRC, many studies have adopted a joint DFRC waveform design. As outlined in Table~\ref{tab:RIS_DFRC_WFD_RA}, two commonly employed approaches are used for joint DFRC waveform design.

\textit{1) Beamforming-based joint design:} 
In beamforming-based joint design, the DFRC BS transmits a superposition signal utilizing beamforming techniques. This approach aims to address each communication user and radar target by employing distinct waveforms and beamformers~\cite{2022_Song,2022_Zhang1,2022_Zhengyu,2022_Zhu,2022_Yan,2022_Liu3,2022_Jiang,2022_wang5,2022_Yu2,2022_Wang9,2022_Wei2,2022_Mai,2022_Zhang4,2022_Yikai,2022_Luo,2023_Sankar,2023_Luo,2023_Xianxin2,2023_Xianxin,2023_Liao,2023_Zhao2,2023_Liu6,2023_Esmaeilbeig,2023_Tian,2023_Xu2,2023_Wu,2023_Li9,2023_Li10,2023_Akshay,2023_Xiao,2023_Prasobh,2023_Zhu2,2023_Kumar}. 

For instance, consider a DFRC BS transmitting a DFRC waveform to $K$ communication users and $T$ radar targets. In this case, the BS transmits the following signal:
\begin{equation}
    \mathbf{x} = \underbrace{\sum_{k=1}^K \mathbf{f}_k s_k}_{\text{Communications signal}} + \underbrace{\sum_{t=1}^T \mathbf{w}_t u_t}_{\text{Radar signal}},
    \label{eq:joint_wf_bf}
\end{equation}
where $\mathbf{f}_k$ and $\mathbf{w}_t$ are the transmit beamformers employed by the BS to serve the $k$-th communication user and the $t$-th radar target, respectively, and $s_k$ and $u_t$ correspond to the transmit waveform intended for the $k$-th communication user and the $t$-th radar target, respectively. While $s_k$ is typically chosen from a set of constellation symbols (e.g., a quadrature amplitude modulation (QAM) constellation), radar waveforms (i.e., $u_t$) are typically modeled as radar pulses. Since communication and radar waveforms are considered as interference to each other, the joint waveform defined in~\eqref{eq:joint_wf_bf} does not achieve full integration between communication and sensing. However, it is a common approach in many papers due to its simplicity, as it eliminates the need to invent entirely new waveforms.

\textit{2) Optimal joint design:} In optimal joint design, a fully integrated waveform is generated by solving an optimization problem~\cite{2022_Wang4,2023_Zhong,2022_Liu4,2022_sun,2023_An,2022_Zhai,2022_Tian}.

An illustration of this approach can be found in~\cite{2022_Zhai}, where a BS with $M_\text{T}$ antennas transmits a joint DFRC signal to serve $K$ single-antenna communication users and $T$ radar targets, aided by an RIS. The signals received by the $K$ users in $L$ consecutive time slots can then be expressed as:
\begin{equation}
    \mathbf{Y} = \mathbf{G} (\boldsymbol{\Phi}) \mathbf{X} + \mathbf{N},
\end{equation}
where $\mathbf{G} (\boldsymbol{\Phi}) = [\mathbf{g}_1 (\boldsymbol{\Phi}),\dots, \mathbf{g}_K (\boldsymbol{\Phi})]^\mathsf{H} \in \mathbb{C}^{K \times M_\text{T}}$ is the channel matrix for the $K$ users with $ \mathbf{g}_k (\boldsymbol{\Phi}) \in \mathbb{C}^{M_\text{T} \times  1}$ indicating the overall channel matrix for the $k$-th user, which is a function of the \text{RIS} matrix $\boldsymbol{\Phi}$. Also, $\mathbf{X} = [\mathbf{x}_1,\dots,\mathbf{x}_L] \in \mathbb{C}^{M_\text{T} \times L}$ is the DFRC waveform matrix for the $L$ time slots, and $\mathbf{N}$ is the additive white Gaussian noise (AWGN) matrix with independent and identically distributed (i.i.d.) entries, each having a zero-mean and a variance of $\sigma_n^2$. By denoting the communication constellation symbols as $\mathbf{S}$ and the ideal radar pulses as $\mathbf{U}$, the waveform and the RIS phase shifts can be co-designed by minimizing the following objective function:
\begin{equation}
    f(\mathbf{X},\boldsymbol{\Phi}) = \gamma ||  \mathbf{G} (\boldsymbol{\Phi}) \mathbf{X} - \mathbf{S}||_\mathsf{F}^2 + (1-\gamma) || \mathbf{U} - \mathbf{S}||_\mathsf{F}^2,
    \label{eq:optm_joint_wf}
\end{equation}
subject to the relevant constraint set. In~\eqref{eq:optm_joint_wf}, the parameter $\gamma \in [0,1]$ serves as a design parameter to prioritize either communication or sensing capability. When $\gamma = 1$, the waveform design objective function~\eqref{eq:optm_joint_wf} becomes $f(\mathbf{X},\boldsymbol{\Phi}) = || \mathbf{G} (\boldsymbol{\Phi}) \mathbf{X} - \mathbf{S}||_\mathsf{F}^2$. This implies that the waveform is designed solely based on the communication constellation matrix $\mathbf{S}$, resulting in excellent communication performance. Conversely, when $\gamma = 0$, the objective function~\eqref{eq:optm_joint_wf} becomes $f(\mathbf{X},\boldsymbol{\Phi}) = || \mathbf{U} - \mathbf{S}||_\mathsf{F}^2$. In this case, the waveform is designed as pure radar pulses, leading to excellent radar performance. Typically, the transmit waveform is co-optimized with other resources to achieve optimal system performance.

Given that RISs offer an additional resource for optimization, the increased flexibility in designing DFRC waveforms often results in superior performance compared to systems without an RIS. In~\cite{2022_Zhai}, the authors demonstrated that the inclusion of an RIS boosted the communication sum rate of the system by approximately 48\% for $\gamma = 0.6$. Furthermore, they reported an enhancement in radar detection probability of around 10\%.

\textbf{Lessons learned:} The exploration of joint DFRC waveform design reveals several key insights. Beamforming-based joint design is favored in the literature for its simplicity, leveraging existing beamforming techniques to serve communication users and radar targets with distinct waveforms, though it does not achieve full integration between communications and sensing. Optimal joint design, on the other hand, generates fully integrated waveforms through solving optimization problems, offering enhanced performance by co-designing the transmit waveform and RIS phase shifts. In this context, RIS can boost system performance, improving communication and radar performance and dynamically adjusting its phases as the propagation environment requires.

\textbf{Challenges, opportunities and open research directions:} A significant challenge in selecting the optimal joint waveform design revolves around the tradeoff between complexity and accuracy. While beamforming-based joint designs benefit from fewer optimization variables, resulting in solutions with lower computational complexity, the need for a separate communication-sensing waveform division might not be essential, as a unified and appropriate waveform could in principle serve both functions. In contrast, optimal joint waveform designs capitalize on this flexibility to find an optimal waveform at the cost of introducing additional computational complexity. A potential compromise between these two approaches could be advantageous. For instance, pre-designed joint waveforms could be developed based on an optimal joint design that optimizes long-term average or worst-case system performance. If these waveforms prove to deliver satisfactory performance in the majority of cases, they could be employed without the need to optimize the design from scratch each time.

\subsection{Optimal Resource Allocation}
\label{sec:RA_DFRC}

\begin{figure*} 
         \centering
    \includegraphics[width=1.8\columnwidth]{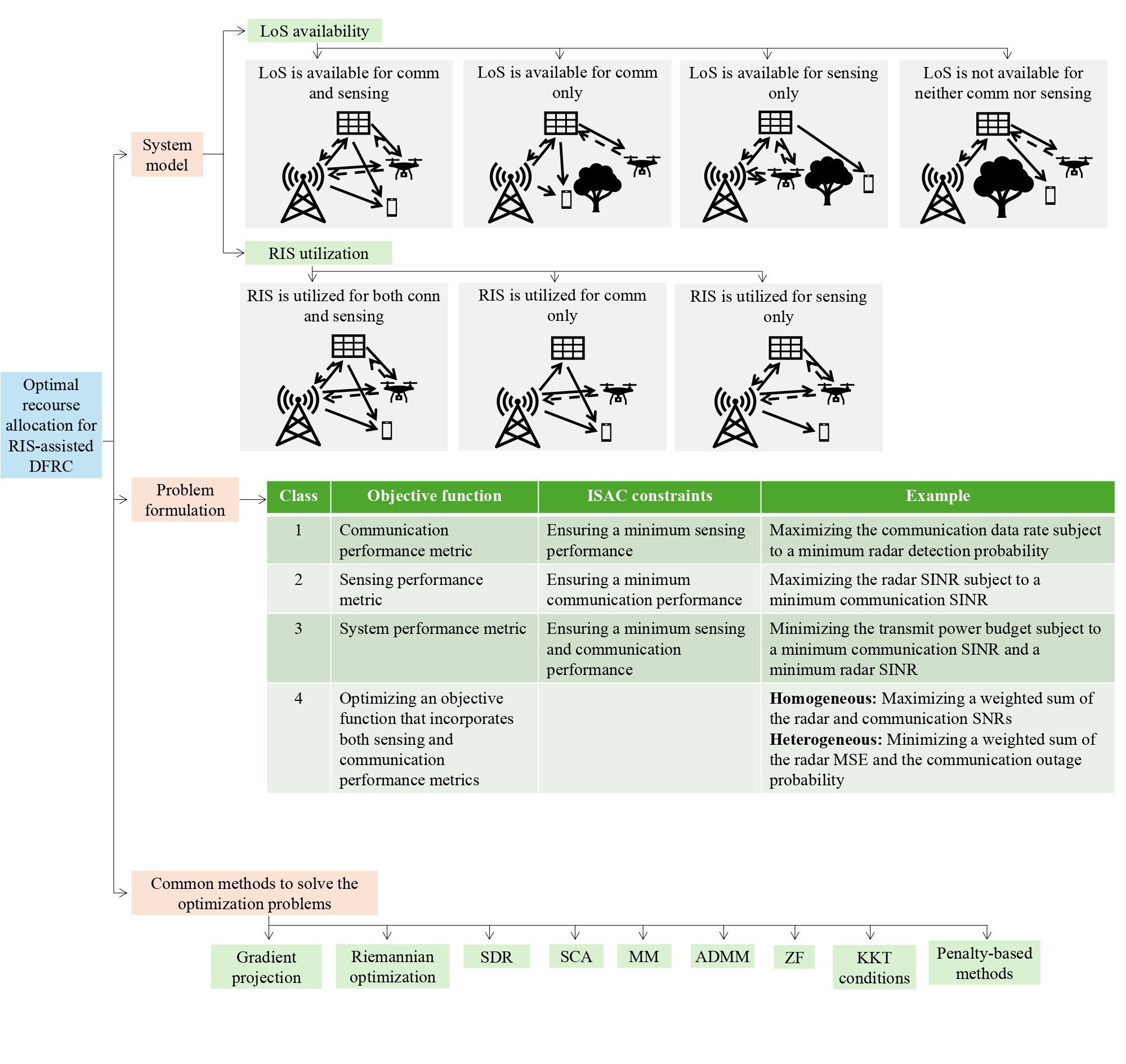}
        \caption{Summary of optimal resource allocation considerations for RIS-assisted DFRC.}
        \label{Fig:RA}
\end{figure*}

The resource allocation problems in RIS-aided DFRC can manifest themselves in various forms, as illustrated in Table~\ref{tab:RIS_DFRC_WFD_RA} and Fig.~\ref{Fig:RA}. This diversity in problem types arises from considerations at three levels, which are system model, problem formulation, and proposed solution.

\textit{1) System model:} The general system model for RIS-assisted DFRC comprises a DFRC transmitter communicating with multiple users while concurrently performing sensing tasks involving multiple targets with the assistance of an RIS.

Several papers have explored diverse scenarios concerning the LoS links between the DFRC BS and the communication users and/or the radar targets. These scenarios can be broadly categorized into three classes:
\begin{itemize}
    \item The first class is characterized by the availability of LoS links to all users and targets, as exemplified in~\cite{2022_Liu3,2022_Zhang4,2023_Tian} and~\cite{2023_Akshay}. While the inclusion of an RIS is not obligatory in this scenario, it can significantly enhance the system performance. For instance, in~\cite{2022_Jiang}, it was noted that an RIS-assisted system can improve the radar SNR by approximately 80\% while maintaining a high communication SNR.
    \item The second case arises when LoS links are available exclusively for either the communication users or radar targets but not both, as in~\cite{2023_Liao,2023_Esmaeilbeig,2023_Li9,2023_Kumar} and~\cite{2022_sun}. In such scenarios, the RIS serves as the exclusive means of communication or sensing for the system functionality with obstructed LoS, concentrating the signal energy on that functionality while maintaining a satisfactory performance for the one with unobstructed LoS.
    \item The third case arises when LoS links are obstructed for all communication users and sensing targets, as demonstrated in~\cite{2022_Song,2023_Prasobh,2023_Xiao} and~\cite{2022_Yu}. In this scenario, the presence of an RIS is crucial, as it functions as the exclusive means for both communication and sensing.
\end{itemize}

Another classification within the system model involves deciding whether to utilize the RIS for both communication and sensing or dedicate it to one of these tasks exclusively. While the former is more common, the latter option can be considered to prioritize one task over the other. For example, in~\cite{2023_Sankar}, the authors proposed an RIS-assisted ISAC system where the RIS is employed solely to enhance communication performance. Motivated by the observation that MIMO radars can achieve optimal sensing performance by directing signals toward specific directions while maintaining signal orthogonality across beams~\cite{2020_Liu}, the authors designed an objective function comprising a weighted sum of beampattern mismatch error and cross-correlation. This method demonstrated significant potential to enhance the worst-case SINR and improve user fairness, albeit with a minor impact on radar performance.

\textit{2) Problem formulation:} Various papers have employed a diverse range of problem formulations to address RIS-assisted DFRC resource allocation problems. Due to the absence of a unified framework for quantifying the performance of both communications and sensing in ISAC, each task is typically assessed based on its own set of performance metrics. Generally, resource allocation problems in RIS-assisted DFRC systems can be categorized into four main classes: 
\begin{itemize}
    \item The first class involves optimizing a communication metric while ensuring a minimum sensing performance, as exemplified in problems found in~\cite{2022_Zhengyu,2022_Yan,2023_Liu6} and~\cite{2023_Xu2}. In these papers, ensuring a specific sensing performance takes precedence, and the communications metric is enhanced subject to that requirement. 
     \item The second class involves optimizing a sensing metric while ensuring a minimum communications performance, as demonstrated in problems presented in~\cite{2022_Yu,2023_Liao,2023_Akshay} and~\cite{2022_Liu4}. In contrast to the previous case, in this scenario, the sensing metric is optimized subject to a specific communication quality constraint.
     \item The third class involves optimizing a system performance metric while ensuring a minimum performance for both sensing and communications. For instance, in~\cite{2022_Yu2}, the authors optimized the system's energy efficiency subject to constraints on the communication user's rate \textit{and} the radar SINR. It is crucial to note that in the first three classes, setting high standards for the metrics used in the constraints may potentially limit the feasible space of the optimization problem and may even lead to infeasible problems. 
     \item The fourth class entails optimizing an objective function that incorporates both sensing and communication performance metrics. In this approach, a new objective function is formulated by combining the performance metrics of both communications and sensing, as expressed in~\eqref{eq:wsum}. In this category, there are two types of objective functions: homogeneous~\cite{2022_Yikai,2023_Esmaeilbeig,2023_Li9,2023_Li10} and heterogeneous~\cite{2023_Zhu2,2022_sun} functions. Homogeneous objective functions involve combining radar and communication metrics that can be measured using the same unit. For instance, combining the radar and communication SNRs~\cite{2023_Li10}. Heterogeneous objective functions, on the other hand, combine two distinct radar and communication metrics. An example of this is the combination of the communication multi-user interference (MUI) energy and the radar output SNR~\cite{2022_sun}. One challenge in formulating heterogeneous objective functions is determining appropriate weights for the distinct radar and communications performance metrics to ensure that one does not become excessively dominant.
\end{itemize}

\textit{3) Proposed solution:} 
After formulating the optimization problem, it is crucial to use appropriate methods to obtain the optimal solution while achieving a tradeoff between performance and complexity. Given that the optimization variables are often coupled in the objective function and constraints, utilizing an AO framework is common in most papers. The gradient projection and Riemannian optimization methods are popular approaches for tackling the RIS optimization sub-problem due to their efficiency in handling the constant modulus (CM) constraint. In practice, the gradient projection method operates by performing the gradient-based update while ignoring the CM constraint, then projecting the updated variable onto the CM constraint by rescaling the magnitudes of elements of the RIS matrix~\cite{1961_Rosen}. Riemannian optimization methods, on the other hand, obtain the Riemannian gradient defined as the gradient along the Riemannian manifold~\cite{2015_Sato}, represented in this case by the CM manifold.

Other common optimization techniques to address RIS-assisted DFRC optimization problems include SDR~\cite{2022_wang5}, SCA~\cite{2022_Yan}, majorization-minimization (MM)~\cite{2022_Liu3}, QCQP~\cite{2022_Jiang}, Dinkelbach’s method~\cite{2022_Yu2}, alternative direction method of multipliers (ADMM)~\cite{2022_Wang9}, zero-forcing (ZF)~\cite{2023_Zhao2}, Karush-Kuhn-Tucker (KKT) conditions~\cite{2022_Tian}, and penalty-based methods~\cite{2023_Luo}.

{\textbf{Lessons learned:} System model considerations are critical in resource allocation for RIS-assisted DFRC, with diverse scenarios for LoS links influencing the necessity and effectiveness of RIS. For instance, RIS enhances ISAC system performance with available LoSs, but becomes crucial when LoS is obstructed. Moreover, the strategy of using RIS for both communication and sensing, or dedicating it to one task, impacts overall system efficiency. Moreover, the problem formulation must be tailored to the system objectives in balancing communication and sensing performance. Optimizing communication metrics while ensuring minimum sensing performance, or vice versa, can lead to efficient resource use, but setting high standards for both metrics might limit feasibility and thus effect optimality. To avoid this, developing combined objective functions that integrate both metrics can be a useful solution, however, it requires careful weight assignment to avoid dominance by one metric. Finally, the algorithms used to solve the optimization problem should be selected based on specific problem characteristics taking into account factors such as capability to converge to a good local solution, convergence speed and computational complexity.}

{\textbf{ Challenges, opportunities and open research directions:} Integrating communication and sensing functionalities into a single system remains challenging due to the conflicting requirements and performance metrics of each. The optimization problems in RIS-aided DFRC are often non-convex and involve coupled variables, making it difficult to find optimal solutions efficiently. Achieving real-time resource allocation and adaptation in dynamic environments is also challenging due to the high computational demands and latency constraints. Additionally, balancing the interference between communication and sensing tasks while optimizing the use of RIS is a significant challenge.} {Optimizing RIS configurations can lead to more energy-efficient systems by focusing signal energy where it is needed most. Furthermore, RIS can provide flexibility in resource allocation, allowing dynamic adaptation to changing environmental conditions and user requirements. Developing RIS-aided systems that can switch between or simultaneously support communication and sensing tasks opens new possibilities for multi-functional infrastructure.}

{Developing a unified framework for quantifying the performance of both communication and sensing in RIS-aided DFRC systems remains an open research area. In addition, incorporating machine learning techniques for dynamic resource allocation and optimization can offer adaptive and real-time solutions. Researching robust and accurate channel estimation techniques that can operate in obstructed LoS scenarios will be vital for improving system performance. Also, developing advanced interference mitigation strategies to balance the requirements of communication and sensing tasks while using RIS is a key area for future research. Another promising direction is investigating practical hardware implementations and the impact of real-world imperfections on RIS performance. Finally, researching ways to scale RIS-aided DFRC systems while reducing computational complexity and ensuring feasibility for large-scale deployments is an ongoing challenge.}

\subsection{Robust System Design}
Robust system design refers to optimizing system performance when perfect CSI knowledge is not available at the transmitter. In practice, channel estimation errors will inevitably occur in the communication part of the system while estimating the direct BS-user channel, the BS-RIS channel, and/or the RIS-user channel. Similarly, channel estimation errors can occur in the sesning part of the system while estimating the direct BS-target channel, and/or the RIS-targe channel.


There are generally two models to characterize the CSI uncertainties:

\textit{1) Bounded error model:} The bounded error model assumes that the error is bounded within a sphere with a specific radius known to the BS, as in~\cite{2023_Luan}. This means that any true channel matrix can be written as a sum of the estimated channels matrix and the channel estimation error matrix. The bounded error model assumes that the Frobenius norms of the error matrix is bounded with a bound know to the BS\footnote{A more general bounded error model confines the error within an ellipsoid. However, it is more common to bound the error within spheres due to simplicity of expression.}. This bounded error model is widely used in the literature, as it can describe the effects of quantization errors, which belong to a bounded region~\cite{CEE11}.

The objective then is to determine the error matrices that result in the worst-case system performance. This performance bound can be determined by utilizing inequalities and/or approximations. Alternatively, the performance bound can be achieved by means of optimization algorithms, i.e., by identifying the values of the error matrices that produce the worst-case performance of the system.

\textit{2) Statistical error model:} In the statistical error model, the entries of the error matrices are assumed to be i.i.d. random variables with a known distribution. The Gaussian distribution is commonly used to characterize the elements of the error matrices. In this system, the goal is to optimize the average performance of the system rather than the worst-case performance. This can be addressed by applying various approximations and/or identities to obtain a deterministic approximation or lower bound. Alternatively, stochastic optimization tools can be leveraged to optimize the average permanence metric.

{\textbf{Lessons learned:}  In practical scenarios with RIS with large dimensions, channel estimation errors are inevitable. To characterize CSI uncertainties, two primary models are typically employed: the bounded error model and the statistical error model. The bounded error model assumes that the error is confined within a sphere with a specific radius known to the BS. The goal here is optimize the worst-case system performance. On the other hand, the statistical error model assumes that the entries of the error matrices are random variables with known distributions. This model focuses on optimizing the average performance of the system rather than the worst-case performance. Both models are important for ensuring reliable system performance under different types of CSI uncertainties.}

\textbf{Challenges, opportunities and open research directions:} While defining these error bounds or distributions helps in reducing the complexity of robust design, in practice, multiple factors, such as the employed channel estimation algorithm, frequency band, users' locations, etc., play a role in error characterization. Hence, it is challenging to obtain the values of the error bounds in the case of the bounded error model and the error distributions in the case of the statistical error model. Furthermore, while robust system design optimizes the average or worst-case system performance, the real system may not be effectively optimized. For example, optimizing the worst-case performance may not necessarily optimize the actual system performance and could lead to more interference between radar and communication signals, impacting the quality of both functions.

Leveraging machine learning techniques can offer opportunities for mitigating the effects of imperfect CSI. Learning-based methods can adapt to complex and dynamic environments, providing intelligent solutions for estimating channels with high precision and optimizing radar and communication functions simultaneously. In addition, developing efficient feedback reduction techniques can address the challenge of limited feedback capacity. Strategies such as compressive sensing or exploiting spatial correlation among the channels can reduce the amount of feedback required for channel estimation and also reduce the channel estimation error.

{
\subsection{Coverage Enhancement}
While use of the mmWave band can enhance both communication and radar performance, signals in the mmWave band experience high sensitivity to blockages. To tackle this issue, the authors of~\cite{2024_Gan1} explored the impact of blockages and leveraged stochastic geometry tools to assess the improvement in coverage achieved by RISs. Specifically, the authors aimed at investigating the  interaction of ISAC dual functions within a single network topology. Based on this, a conditional coverage probability was obtained for a two-user scenario using a beampattern model and association strategies. By employing a distance-dependent thinning method, the marginal coverage rate for the network was computed. Notably, results reported by the authors demonstrated that RIS deployment, particularly with a BS density of \unit[40]{$\text{km}^{-2}$}, can significantly enhance the overall ISAC coverage rate, with a substantial increase from 67.1\% to 92.2\%.}

{\textbf{Lessons learned:} By utilizing stochastic geometry along with accurate statistical models, the coverage enhancement provided by RIS to ISAC systems can be effectively quantified; this quantification, in turn, leads to a deeper understanding of how various factors, such as environmental conditions, deployment configurations, and signal processing strategies, influence coverage enhancement. Such insights are crucial not only for optimizing the performance of the system but also for tailoring the deployment and configuration of RIS to meet the specific requirements of ISAC, thus enabling more efficient utilization of available resources and ensuring the system's robustness under diverse conditions.}

{\textbf{Challenges, opportunities and open research directions:} RISs open up opportunities for significant improvements in coverage by establishing controllable virtual links between BSs, targets and users. Future research directions include optimizing the placement and configuration of RISs to maximize their effectiveness in mitigating blockage effects, developing adaptive algorithms for real-time user association, and investigating the integration of RISs with existing network architectures. Additionally, exploring the tradeoffs between deployment costs and performance gains will be essential for realizing the full potential of RIS technology in enhancing coverage within ISAC frameworks.}

\subsection{Channel Estimation} \label{sec:CE}
{Channel estimation for RIS-assisted ISAC involves estimating several channels: the BS-user channels, BS-RIS channel, RIS-user channels, BS-target channels, and RIS-target channels. While the first three can be estimated using methods employed in RIS-assisted communication systems (i.e., by transmitting pilots), the last two are more challenging to estimate due to the passive nature of the target and its inability to transmit or process pilot signals.} In this subsection, we discuss two strategies to address this challenge.

\begin{figure*}
  \centering
  \begin{tabular}{c c}
    \includegraphics[width=0.8\columnwidth]{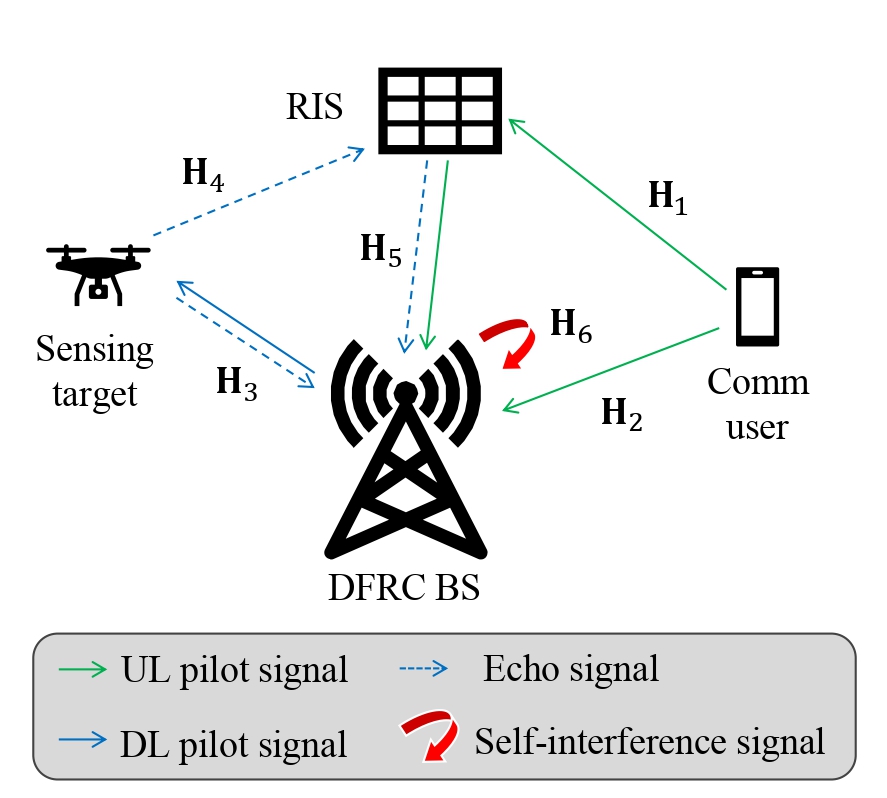} &
      \includegraphics[width=0.8\columnwidth]{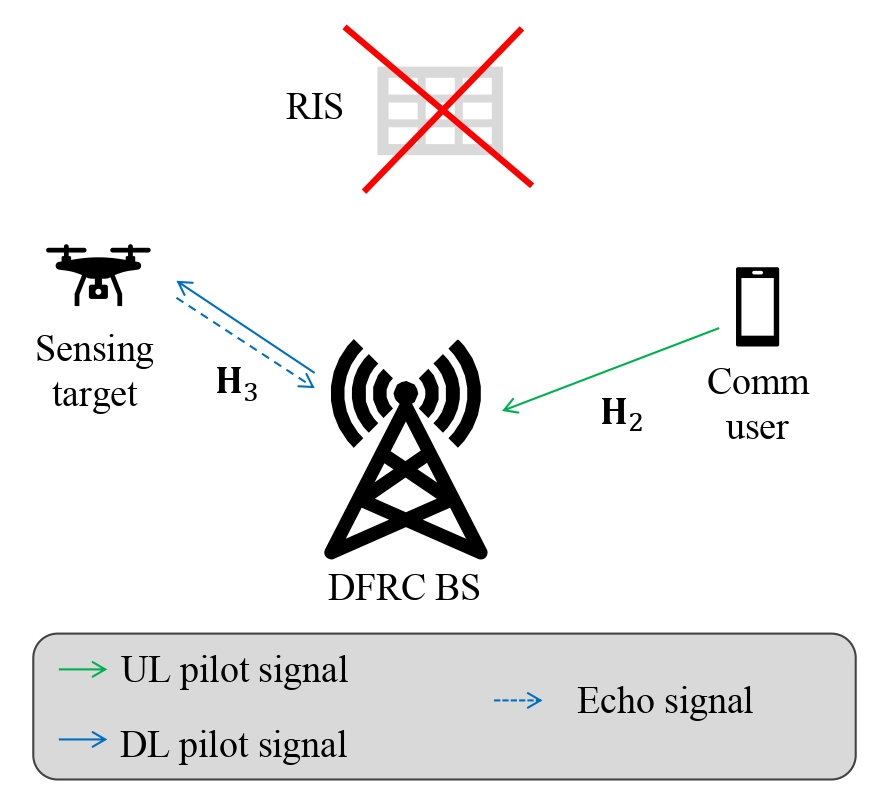}  \\
      \scriptsize (a)   &
      \scriptsize (b) \\
    \includegraphics[width=0.8\columnwidth]{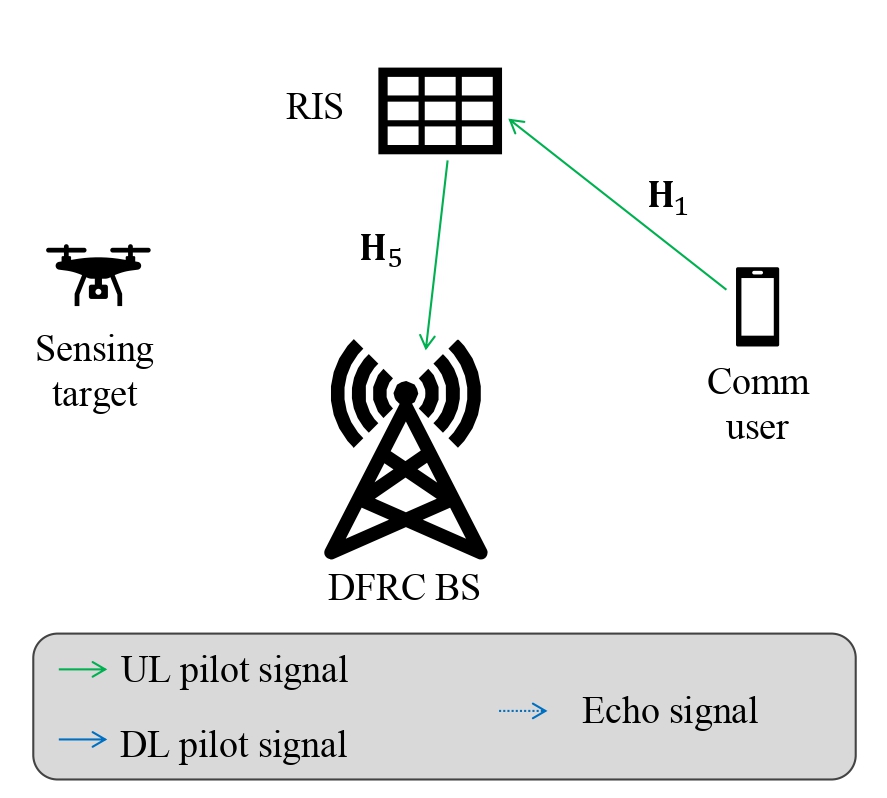} &
      \includegraphics[width=0.8\columnwidth]{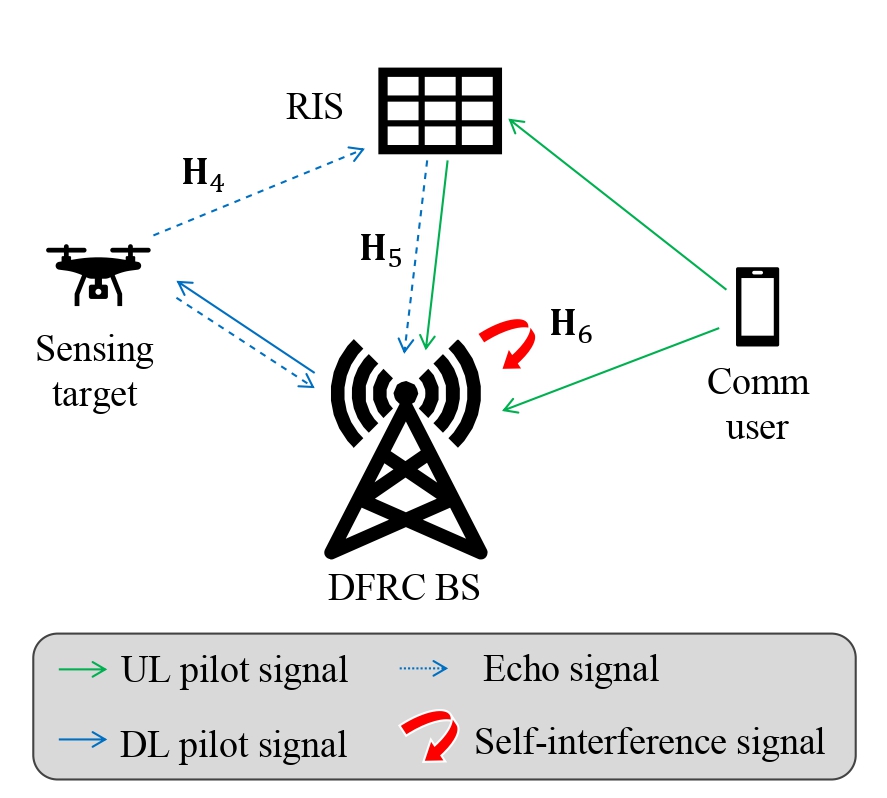}  \\
      \scriptsize (c)   &
      \scriptsize (d) \\
  \end{tabular}
    \medskip
  \caption{Channel estimation framework for RIS-assisted ISAC proposed in~\cite{2022_Liu7,2023_Liu7,2023_Liu4}. (a) System model. (b) Stage 1: Estimating the direct channels. (c) Stage 2: Estimating the reflected communication channel. (d) Stage 3: Estimating the reflected sensing channel.}
  \label{fig:CE}
\end{figure*}

\textit{1) Joint channel estimation and target sensing:} In joint channel estimation and target sensing, the BS aims to estimate the channel matrices of the communication users and sense the locations of the targets through pilot transmission. One way to achieve this is by estimating the communication channels using pilots while simultaneously processing the echoes received from the sensing targets to estimate their channels. Estimating the channel matrices is valuable for channel equalization, and location information can be deduced from the estimated sensing matrices~\cite{2023_Li6,2023_Chen3,2023_Cao,2023_Zijian}.

In particular, the authors of~\cite{2022_Liu7,2023_Liu7,2023_Liu4} considered a full-duplex RIS-assisted DFRC system where a BS communicates with a single user and senses a target with the help of an RIS. As illustrated in Fig.~\ref{fig:CE}(a), the communication user transmits UL pilots, while simultaneously, the BS transmits DL pilots toward the target and captures the reflected signals for location sensing. This setup results in six different channels, as depicted in Fig.~\ref{fig:CE}(a).

To streamline the channel estimation procedure and reduce dimensionality, the authors proposed a three-stage approach spanning the channel coherence time. In the first stage, the RIS is turned off to facilitate the estimation of the direct BS-target and BS-user channels, as illustrated in Fig.~\ref{fig:CE}(b). In the second stage, the RIS is turned on and the BS halts its DL pilot transmission to enable the estimation of the user-RIS-BS channel, as depicted in Fig.~\ref{fig:CE}(c). Finally, in the third stage, both DL and UL pilot transmissions are conducted concurrently to estimate the reflected sensing and self-interference channels, as shown in Fig.~\ref{fig:CE}(d). For each stage, a deep learning model is employed to estimate the relevant channels.

\textit{2) Joint data transmission and target sensing:} {While RIS can enhance the performance of ISAC, ISAC can also mitigate some limitations associated with using RIS for communication, one of which is channel estimation. As ISAC BS can transmit data to users while simultaneously utilizing the received echo signals to estimate their AoD and AoA, thus enabling the construction of channel matrices based on this information.} Based on this and to address the challenge of reduced communication rate caused by allocating specific time slots for pilot transmission, researchers in~\cite{2022_Liu2,2023_Yang,2023_Hu,2024_Chen} and~\cite{2023_Qian} proposed a novel approach that eliminates the use of dedicated pilots, opting instead for useful data symbols. The system model in these papers focuses on acquiring the CSI of the communication user through sensing, leveraging the assistance of distributed RISs. The proposed two-phase protocol operates as follows: in the first phase, the user transmits an UL data stream to the BS via an RIS. Simultaneously, two RISs, dedicated exclusively to sensing purposes, perform location sensing relying on limited CSI. In the second stage, the BS processes the received echo signal, estimates the user's location, and designs the beamformer based on this information. The key advantage of this method lies in replacing the overhead of dedicated pilot transmission with useful data signals, thereby enhancing communication performance. 

{\textbf{Lessons learned:} In RIS-assisted ISAC, channel estimation is critical for effective operation. However, it presents challenges due to the passive nature of targets and the RIS. To address these challenges, two strategies have been proposed. The first involves joint channel estimation and target sensing, where the BS uses pilot signals to estimate communication channels while simultaneously processing echoes from sensing targets. The second strategy focuses on joint data transmission and target sensing. This method eliminates the need for dedicated pilot transmissions, replacing them with useful data symbols to enhance communication performance. Both strategies are examples of the innovative ways in which an RIS can aid channel estimation in ISAC systems.}

{\textbf{Challenges, opportunities and open research directions:} One significant challenge of RIS-assisted ISAC is the complexity of accurately estimating channels in the presence of passive targets, which limits the effectiveness of conventional pilot-based channel estimation methods. Additionally, achieving real-time processing and adaptive algorithms for channel estimation and target sensing remains difficult due to the dynamic nature of environments and the need for high accuracy. However, these challenges also create opportunities for developing advanced techniques, such as deep learning algorithms, for efficient channel estimation and improved sensing capabilities. Open research directions include exploring new methodologies for joint channel estimation and data transmission, optimizing the use of distributed RIS configurations, and investigating alternative pilot-free approaches to enhance communication performance while accurately sensing targets. Furthermore, examining the tradeoffs between system complexity, cost, and performance can lead to more practical implementations of RIS-assisted ISAC in real-world applications.}

\section{RIS as an Architectural Enabler for DFRC}  \label{sec:RIS_DFRC2}

{This section presents research works where RISs are used more intelligently for purposes beyond simply directing signals toward specific locations. The works discussed in this section introduce greater advantages of employing RISs for DFRC compared to those presented in the previous section.}

\subsection{RIS Partitioning}

\begin{figure*}
  \centering
  \begin{tabular}{c c}
    \includegraphics[width=0.95\columnwidth]{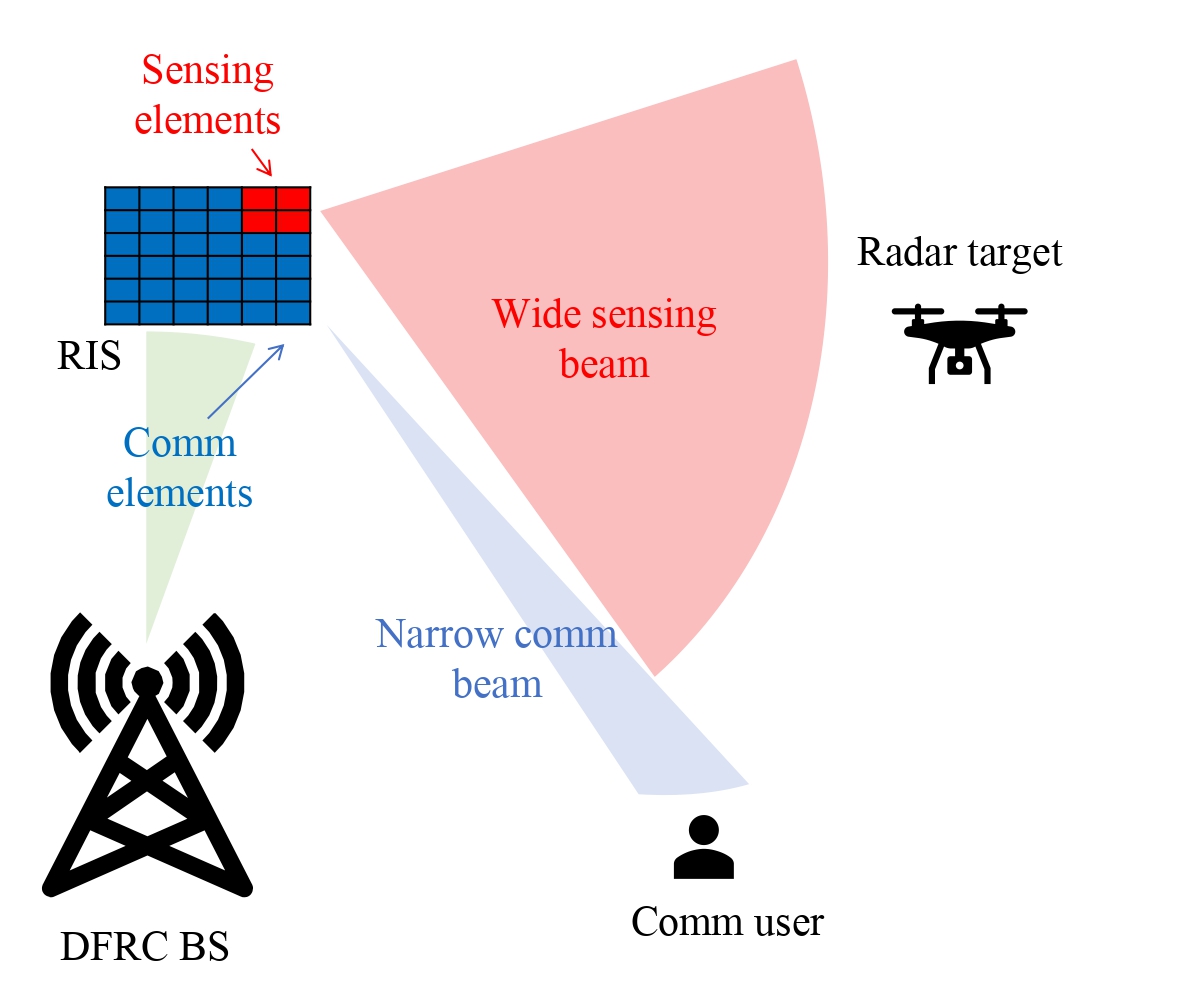} &
      \includegraphics[width=0.95\columnwidth]{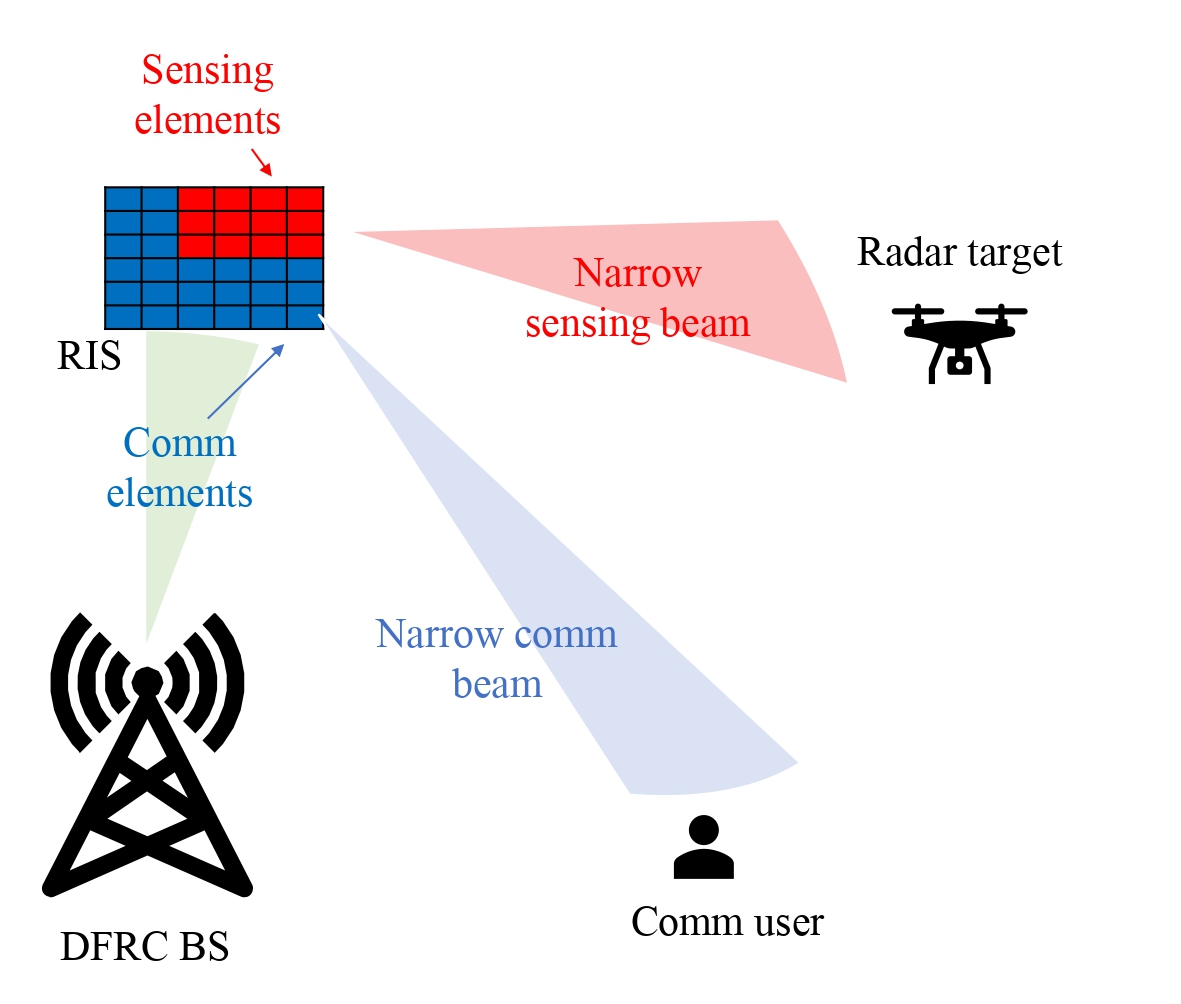}  \\
      \scriptsize (a)    &
      \scriptsize (b)   \\
  \end{tabular}
    \medskip
  \caption{RIS partitioning scheme proposed in~\cite{2021_Prasobh}. (a) Surveillance stage, and (b) Localization stage.}
  \label{fig:RIS_port}
\end{figure*}

One of the early contributions in the field of RIS-assisted ISAC was presented in the work of~\cite{2021_Prasobh}. This approach, though slightly different to that of classical resource allocation methods, provided an intriguing perspective. In~\cite{2021_Prasobh}, the transmission protocol was divided into two stages: the surveillance stage and the localization stage.

During routine surveillance operations, the authors proposed a unique strategy in which a portion of the RIS is dedicated to transmitting wide beams for sensing as shown in Fig.~\ref{fig:RIS_port}(a). Simultaneously, the majority of the RIS elements are employed for communication purposes. The goal of this stage is to detect the presence of a target without causing a significant degradation in the communication performance.

Once a target is detected using the wide beams in the surveillance stage, the focus shifts to obtaining precise location information during the localization stage. In this stage, the transmitted beams gradually sharpen through the utilization of a hierarchical codebook. This progression involves reserving larger portions of the RIS elements for localization, as depicted in Fig.~\ref{fig:RIS_port}(b). Consequently, the number of RIS elements used for communications is reduced, and as a result, the beams serving the communication users in the localization stage are wider than those in the surveillance stage. The significance of the proposed codebook lies in diminishing the necessity for exhaustive search across the entire region of interest, as feedback from echo signals emitted by the target aids in the process. The authors proposed a method to determine the optimal portion of the RIS elements needed for communication and sensing in each stage. This strategy compensates for the varying array gain, ensuring the desired performance in terms of the probability of target localization error.

{\textbf{Lessons learned:} Partitioning the RIS to deal with different stages, each with its own goals, in ISAC can yield a better system performance. For instance, when surveilling a target, a small portion of the RIS can be dedicated to generate wide beams for sensing, while the major portion supports communication. Moreover, hierarchical codebooks can be utilized to reduce the need for exhaustive search methods. This strategy underscores the necessity of algorithms that dynamically adjust resource allocation to balance tradeoffs between communication and sensing performance.}

{\textbf{Challenges, opportunities and open research directions:} The two-stage protocol proposed by~\cite{2021_Prasobh} dedicates RIS elements to support communication or sensing dynamically, yet the optimization of these allocations in real-time remains a significant technical hurdle. A critical technical innovation introduced by the authors is the hierarchical codebook design for beamforming. This design allows for the gradual sharpening of beams during the localization stage, thereby reducing the need for exhaustive search and enhancing efficiency. However, optimizing these codebooks to balance computational complexity and accuracy presents an ongoing research challenge. Moreover, interference management between sensing and communication tasks demands sophisticated strategies to ensure that the intensified focus on one task does not detrimentally impact the other, particularly in scenarios with overlapping operational requirements.}

{Research into practical implementations for such an element division strategy is also critical, exploring hardware constraints, scalability issues, and the impact of real-world conditions on system performance. Moreover, advancing interference mitigation techniques and refining channel estimation methods in obstructed LoS environments are pivotal to enhancing system reliability and accuracy.}
{
\subsection{Automatic RIS Phase-Shift Update}
In environments like smart homes and factories, users often require both high-speed data transmission and precise positioning. The study in~\cite{2024_Ohyama} explored this scenario in indoor settings. In such a scenario, the RIS typically sweeps the reflected beam across a large area by periodically changing phase shifts for localization. However, this approach often results in directing the beam to areas without users, thereby reducing system efficiency. To solve this problem, the authors proposed a method that automatically updates the phase-shift switching schedule to ensure that the RIS directs the reflected beam only towards areas with users. A notable feature of the proposed RIS is its ability to operate without external control, making it more efficient than conventional RIS systems.}

{This automatic update method relies on embedded sensors on the RIS to monitor received signal strength and employs a threshold-based decision-making approach, reducing both cost and power consumption. Additionally, the updated phase-shift schedule integrates a fingerprint-based localization technique, making the system suitable for communication and localization tasks in environments like smart factories. The study also proposed a quantum computing based method to optimize resource allocation and improve localization accuracy while enhancing energy efficiency.}

{\textbf{Lessons learned:} By employing an automatic phase-shift switching schedule in RIS-assisted ISAC systems, the system can dynamically focus the reflected beam on areas with users, eliminating inefficiencies caused by directing beams to empty locations. This approach, which operates without external control, enhances system performance while minimizing costs and energy consumption. Additionally, the integration of fingerprint-based localization and quantum computing for resource allocation further improves localization accuracy and energy efficiency, demonstrating the potential of combining advanced technologies for smarter, more efficient communication systems.}

{\textbf{Challenges, opportunities and open research directions:} A significant challenge in automatic update logarithms is the complexity of designing an RIS-aided system that can automatically and efficiently update its phase-shift schedule without relying on external control, while maintaining low costs and power consumption. Another challenge is ensuring that the system can effectively localize users in dynamic and cluttered indoor environments, such as smart factories. However, these challenges also present opportunities, particularly in refining automatic update mechanisms using advanced signal processing and machine learning. Open research directions include improving the scalability of RIS systems for larger and more complex environments, developing more sophisticated localization algorithms that can adapt to varying user movements, and enhancing the system’s ability to operate seamlessly in multi-operator networks, where resource sharing and interference management become critical.}

{
\subsection{User Grouping}
User channels of RIS-assisted ISAC experience significant channel correlation when the antennas at the BS or RIS are densely deployed. This correlation can lead to a considerable degradation in communication performance. Traditionally, user grouping and spatial-domain processing are the two primary strategies used to address this correlation issue. However, spatial-domain processing is less effective in ISAC, where communication and sensing inevitably compete for resources.}

{To tackle this issue, a user grouping strategy was proposed in~\cite{2024_Jinsong} to separate users with strong channel correlation into different groups, thereby enhancing both communication and sensing performance. More specifically, the authors first formulated an optimization problem that integrates joint active and passive beamforming with adaptive user grouping, aiming to maximize the weighted sum rate of communications while also optimizing the received sensing signal power. Further, the authors introduced an algorithm for adaptively assigning users to different groups, effectively minimizing channel correlation within each group. In addition, the authors derived the asymptotically optimal number of user groups. A seven-fold increase in the weighted sum rate was demonstrated compared to the case with no grouping.
}

{\textbf{Lessons learned:} User grouping for RIS-assisted ISAC systems demonstrates that effectively managing channel correlation among users can significantly enhance overall system performance. By separating users with strong channel correlation into distinct groups, both communication and sensing performance can be improved, addressing the limitations of traditional methods. Additionally, deriving the optimal number of user groups emphasizes the need for tailored approaches in resource allocation, showcasing how targeted algorithms can lead to substantial improvements. This underscores the potential of innovative strategies in overcoming challenges posed by dense antenna deployments and resource competition in ISAC systems.}

{\textbf{Challenges, opportunities and open research directions} User grouping in RIS-assisted ISAC systems to address channel correlation faces challenges such as accurately assessing real-time channel conditions and the computational complexity of determining optimal groupings, especially in high-density antenna deployments. Traditional methods often struggle to balance the performance needs of both communication and sensing tasks. However, there are opportunities for innovation by leveraging machine learning techniques to create adaptive algorithms that optimize user assignments based on current conditions. Future research directions could explore hybrid approaches that integrate user grouping with beamforming optimization, assess the impact of user distribution patterns, and investigate multi-objective optimization frameworks that enhance both communication capacity and sensing accuracy. Additionally, real-world testing in diverse environments can help validate these strategies and provide insights for further improvements.}

\subsection{Multiple RISs}
{While employing a single RIS for DFRC can enhance the overall performance and compensate for the absence of the LoS link, deploying more than one RIS can further improve system performance in multiple ways as listed blow.}

{\textit{1) Performance improvement:} Since an RIS constitutes an additional resource that can be configured to enhance system performance, employing multiple RISs yields better performance compared to deploying a single RIS. Building on this, a number of research works have investigated the additional gains resulting from employing multiple RISs distributed in different physical locations within the cell~\cite{2024_Chunjie,2024_Elhag}.}

The authors of~\cite{2022_Wei} and~\cite{2023_Wei} proposed the use of multiple RISs to support a wideband DFRC system. Specifically, an OFDM waveform was employed for the dual purpose of detecting a moving target and simultaneously communicating with multiple users. To handle the unknown Doppler shift at the radar receiver, a Doppler filter bank was introduced. The optimization problem then targets optimizing the transmit beamformer, phase-shift matrix, and Doppler filter bank. This optimization aims to maximize the average SINR across all sub-carriers while ensuring that the average SINR among all users remains above a predetermined threshold, thereby guaranteeing the quality of service for communication users. The numerical findings demonstrated that the multi-RIS-assisted DFRC system significantly outperforms its narrowband, non-RIS-assisted, and single-RIS-assisted counterparts.

{\textit{2) Task division:} An alternative use case of employing multiple RISs is to allocate specific tasks for each of them.} An illustrative case of this situation is presented in~\cite{2023_Sankar} where two RISs are used, with one dedicated for communication and the other for sensing. Although the authors initially explored the use of a single RIS exclusively for communication, the adoption of two RISs demonstrated improvements in both communication and sensing performance. This enhancement is particularly notable when there is no direct path between the sensing targets and the DFRC transmitter. {Other examples of utilizing multiple RISs for task division involve utilizing one RIS for communication only and the other for both communication and sensing~\cite{2024_Fei}, as well as using one RIS for UL transmission and the other for DL to minimize self-interference at the BS~\cite{2024_Li}.}

\begin{figure} 
         \centering
         \includegraphics[width=0.8\columnwidth]{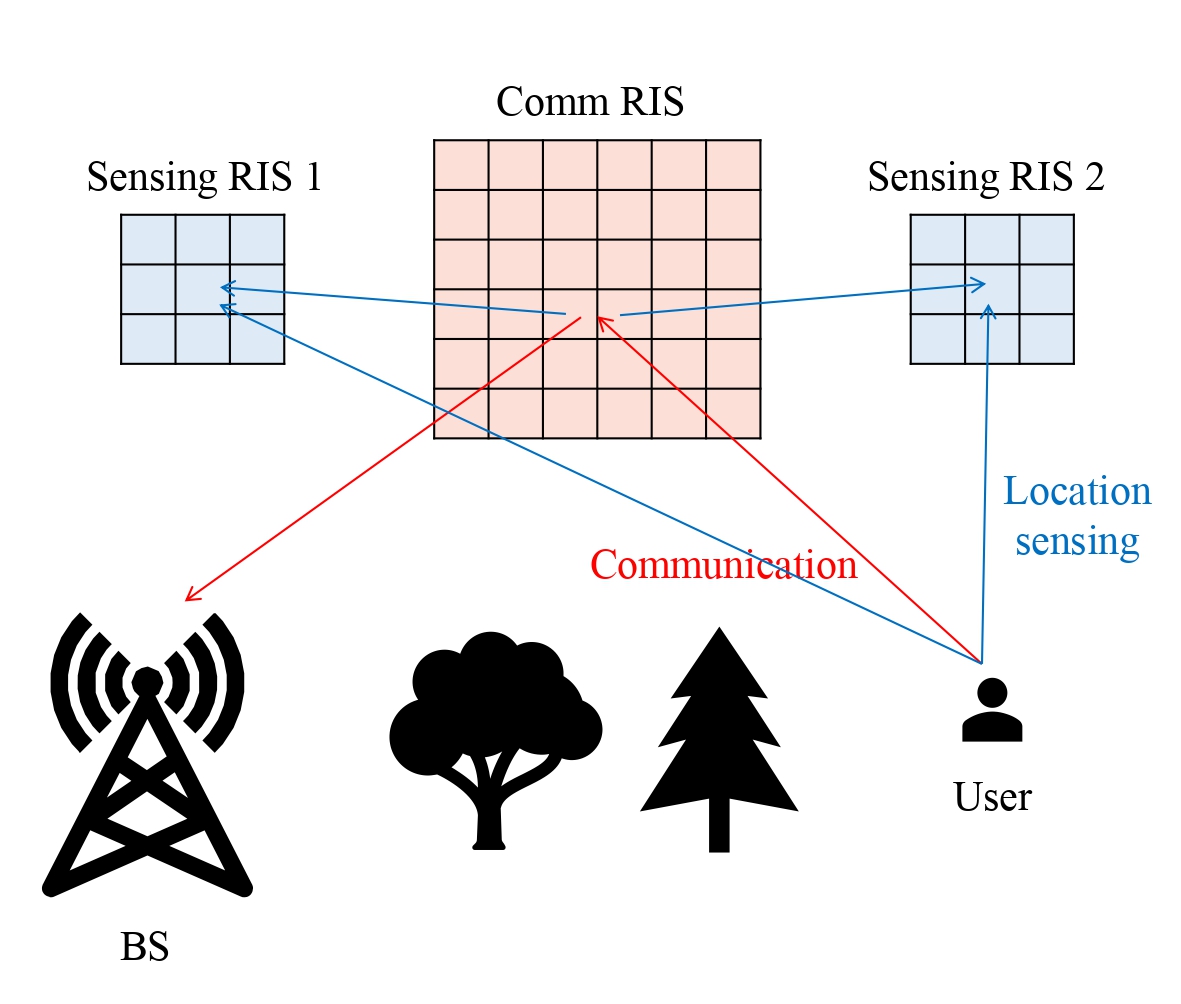}
         \vspace{0.05cm}
        \caption{Sensing-assisted beamforming system model proposed in~\cite{2022_Liu2,2023_Qian,2022_Yu}.}
        \label{fig:DRIS}
\end{figure}

{\textit{3) Sensing-assisted communication:} Since multiple RISs can receive the echo signals reflected from targets at different locations, they can be viewed as localization anchors. These signals can then be processed to estimate the location of the users and reconstruct their channels.} The utilization of distributed RISs was proposed in~\cite{2022_Liu2,2022_Yu} and~\cite{2023_Qian} within a sensing-assisted communication framework. The system model outlined in these papers involves a large-sized reflecting sub-surface and multiple small-sized sensing sub-surfaces, as shown in Fig~\ref{fig:DRIS}. To enhance both the communication and location sensing performance, the authors developed a two-phase ISAC transmission protocol. In the first phase, the user transmits an UL communication signal to the BS via the large-sized reflecting sub-surface. Simultaneously, the two sensing sub-surfaces receive these signals to perform location sensing. The transmission design in this phase relies on limited and imperfect CSI knowledge. The reflected signals from the sensing RISs are then exploited to obtain accurate coarse-grained sensing information, which is then used in the second stage to design the beamformer.

This protocol aims at minimizing the degradation of the communication functionality even when perfect CSI is not available. It then utilizes location sensing as a service to construct the communication channel in the second stage. Notably, through the use of these sensing-based beamforming techniques, the protocol achieves communication performance nearly identical to that of an RIS-aided communication system with perfect CSI knowledge, although the first stage in the protocol relies on limited CSI knowledge. Employing a similar system model but with semi-passive distributed RISs with a few active elements acting as location sensors, the authors of~\cite{2022_Yu3,2023_Yuan,2023_Hu} showed that better system performance can be realized by conducting location sensing in the RIS instead of the BS. {In a similar manner, the exploitation of multiple RISs for sensing-assisted communications was also targeted in other works such as~\cite{2024_Ma,2024_Peng1,2024_Peng} and~\cite{2024_Lyu}.}

{\textbf{Lessons learned:} The use of multiple RISs has been shown to enhance overall system performance beyond what is achievable with a single RIS as each acts as an additional resource. Moreover, multiple RISs can be used by allocating specific tasks to each RIS to better manage the system. For instance, one RIS can be dedicated solely to communication while another can handle sensing tasks. Furthermore, multiple distributed RISs can also facilitate sensing-assisted communication by acting as localization anchors that receive reflected echo signals from different locations. These signals can be processed to estimate user locations and reconstruct their channels. The sensing-assisted beamforming approach effectively addresses communication-sensing tradeoffs, optimizing both communication and sensing performance. Notably, this approach achieves communication performance nearly identical to that of an RIS-aided communication system with perfect CSI knowledge without the need for pilot transmission.}

\textbf{Challenges, opportunities and open research directions:} Deploying and managing multiple RISs can increase the overall complexity and cost of the system. This includes hardware costs, installation costs, and maintenance efforts associated with multiple RIS units. Also, coordinating the actions of multiple RISs to avoid interference and optimize their combined effect is a non-trivial task. Ensuring that the RISs work together seamlessly without causing detrimental effects to each other or other communication systems is a significant challenge in this domain. In addition, with multiple RISs, accurate and efficient channel estimation becomes more challenging. Coordinating the estimation of channels between different RISs and the various communication and radar links requires sophisticated channel estimation algorithms and large pilot overhead. Moreover, allocating resources optimally and maintaining synchronization among different RISs and for different functions (radar and communication), is more complex in the case of multiple RISs.

While employing RISs for DFRC introduces added complexity, it also offers various opportunities. First, multiple RISs provide the opportunity to exploit spatial diversity and create favorable multipath conditions for both radar and communication purposes, which have great potential in enhancing the performance of both systems. Furthermore, the use of multiple RISs allows a better optimization of radar and communication functionalities. Coordinated optimization can lead to improved overall system performance and better tradeoffs between conflicting objectives.

Also, multiple RISs can be strategically placed to enhance coverage in specific areas, providing better service for both communication and radar applications. This can be particularly beneficial in scenarios with challenging propagation environments. Distributing the processing tasks among multiple RISs presents another opportunity as it can lead to more efficient and scalable DFRC systems. In this setup, each RIS can handle specific functions independently, reducing the burden on individual units. Finally, multiple RISs enable greater flexibility of dynamic adaptation to changing environmental conditions. This results in systems capable of adjusting the configuration of RISs in real time to optimize performance based on the current requirements and channel conditions.

\begin{table}
\footnotesize
\centering
\caption{References with different RIS types for RIS-assisted DFRC systems.}
\begin{tabular} {|m{1.5cm}| m{0.9cm} | m{0.9cm} |m{0.9cm} |m{0.9cm} |m{0.9cm} |} 
 \hline 
 References &  STAR-RIS & Semi-passive RIS  & Active RIS & BD-RIS & Hybrid RIS \\  [0.5ex] 
 \hline 
 \hline 
 \cite{2023_Ziheng,2022_Wang10,2023_Wang,2023_Liu3,2023_Sun,2023_Wang5,2023_Zhu4,2023_Liu5,2024_Shuang,2024_Zhengyu,2024_Wei,2024_Hanxiao,2024_Zhang,2024_Xu,2024_Wang2} & \checkmark &   &  &  & \\
 \hline 
\cite{2022_Yu3,2023_Hu,2023_Asif,2023_Haider}  &  & \checkmark  &  &  & \\
 \hline 
\cite{2023_yu2,2023_Zhu3,2023_Hao,2023_Chen4,2023_Zhang,2023_Li,2023_Zhu,2023_Salem,2024_Yang,2024_Miaomiao,2024_Wang} &   &   &  \checkmark &  & \\
\hline 
 \cite{2023_Zhang6,2023_Zhang2} & \checkmark  &  \checkmark &   &  & \\
 \hline 
  \cite{2023_Zhang5} & \checkmark &   &  \checkmark  &  & \\
  \hline 
    \cite{2024_Tara,2024_Guang} &  &   &   & \checkmark & \\
 \hline
   \cite{2023_Wang3} & \checkmark  &  \checkmark &   & \checkmark & \\
   \hline
   \cite{2024_Hao,2024_Zhao,2023_Liao2,2023_Liu8,2022_Sankar,2023_Florian} &   &   &   & &  \checkmark \\
 \hline
 \cite{2024_Magbool1} &   &  \checkmark & \checkmark  & &  \\
 \hline
\end{tabular}
\label{tab:RIS_types}
\end{table}

\subsection{STAR-RISs} \label{ssec:other_RIS}
Several papers proposed the use of different types of RISs for DFRC. Table~\ref{tab:RIS_types} presents these papers together with the relevant RIS type  used in each of them.

Exploiting the comprehensive spatial coverage of STAR-RISs, researchers have explored the application of STAR-RISs in DFRC. In general, there are three use cases of STAR-RISs for DFRC.

\begin{figure*}
  \centering
  \begin{tabular}{c c}
    \includegraphics[width=0.95\columnwidth]{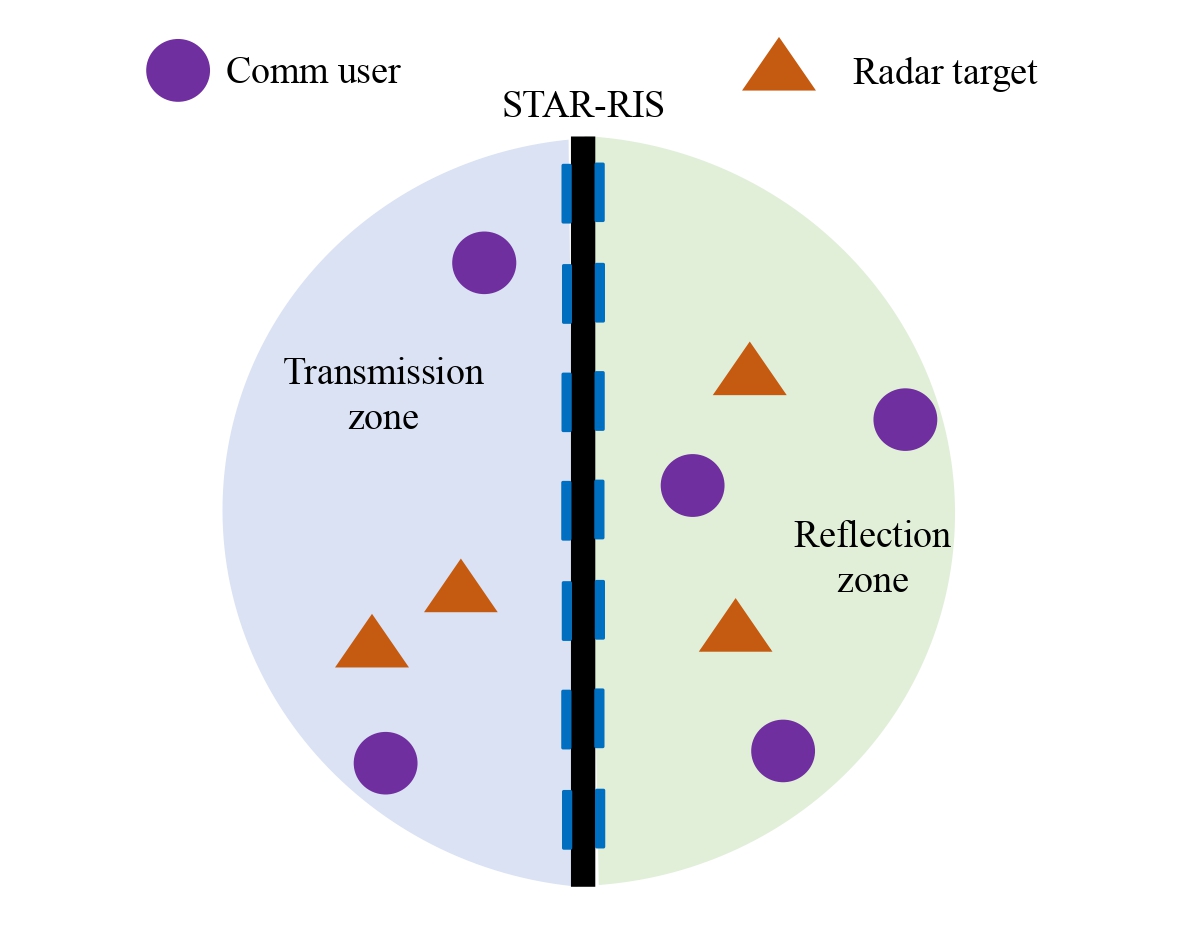} &
      \includegraphics[width=0.95\columnwidth]{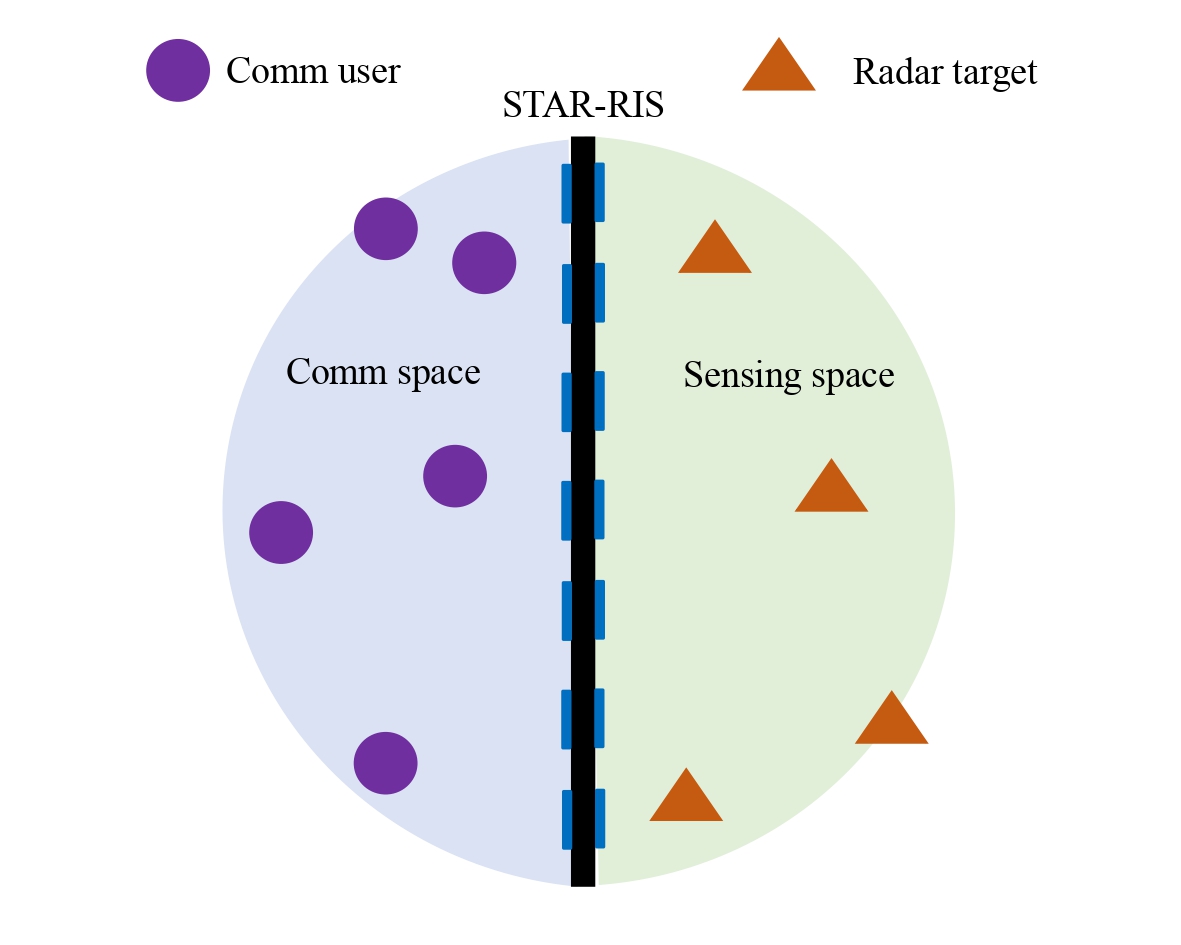}  \\
      \scriptsize (a)    &
      \scriptsize (b)   \\
    \includegraphics[width=0.95\columnwidth]{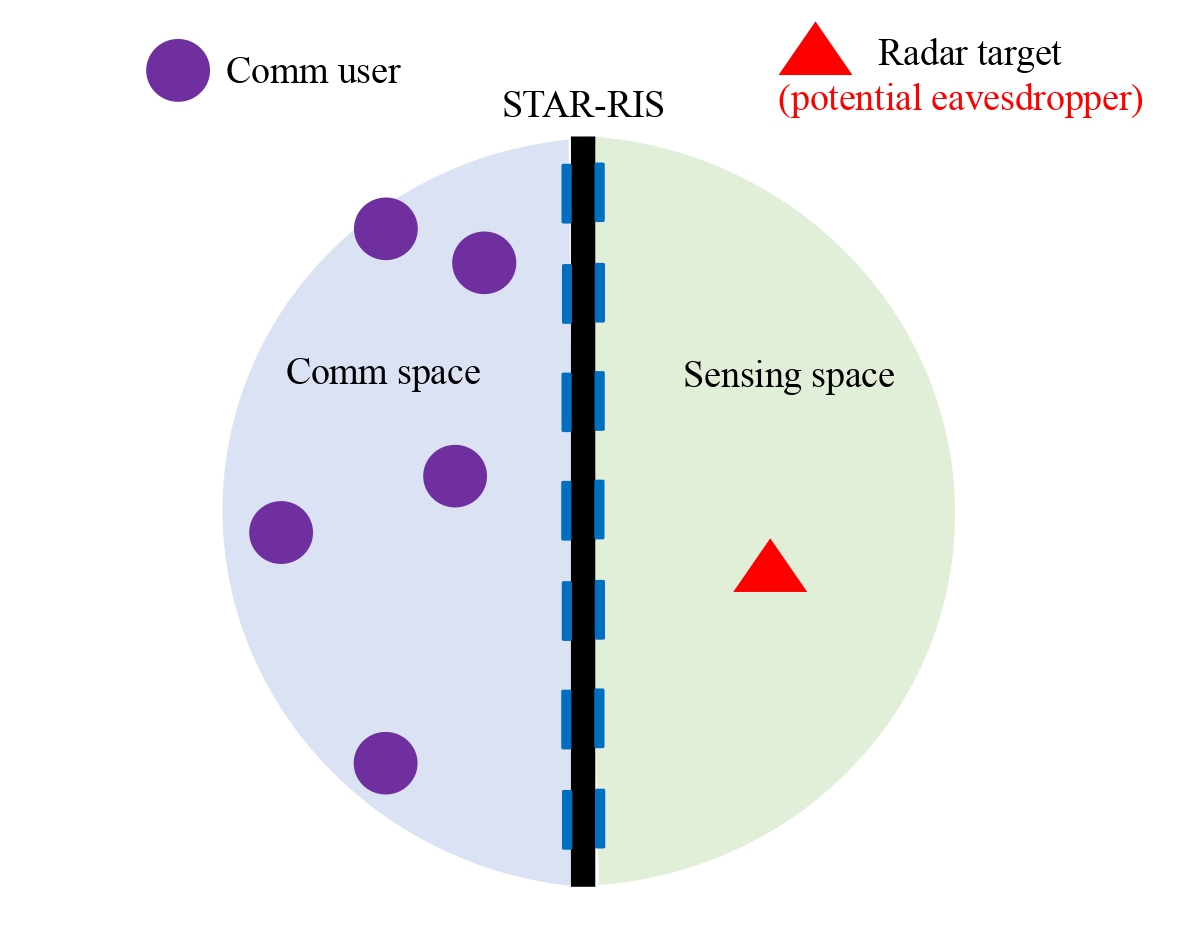} &
      \includegraphics[width=0.95\columnwidth]{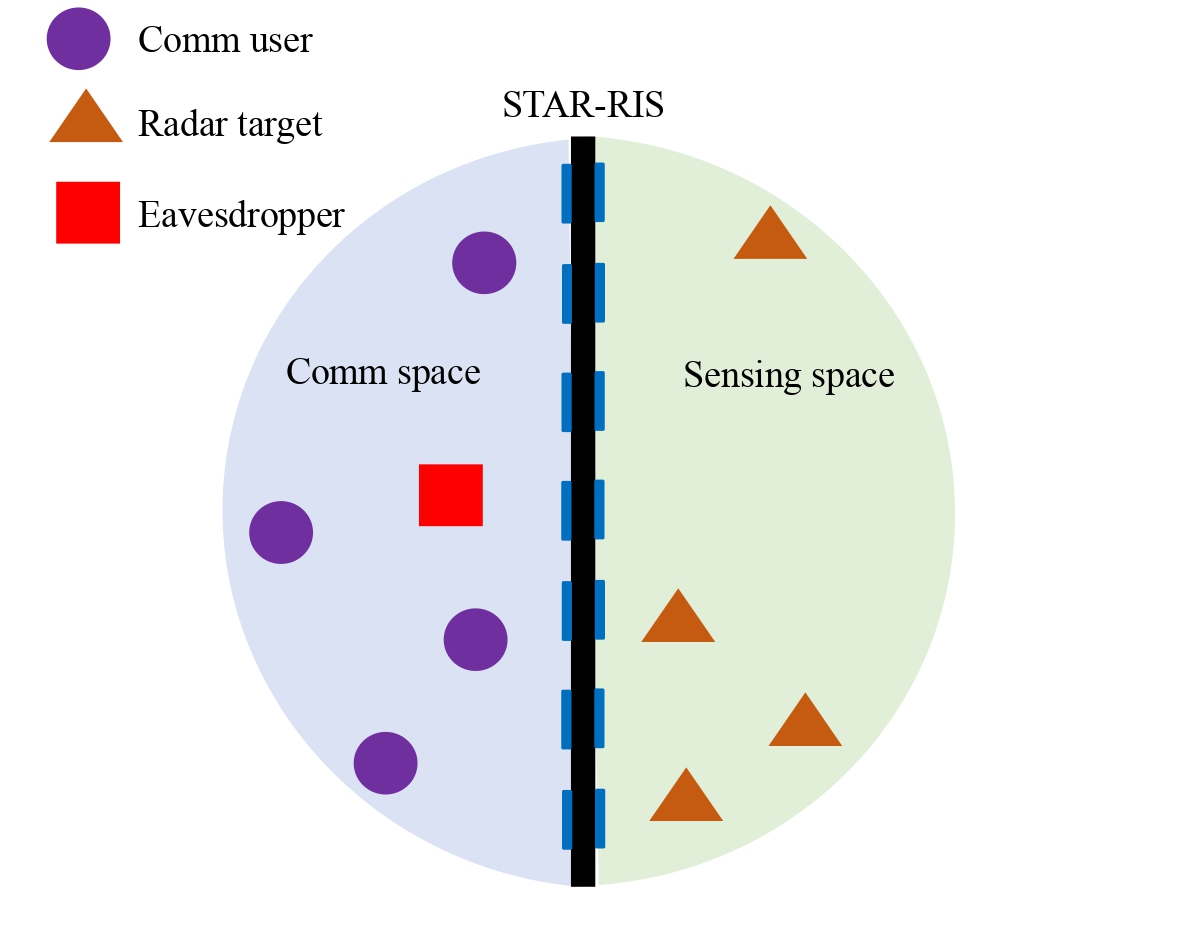}  \\ \vspace{-0.15cm}
      \scriptsize (c)   &
      \scriptsize (d) 
  \end{tabular}
    \medskip
  \caption{The main use cases of STAR-RIS in DFRC systems. (a) STAR-RIS for converge extension in DFRC. (b) STAR-RIS for space partitioning in DFRC. (c) STAR-RIS for PLS in DFRC when the target is a potential eavesdropper. (d) STAR-RIS for PLS in DFRC with an eavesdropper in the communication space.}
  \label{STAR_RIS_DFRC}
\end{figure*}

{\textit{1) Coverage extension:}} The first application is to allow a $360^\circ$ coverage for DFRC instead of the $180^\circ$ coverage offered by conventional RISs, as shown in Fig.~\ref{STAR_RIS_DFRC}(a)~\cite{2023_Ziheng,2024_Zhang,2024_Xu}. A basic model for this case was outlined in~\cite{2023_Ziheng}, featuring multiple communication users and radar targets distributed on both sides of the STAR-RIS. 

{\textit{2) Task-based zone division:}} The second use case of STAR-RISs in DFRC systems is to exploit one of the zones for communication and the other for sensing~\cite{2022_Wang10,2023_Wang,2023_Liu3,2024_Shuang,2024_Hanxiao,2024_Wang2}, resulting in distinctive communication and sensing spaces, as shown in Fig.~\ref{STAR_RIS_DFRC}(b). This division based on tasks provides better controllability over prioritizing either communication or sensing capabilities by selecting appropriate values of the amplitude parameters to satisfy the energy splitting condition. Additionally, this space division allows better control over interference. In this situation, STAR-RIS can be considered as a signal processing entity which filters out the communications signal from the sensing space and the sensing signal from the communications space.

{\textit{3) PLS enhancement:}} The third use case of STAR-RIS is to enhance PLS. This can be achieved by multiple means, depending on the possible location of the eavesdropper~\cite{2023_Sun,2023_Wang5,2023_Zhu4,2023_Liu5,2024_Zhengyu,2024_Wei}. The approach of~\cite{2023_Sun} attempted to limit the information leakage from the communication domain to the sensing domain, thereby minimizing the risk of a security threat from an eavesdropper in the sensing domain, as depicted in Fig.~\ref{STAR_RIS_DFRC}(c). The problem is formulated to maximize the sensing beampattern gain while ensuring constraints on the communication users' SINRs. At the same time, the problem formulation ensures that the information leakage, defined as the useful communication SINR in the sensing domain, remains below a certain threshold. Similarly, the presence of public and secure communication users was discussed in~\cite{2023_Wang5}, with secure transmission only granted for the secure users. This scenario is suitable when the information of some users is not sensitive, allowing the removal of some security constraints to enhance the system performance.

A more challenging scenario was presented in~\cite{2023_Zhu4,2023_Liu5} with a potential eavesdropper in the communication space, as shown in Fig.~\ref{STAR_RIS_DFRC}(d). Since both the legitimate user and the eavesdropper are located in the same space, the problem is formulated to maximize the average long-term security rate of the legitimate user, while imposing thresholds on the communication and radar SINRs.

{\textbf{Lessons learned:} The $360^{\circ}$ spatial coverage of STAR-RISs allows for simultaneous support of multiple communication users and radar targets. Moreover, STAR-RISs facilitate task-based zone division, where one zone is used for communication and the other for sensing. This division allows for better control and prioritization of communication and sensing functions by adjusting amplitude parameters and managing interference. Furthermore, STAR-RISs enhance PLS by minimizing information leakage between communication and sensing domains, crucial for protecting against eavesdroppers in the sensing domain. In addition, STAR-RISs can handle scenarios with eavesdroppers in the communication space. }

{\textbf{Challenges, opportunities and open research directions:} One of the primary challenges in the deployment of STAR-RISs is the complex beamforming optimization required for both communication and sensing functions, especially with the additional non-linear and non-convex constraints STAR-RISs impose. Although task-based zone division in STAR-RISs helps mitigate interference by separating communication and sensing spaces, maintaining optimal performance consistently in real-world scenarios remains difficult.} {Despite these challenges, STAR-RISs offer numerous opportunities. One significant opportunity lies in the potential for enhanced network performance. By extending coverage, improving signal strength, and reducing interference, STAR-RISs can make communication networks more reliable and efficient, particularly in high-mobility scenarios such as intelligent transportation systems. Moreover, the integration of STAR-RISs provides a platform for developing more advanced and dynamic security protocols. Enhanced PLS can be achieved, protecting sensitive information in communication networks and mitigating the risk of eavesdropping.}

{Dynamic interference management is a critical research area. More sophisticated techniques that can adapt to high-mobility environments and varying user densities are required to maintain optimal performance. Furthermore, robust and scalable hardware design is another open research direction. Investigating the design and deployment of STAR-RIS hardware that can perform reliably in various environmental conditions and integrate seamlessly with existing infrastructure is crucial for the technology's success. Finally, the incorporation of machine learning techniques offers further opportunities. These technologies can be used to optimize beamforming, interference management, and security measures, leading to smarter and more adaptive DFRC systems.}

\subsection{Semi-Passive RISs} 
Semi-passive RISs are RISs equipped with a limited number of active sensors. In contrast to hybrid passive and active RISs, where the active elements serve both communication and sensing purposes, the active sensors in semi-passive RISs are exclusively deployed to fulfill the sensing functionality.

{\textit{1) Sensing-assisted communications:}} The authors of~\cite{2023_Asif,2023_Hu,2022_Yu3} and~\cite{2023_Haider} investigated the use of semi-passive RISs in a sensing-assisted communications framework. In particular, a two-stage procedure was proposed. In the first stage, multiple semi-passive RISs are employed to sense the users' locations and simultaneously provide limited communication performance based on outdated channel knowledge. The channel matrix is then constructed based on the sensed locations. In the second stage, communication is performed via a passive RIS. This method eliminates the need for dedicated pilot symbols for channel estimation, resulting in higher data rates.

{\textit{2) STAR-RISs and BD-RISs:}} The use of semi-passive STAR-RISs was proposed in~\cite{2023_Zhang6} and~\cite{2023_Zhang2}. Given a fixed number of RIS elements, the authors in~\cite{2023_Zhang6} investigated the optimal ratio of active sensors to passive elements in a semi-passive STAR-RIS-assisted system. While deploying additional sensors has the potential to enhance the resolution of echo processing, increasing the number of passive elements can introduce more spatial degrees of freedom, benefiting both communication and sensing performance. Consequently, a tradeoff exists between the number of active sensors and passive elements. Two main findings are observed in~\cite{2023_Zhang6}. First, the STAR-RIS architecture ensures that the sensing performance is not compromised by reducing the number of receive antennas at the BS, as long as the QoS requirements are met. Second and more importantly, when considering a constant number of STAR-RIS elements, enhancing the sensing performance by deploying additional passive elements is more attractive than increasing the number of active sensors. The use of semi-passive BD-RISs was proposed in~\cite{2023_Wang3} to eliminate the need for a one-to-one connection between the receive and transmit parts of the RIS elements, resulting in a non-diagonal RIS matrix.

{\textbf{Lessons learned:} Research has demonstrated the potential of semi-passive RISs for sensing-assisted communications, eliminating the need for dedicated pilot symbols for channel estimation and thereby increasing data rates. The optimal balance between active sensors and passive elements in semi-passive STAR-RIS systems has been explored, revealing that while additional sensors can enhance echo processing resolution, increasing passive elements offers more spatial degrees of freedom, benefiting both communication and sensing. Thus, enhancing sensing performance through additional passive elements is more advantageous than increasing active sensors.}

{\textbf{Challenges, opportunities and open research directions:} One of the main challenges of semi-passive RISs is determining the optimal ratio of active sensors to passive RIS elements. This difficulty arises because the ratio can be influenced by factors such as the number and locations of users and the imposed communication and/or sensing constraints, making an average evaluation of this ratio more practical than an instantaneous one. Another challenge is the hardware design for elements that can switch from being reflective RIS elements to radar sensors. While a sensor can be considered an absorptive RIS element, the practical implementation of such hardware must be examined carefully in real-world scenarios.}

\subsection{Hybrid and Active RISs} 
{Active and hybrid RISs can be utilized for DFRC in several ways.}

{\textit{1) Performance improvement:}} The issue of double path loss becomes more pronounced in ISAC systems when the direct LoS path between the BS and the sensing targets is obstructed. In such scenarios, the signal must travel from the BS to the target through the RIS and return in the opposite direction after being reflected from the target. This results in a quadruple path loss, yielding a significantly weakened signal at the radar receiver. To address this issue, several research studies proposed the implementation of active RISs, which have the capability of amplifying the impinging signals~\cite{2023_yu2,2023_Zhu3,2023_Hao,2023_Chen4,2023_Zhang,2023_Li,2023_Zhu,2023_Zhang5}. A thorough comparison between active and passive RIS-assisted ISAC systems is presented in~\cite{2023_Rihan2}. The main conclusion from this study was that active RISs can achieve significantly better system performance than their passive counterparts under the same power budget constraint taking advantage of splitting the power between the transmitter and the RIS.

Active RISs, while offering superior performance compared to their passive counterparts, tend to consume more power, potentially impacting the overall energy efficiency of the system. To mitigate this challenge, employing hybrid RISs, predominantly composed of passive elements but with a few active elements, for ISAC was introduced in~\cite{2023_Liao2,2023_Liu8,2022_Sankar,2023_Hui2,2024_Hao,2024_Zhao,2023_Florian}. In contrast to semi-passive RISs, where active sensors are utilized solely for sensing purposes, the active elements in hybrid RISs serve a dual purpose, enhancing both communication and sensing performance.

{\textit{2) PLS enhancement:}} The application of active RISs for enhancing PLS has been explored in the literature, as highlighted in several key studies~\cite{2023_Salem,2024_Yang,2024_Miaomiao,2024_Wang}. These works demonstrate that the unique signal amplification capabilities of active RISs can significantly bolster the secrecy rate, which is a critical metric in PLS. For example, it has been shown in~\cite{2023_Salem} that by leveraging the amplification properties of active RISs, a remarkable improvement in the secrecy rate, amounting to approximately 220\%, can be achieved when compared to the traditional passive RIS approach. Notably, this enhancement is realized without exceeding the same power budget, underscoring the efficiency and potential of active RISs in securing wireless communications against eavesdropping threats. This substantial improvement illustrates the pivotal role that active RISs can play in advancing the security of next-generation wireless networks. 

{\textbf{Lessons learned:} Active RISs have been shown to effectively mitigate the issue of quadruple path loss in ISAC, which occurs when the direct path between the BS and sensing targets is obstructed. By amplifying the impinging signals, active RISs can substantially improve the signal strength at the radar receiver, outperforming passive RISs under the same power budget. However, the increased power consumption of active RISs presents a challenge to energy efficiency. To address this, hybrid RISs, which combine mostly passive elements with a few active ones, have been proposed. These hybrid RISs offer a balanced tradeoff by enhancing both communication and sensing performance without significantly increasing power demands. Furthermore, active RISs have been demonstrated to significantly enhance PLS by boosting the secrecy rate, making them a promising solution for securing wireless communications in next-generation networks. This body of research highlights the critical role that active and hybrid RISs can play in improving both the performance and security of ISAC systems.}

{\textbf{Challenges, opportunities and open research directions:} On one hand, while active RISs can significantly enhance system performance by amplifying weak signals and mitigating the issue of quadruple path loss, they also increase power consumption, potentially affecting overall energy efficiency. Hybrid RISs, which combine passive and active elements, offer a promising compromise by balancing performance and energy efficiency, yet they introduce complexity in design and management. Moreover, active RISs have shown substantial potential in improving PLS by enhancing the secrecy rate, demonstrating a significant advancement over passive RISs. This improvement, however, comes with the challenge of optimizing power usage and system design to fully leverage the benefits of active RISs without compromising energy efficiency. Future research directions could focus on optimizing hybrid RIS configurations, exploring energy-efficient amplification techniques, and developing robust models to balance performance with power consumption, thereby advancing the integration of active RISs in secure and efficient wireless communications.}

\subsection{BD-RISs}
{BD-RISs provide better beamforming gain at the RIS, as they enable all-to-all connections within the RIS globally or within specific groups. As a result, they have been shown to offer improved performance for ISAC, as demonstrated in several research studies. }

{\textit{1) Conventional DB-RIS:} The work of~\cite{2024_Guang} examined the problem of transmit power minimization in BD-RIS-assisted ISAC, while adhering to QoS constraints for both communication and sensing performance. The authors demonstrated that a reduction of approximately 25\% in the transmit power can be realized when a BD-RIS is used instead of a diagonal one, which is a significant advantage in massive MIMO networks. In~\cite{2024_Tara}, the authors aimed to maximize the weighted sum of radar and communication SNRs for an ISAC system with a BD-RIS. Thanks to the extra degrees of freedom provided by BD-RIS, they reported an increase in the communication SNR from \unit[-2.5]{dB} to \unit[2.5]{dB}, and an increase in the radar SNR from \unit[0]{dB} to \unit[12.5]{dB}, when a BD-RIS is used compared to a diagonal RIS.}

{\textit{2) Semi-passive STAR-DB-RIS:} A more advanced RIS was studied in~\cite{2023_Wang3}, which incorporates the beyond-diagonal structure of the RIS matrix along with the capability to simultaneously transmit and reflect signals (i.e., STAR-RIS). To improve sensing performance, the authors also embedded radar sensors at the RIS. Their goal was to maximize the minimum signal-to-clutter-plus-noise ratio of the sensing target, subject to communication QoS constraints. The study showed a 37.5\% improvement in the minimum signal-to-clutter-plus-noise ratio for the BD-RIS case, bringing it very close to the performance of a radar-only system, where communication QoS constraints are not imposed. This highlights the advantage of BD-RIS in improving the tradeoff between communication and sensing performance in ISAC.}

{\textbf{Lessons learned:} BD-RISs provide key benefits for ISAC systems by enhancing beamforming gain and reducing the transmit power needed, making them well-suited for massive MIMO networks. They also improve both communication and radar performance compared to diagonal RIS. Additionally, advanced BD-RIS designs further enhance sensing performance, approaching the effectiveness of radar-only systems. Overall, BD-RIS strikes a better balance between communication and sensing, outperforming standard RIS structures.}

{\textbf{Challenges, opportunities and open research directions:} One key challenge is the complexity of designing and optimizing RIS configurations for both communication and sensing tasks, especially in dynamic environments. The need for real-time adaptation and resource allocation also requires more advanced algorithms and hardware capabilities. However, the opportunities are significant, including the potential for greater energy efficiency, enhanced signal quality, and improved integration of communication and sensing functions. Open research directions include exploring more efficient algorithms for joint optimization, improving scalability for massive MIMO, and developing practical implementations of advanced RIS designs like STAR-RIS. Additionally, the tradeoffs between performance, complexity, and cost need further investigation to make BD-RIS a more viable solution in real-world applications.}

\section{RIS for Advanced DFRC Operations}  \label{sec:RIS_DFRC3}
{Several research works investigated the use of RISs to support advanced DFRC operations. This section offers a summary of these works.}

\subsection{Full-Duplex Systems}
A significant body of research on RIS-assisted ISAC has predominantly focused on half-duplex configurations. Transmitting and receiving using different frequency bands (i.e., frequency-division multiplexing (FDM)) poses a challenge in ISAC as both communication and sensing performance are highly reliant on the frequency band~\cite{2022_Zhang3}. Alternatively, transmitting and receiving at different times, known as time-division multiplexing (TDM), is commonly considered in ISAC system models, but it still imposes limitations on system capabilities such as degrading the communication rate. On the other hand, full-duplex systems, where the transmission, reception and echo processing happens in the same time-frequency resources, can offer improved system performance. This has spurred the interest of several researchers to investigate the performance of full-duplex RIS-assisted ISAC systems~\cite{2023_Nhat,2022_Shao,2023_Yuan2,2024_Bayraktar}.

For example, the authors in~\cite{2023_Yuan2} investigated maximizing the data rates for the communication users while ensuring high-quality radar sensing in a full-duplex RIS-assisted ISAC system. Additionally, this paper investigates a comprehensive signal propagation model that fully considers the impact of the RIS on both the forward and reflected radar probing signals. By employing the PDD method along with the MM algorithm, the authors found that the full-duplex system is capable of increasing the communication sum rate of the system by 100\% and 125\% compared to the corresponding half-duplex systems with and without RIS, respectively. More interestingly, the full-duplex RIS-assisted system is capable of achieving a better sum rate even with a high radar SNR requirement than its half-duplex counterpart, resulting in excellent communication and sensing performance.

{\textbf{Lessons learned:} Full-duplex ISAC, which allows simultaneous transmission, reception, and echo processing in the same time-frequency resources, demonstrate enhanced performance potential. Recent studies have explored full-duplex RIS-assisted ISAC systems, with findings indicating substantial improvements in communication and sensing performance. Research shows that full-duplex configurations can significantly enhance the communication sum rate compared to half-duplex systems, even under high radar SNR requirements. These advancements highlight the advantages of full-duplex systems in achieving superior communication and sensing performance, prompting further investigation into their practical applications and implications in ISAC.}

\textbf{Challenges, opportunities and open research directions:} Full-duplex systems require sophisticated signal processing techniques to effectively separate the transmitted and received signals. This complexity increases in the presence of an RIS and the dual functions of DFRC, necessitating advanced algorithms for self-interference cancellation and signal enhancement. Practical implementation of full-duplex systems involves challenges in designing transceiver hardware that can handle simultaneous transmission and reception without significant degradation in performance. One accentuated hardware consideration is the increased power consumption requirement compared to half-duplex systems. This can be a concern, especially in energy-constrained scenarios or battery-operated devices.

The opportunities of full-duplex RIS-assisted DFRC systems include enhanced spectral efficiency. By enabling simultaneous transmission and reception, full-duplex systems can achieve higher spectral efficiency, leading to improved sensing and communication performance. Full-duplex DFRC also enhances the reliability of DFRC by allowing for real-time feedback and re-transmission in case of errors, improving the quality of service for communication and sensing applications. Moreover, channel reciprocity can be exploited in the full-duplex system to minimize the complexity of channel estimation, target detection, and resource allocation.

\subsection{Wideband Systems}
Using a CCD, the authors in~\cite{2023_Liao,2023_Wei,2022_Wei3,2024_Wang1} and~\cite{2024_Zhang1} focused on employing a wideband OFDM waveform in RIS-assisted ISAC systems. {These works studied RIS-assisted wideband ISAC from three different perspectives.}

{\textit{1) Resource allocation:} The resource allocation problem for wideband ISAC was studied in~\cite{2023_Liao,2023_Wei} and~\cite{2022_Wei3}.} In~\cite{2023_Liao}, the authors investigated a wideband RIS-assisted ISAC system in a clutter-rich environment caused by the NLoS signals, an aspect that is often overlooked in RIS-assisted system research. The BS aims to simultaneously detect multiple targets and communicate with multiple users with the aid of an RIS. A joint RIS-assisted ISAC design was proposed to maximize the minimum sensing beampattern gain while ensuring a certain QoS requirement for each communication user, with the objective of reducing clutter interference caused by the NLoS signal components. The results show that the presence of an RIS can increase the minimum beampattern gain by a factor of six, thanks to the RIS's capability of reducing clutter interference.

{\textit{2) Resolving the scaling ambiguity of the NLoS sensing:} The problem of parameter estimation for an RIS-assisted sensing system with OFDM waveforms was examined in~\cite{2024_Wang1}. Wideband sensing results in scaling ambiguity when the channel matrix has a rank of one caused by having only one degree of freedom to simultaneously determine both the path gain and the angles of the target. To overcome this scaling ambiguity, the authors proposes a two-phase approach. The received echo signals from the two phases are collected and represented as third-order tensors. A method based on canonical polyadic decomposition is then developed to estimate the parameters of each target, including the direction of arrival, Doppler shift, and time delay.}

{\textit{3) Cloud radio access network (C-RAN):} An RIS-assisted ISAC C-RAN was studied in~\cite{2024_Zhang1}, where multiple remote radio heads transmit OFDM ISAC signals to illuminate targets while simultaneously communicating with users cooperatively. The goal was to minimize the error between the actual and desired space-frequency beampatterns of the remote radio heads while meeting user communication rate targets and fronthaul capacity requirements.}

{\textbf{Lessons learned:} Studies show that RIS can significantly enhance sensing performance by reducing clutter interference and improving beampattern gains. Moreover, the issue of scaling ambiguity in parameter estimation for RIS-assisted sensing systems can be addressed by employing a two-phase sensing approach. This method allows for more accurate estimation of target parameters, such as direction of arrival, Doppler shift, and time delay, by using canonical polyadic decomposition on third-order tensors. In addition, RIS deployment in C-RAN can optimize space-frequency beampatterns for remote radio heads, effectively balancing communication and sensing performance while meeting fronthaul capacity constraints.}

\textbf{Challenges, opportunities and open research directions:} Wideband operation introduces increased complexity in signal processing and hardware design. Handling a broader spectrum of frequencies requires more sophisticated algorithms and flexible hardware architectures. In wideband scenarios, accurate channel estimation becomes more challenging due to frequency-selective fading and the beam-squint effect. When ignored, frequency-selective fading can also affect the performance of the system, as the reflection patterns of the RIS elements are frequency-dependent. Furthermore, wideband operation often demands higher power consumption compared to narrowband systems. Managing power requirements becomes a critical consideration, especially in energy-constrained scenarios.

On the other hand, various opportunities can be realized from wideband RIS-assisted DFRC systems. The most obvious one is the enhanced spectrum utilization. Wideband operation allows for more efficient use of the spectrum, aligning with the fundamental goal of ISAC. Moreover, the use of wideband signals in radar applications enhances range and Doppler resolution. This improvement can lead to enhanced target discrimination and tracking capabilities, especially in complex and dynamic environments.

\subsection{Physical Layer Security} \label{sec:pls}
Sharing the waveform, beamformers, and spectrum in DFRC can introduce several security threats. Since the DFRC BS transmits a combined radar and communications signal, the risk of information leakage to illegitimate users exists in these systems. The capability of RISs to enhance PLS in communication systems has been investigated in several studies. Motivated by this, several research works have investigated the capability of RISs to enhance PLS in communication systems~\cite{2019_Chen,2021_Zhang3,2021_Jiayi,2021_Feng,2022_Xu}. While some of these works were already discussed in Section~\ref{sec:RIS_DFRC2}, we overview two additional aspects in this subsection.

\textit{1) Propagation manipulation:} Since RISs can adjust the phase of the incident signals, they can direct the signal towards specific desired locations. This implies that RISs can be used as an additional resource to optimize a system security measure. One of the most common security measures for communications is the secrecy rate, which is defined as the amount of information that can be reliably transmitted per channel use that is also inaccessible to potential attackers. The secrecy rate can be expressed as the difference between the rate of the legitimate user and the rate of the eavesdropper. If this difference is negative, the secrecy rate is set to zero.

The secrecy rate optimization problem can manifest itself in various forms depending on the potential eavesdropper's location~\cite{2023_Hua2,2022_Chen2,2023_Chen2,2023_Zhao3,2024_Zhang3,2024_Sun}. For instance, in~\cite{2023_Hua2}, the authors considered a scenario where the target is a potential eavesdropper, and attempted to minimize the information accessible to the target by imposing a constraint on the maximum information leakage. The scenario of a potentially eavesdropping target was also addressed in~\cite{2022_Chen2} with the objective of maximizing the system's secrecy rate. This was further studied with partial CSI knowledge at the BS in~\cite{2023_Chen2}. Another system model was explored in~\cite{2023_Zhao3}, assuming the potential eavesdropper to be one of the communication users. This scenario is more challenging, as the security measures should be carefully imposed, while at the same time they should not severely affect the communication performance. To address this problem, the authors aimed to minimize the maximum eavesdropping SINR while ensuring QoS requirements for the communication users. 

\textit{2) Artificial noise:} Another approach to enhance the security of DFRC systems is to incorporate a noise-like signal into the transmit waveform~\cite{2022_Mishra,2023_Li7,2021_Fang,2023_Zhang7,2024_Su,2024_Li3}. Thus, the transmit communication signal takes the following form:
\begin{equation}
    \mathbf{x}_c = \mathbf{W} \mathbf{s} + \mathbf{W}_n \mathbf{n}_a,
\end{equation}
where $\mathbf{W}$ and $\mathbf{W}_n$ are the information and the artificial noise precoders, respectively, $\mathbf{s}$ is the information symbol vector, and $\mathbf{n}_a$ is the artificial noise vector with elements drawn from a specific distribution, usually Gaussian. 

While the artificial noise affects both the legitimate user and the eavesdropper, the precoder of the noise is co-designed with the information precoder and the RIS phase shifts to maximize a system security metric while ensuring a threshold on the communication performance. For example, it was shown in~\cite{2022_Mishra} that the presence of an RIS leads to a seven-fold increase in the secrecy rate for DFRC systems when the artificial noise approach is employed.

{\textbf{Lessons learned:} In DFRC systems, sharing the waveform, beamformers, and spectrum introduces security risks, such as potential information leakage to unauthorized users. RISs offer a promising solution to enhance PLS by manipulating signal propagation or incorporating artificial noise into transmissions. RISs can optimize a security performance measure such as the secrecy rate by adjusting the signal to favor legitimate users while minimizing the risk of interception by eavesdroppers. Research has also showen that incorporating artificial noise into the transmission can confuse eavesdroppers while maintaining communication and sensing performance. Co-designing the artificial noise with RIS phase shifts and information precoders can significantly boost system security, with studies showing a dramatic increase in secrecy rates when RISs are employed.}

\textbf{Challenges, opportunities and open research directions:} The combination of radar and communication signals in DFRC systems introduces the risk of information leakage. Securing the system against eavesdropping and unauthorized access becomes more complex due to the multifunctional nature of the signals. Additionally, the dynamic and often unpredictable nature of the environment, including user mobility and changing channel conditions, poses challenges for designing robust security mechanisms. {While the utilization of RISs in enhancing PLS is promising,~\cite{2024_Beiyuan} has described an RIS-based attack scheme that can compromise PLS in the presence of a malicious RIS that can manipulate both radar and communication signals. In this scheme, the attacker uses a malicious RIS and optimizes the phase shifts of the RIS to either reduce the radar echo power received by the BS or enhance the wiretap link for a pilot spoofing attack, based on statistical channel state information. The authors demonstrated therough simulations that such as attack can significantly lower the secrecy rate of the legitimate user while remaining undetected by the DFRC BS.}

On the flip side, machine learning algorithms can be employed to enhance PLS in RIS-assisted DFRC systems. These algorithms can adaptively learn and optimize security measures based on real-time observations, improving the system's ability to detect and respond to security threats. Furthermore, RISs can be utilized for secure key distribution, enabling the establishment of cryptographic keys for communication data encryption. However, realizing these opportunities requires addressing several technical challenges. Efficient algorithms for secure beamforming, dynamic adaptation of security measures, and robust key management in the presence of RISs need to be developed. Moreover, the integration of security mechanisms should not compromise the primary functions of radar and communications, necessitating a careful balance between security and performance.

{
\subsection{Covert ISAC}
The concept of covert communications refer to the practice of transmitting information in a way that hides not only the content of the message but also the existence of the communication itself. The goal is to prevent detection by adversaries or unauthorized parties, ensuring that even the fact that communication is taking place remains concealed. This is typically achieved by transmitting signals below the detection threshold of monitoring systems, often by utilizing noise-like signals, spreading transmission power across wide bandwidths, or exploiting environmental interference. Covert communications are especially important in sensitive or high-security environments, as they protect both the privacy of the information and the confidentiality of the communication event.}

\begin{figure} 
         \centering
         \includegraphics[width=0.8\columnwidth]{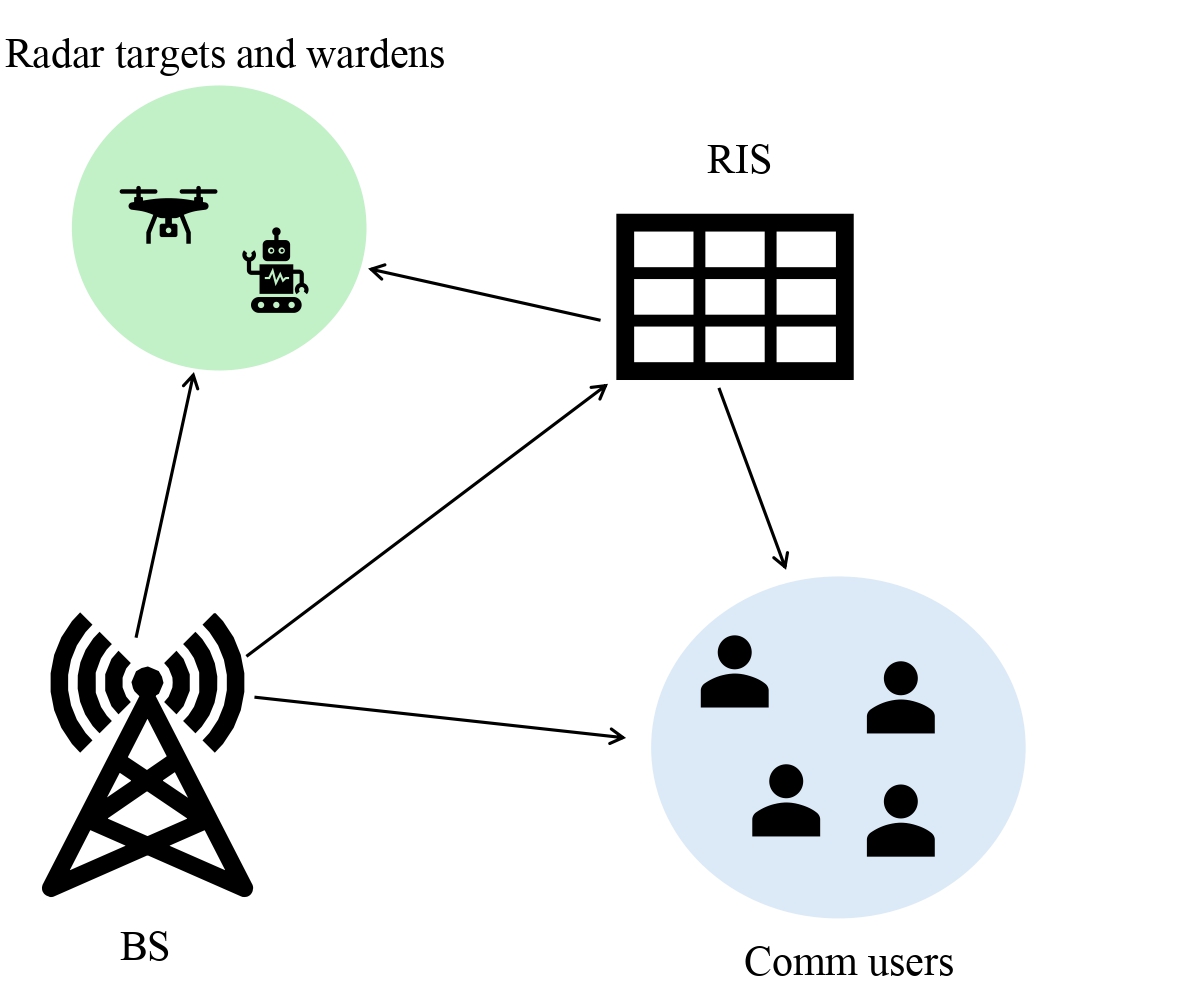}
         \vspace{0.05cm}
        \caption{System model of RIS-assisted covert ISAC presented in~\cite{2024_Ghosh}.}
        \label{fig:covert}
\end{figure}

{RIS-assisted covert ISAC was investigated in~\cite{2024_Ghosh} with covert communications between a DFRC BS and multiple legitimate users as shown in Fig.~\ref{fig:covert}. The DFRC aims also at detecting multiple targets, co-located with a warden attempting to detect the communication. The authors maximized the worst-case data rate among the communication users while satisfying radar detection and covertness constraints. To achieve this, the authors proposed overlaying artificial noise with the message signal to ensure the warden’s received signal statistics remain largely unchanged when communication occurs. They formulated an optimization problem to determine the RIS's passive beamforming and the transmitter’s active precoding, solving it using AO and a modified stochastic gradient descent approach. Through simulations, the authors demonstrated a significant improvement (i.e., by a factor of approximately 3 or 4) in the data rate for the RIS-assisted system compared to its non-RIS-assisted counterpart, highlighting the considerable advantage of RIS in improving system permanence and simultaneously confusing potential wardens.}

{\textbf{Lessons learned:} Covert communications aim to hide the existence of communications using techniques such as noise-like signals or wideband transmission to remain undetected. In RIS-assisted systems, as demonstrated by recent research, covert communication between a DFRC BS and multiple users can be significantly enhanced. By overlaying artificial noise with the message signal, the system ensures that wardens trying to detect the transmission do not notice a significant change in the received signal, maintaining covertness. Optimizing RIS and transmitter beamforming further improves performance, with simulations showing a three- to four-fold increase in data rates compared to non-RIS systems. }

{\textbf{Challenges, opportunities and open research directions:} One key challenge is maintaining a balance between radar detection performance and communication covertness, as increasing communication efficiency often risks exposure to eavesdroppers or wardens. Additionally, the dynamic and unpredictable nature of real-world environments, including mobility, interference, and rapidly changing channel conditions, complicates the design of robust and adaptive covert communication systems. On the opportunity side, RIS provides a promising avenue to enhance covertness by intelligently manipulating signal propagation, allowing for more flexible control over the communication channels. Machine learning and artificial intelligence algorithms also offer potential to adaptively optimize covert communication strategies in real-time. Open research directions include exploring more sophisticated optimization techniques for joint beamforming, artificial noise generation, and precoding schemes, especially in multi-user and multi-target scenarios. Furthermore, addressing the security vulnerabilities posed by adversarial RIS attacks, developing energy-efficient covert communication protocols, and investigating the potential of quantum communications for ultra-secure covert transmissions are promising future areas of study.}

{
\subsection{Jamming} \label{sec:jamming}
Jamming is a deliberate act of disrupting or obstructing communication signals to prevent effective communication. A number of works have proposed anti-jamming techniques for RIS-assisted ISAC.}

{\textit{1) Vehicular networks: } The authors of~\cite{2024_Yao} proposed an anti-jamming technique for communication in RIS-aided vehicular networks using reinforcement learning. The goal was to maximize the data rate by jointly designing the transmit beamforming at the roadside unit and the RIS reflection coefficients. The anti-jamming strategy is based on training the reinforcement learning model to reduce the deviation caused by distribution changes while achieving fast convergence and avoiding local optima.
}

{
\textit{2) Unmanned aerial vehicles (UAVs):} In a similar manner, an anti-jamming method for RIS-UAV (i.e., an RIS deployed at a UAV) assisted DFRC was proposed in~\cite{2024_Xu1}. The authors of that work presented two security solutions for RIS-UAV enabled DFRC to tackle jamming and eavesdropping threats. The anti-jamming solution involves deploying the RIS-UAV to establish a clear LoS link for blocked targets, utilizing passive beamforming to counteract malicious jammers. This aerial RIS also introduces artificial noise to suppresses eavesdropping attempts while at the same time ensures precise target sensing. 
}

{\textbf{Lessons learned:} The studies on anti-jamming techniques for communication in RIS-aided vehicular networks and RIS-UAV enabled DFRC highlight the effectiveness of using intelligent surfaces to enhance system resilience. In both cases, the anti-jamming strategies focused on optimizing transmission through joint beamforming design and reflection coefficient adjustments, enabling fast convergence and avoiding local optima. Additionally, integrating aerial platforms like UAVs for LoS communication and employing artificial noise significantly improved both the transmission rate and security against eavesdropping.}

{\textbf{Challenges, opportunities and open research directions:} One key challenge in anti-jamming communication for RIS-aided DFRC is the complexity of jointly optimizing transmit beamforming and reflection coefficients in highly dynamic environments, especially when factoring in mobility and unpredictable interference. Opportunities lie in further leveraging advanced machine learning techniques, such as deep reinforcement learning, to enhance real-time adaptability and decision-making in hostile environments. Another research direction is the integration of RIS with other emerging technologies like quantum communication to enhance security and data rate performance. Additionally, addressing energy efficiency and ensuring reliable performance in multi-user, multi-target scenarios remain critical areas for future exploration. }

\subsection{Non-Orthogonal Multiple Access}

Non-orthogonal multiple Access (NOMA) can provide better spectral efficiency by allowing users to share the same time-frequency resources via multiplexing them in the power domain or the code domain. While several researchers considered code-domain NOMA as a variant of code-division multiple access (CDMA), power-domain NOMA has gained increasing popularity in the last decade due to its ability to provide higher spectral efficiency. In power-domain NOMA, each user is allocated a power level depending on its channel strength, and a successive interference cancellation (SIC) technique is applied at the receivers to control the interference caused by the resource sharing among users~\cite{2022_Magbool}.

Since ISAC and NOMA share the goal of using the spectrum more efficiently, the integration of the two technologies has been studied in a number of works. These works can be put in three main categories.

{\textit{1) PLS:} The works of~\cite{2024_Dongdong,2024_Dongdong1} and~\cite{2024_Jiang}  investigated the use of RIS to enhance the PLS of RIS-assisted DFRC systems with NOMA transmission, leveraging the additional degrees of freedom provided by the RIS. The findings in~\cite{2024_Jiang} demonstrated that the RIS-assisted scheme can provide a 21\% improvement in the sum secrecy rate for a NOMA system, under the same sensing QoS, compared to a non-RIS-assisted system. The authors also reported a 100\% increase in the secrecy rate for RIS-assisted DFRC with NOMA transmission compared to a system utilizing orthogonal transmission.}

{\textit{2) Performance analysis:} The performance of RIS-assisted NOMA for DFRC was analyzed in~\cite{2024_Parihar}. More specifically, ~\cite{2024_Parihar} presented an analytical framework for assessing DL transmissions in MIMO heterogeneous networks. The spatial distribution of NOMA-enabled BSs and users was molded using independent homogeneous Poisson point processes. Active RISs, also distributed according to homogeneous Poisson point processes, are utilized to overcome blockages for user equipment when direct links from BSs are unavailable. The study derived approximate and asymptotic outage probability expressions for two cases: one where transmission occurs directly from the BS to a blocked user, and another where the transmission is routed through an active RIS. The analysis highlighted the advantages of the proposed active RIS-assisted ISAC with NOMA over that with traditional orthogonal multiple access. For instance, the outage performance of the active RIS-NOMA system can be notably enhanced by increasing the number of RIS elements compared to a less significant improvement in the case of orthogonal multiple access. }

{\textit{3) Performance optimization:}} The authors of~\cite{2022_Zuo,2023_Zuo,2023_Lyu,2024_Nassar} considered an RIS-assisted DFRC system with NOMA applied to achieve better communication performance. Although the use of NOMA targets mainly improves the communication performance, a more efficient use of the spectrum for the purpose of communication allows better spectrum utilization for sensing also. For example, the authors of~\cite{2022_Zuo} reported a 300\% improvement in the radar minimum beampattern gain caused by the use of NOMA for communications.

Rate-splitting multiple access (RSMA) is a generalized version of power-domain NOMA. In RSMA, the transmitted data is divided into distinct layers. These layers are then transmitted concurrently over the same time-frequency-code resources. At the receiving end, each user aims to decode its designated layer while treating the other layers as interference. This approach is known as \textit{rate splitting} and involves dividing the total rate among the users~\cite{2022_Mao}. The fundamental concept behind RSMA is to leverage the interference among users constructively rather than considering it as a hindrance. This strategy has the potential to enhance the spectral efficiency and increase the overall throughput of the system~\cite{2020_Dizdar}. 

The use of the RSMA transmission scheme for RIS-assisted DFRC was investigated in~\cite{2023_Chen} with the goal of maximizing the SNR for target detection. Simulation results showed that a gain of approximately \unit[10]{dB} can be attained through the combined effect of using RISs and RSMA, compared to classical space-division multiple access (SDMA) systems that do not utilize RISs for transmission. It is worth mentioning, nevertheless, that this advantage comes as the cost of increased transmitter and receiver complexity. 

{\textbf{Lessons learned:} NOMA allows users to share the same time-frequency resources by multiplexing in the power domain, which has proven particularly effective in achieving higher spectral efficiency. The studies show that incorporating RIS can significantly enhance PLS, with notable improvements in secrecy rates compared to non-RIS-assisted systems and orthogonal multiple-access transmission. Additionally, performance analyses indicate that the use of active RISs can improve outage performance, especially as the number of RIS elements increases. The research further underscores that employing NOMA not only enhances communication performance but also contributes to better spectrum utilization for sensing purposes.}

{\textbf{Challenges, opportunities and open research directions:} One major challenge RIS-aided DFRC with NOMA is the increased complexity in the design and optimization of communication systems that leverage both technologies, particularly in the context of resource allocation and beamforming. The reliance on advanced techniques like SIC and rate-splitting can lead to significant computational overhead, which may hinder real-time performance. Additionally, varying channel conditions and user mobility in practical scenarios pose challenges in maintaining reliable communication and sensing performance. Ensuring seamless coordination between the communication and sensing functions while managing interference effectively also represents a critical hurdle.} {Research into new algorithms for beamforming that account for imperfect channel state information and real-time adjustments to changing conditions will also be crucial for advancing these integrated systems. Overall, the interplay between NOMA and ISAC offers a rich landscape for innovative solutions that can enhance the efficiency and reliability of future communication and sensing applications.}

\subsection{Intelligent Transport}
The increasing demand for intelligent transport applications, encompassing vehicle-to-vehicle (V2V), vehicle-to-infrastructure (V2I), and vehicle-to-everything (V2X) communications, highlights the significant role of ISAC in such scenarios. A key distinguishing feature that makes ISAC well-suited for intelligent transport applications is the predictability of motion. Unlike general communication scenarios, vehicles, trains, and buses follow pre-defined routes. This predictability facilitates the development of low-complexity solutions for both communication and sensing aspects within the context of intelligent transport applications. {There are two categories of studies for RIS-assisted DFRC assisted by RIS for intelligent transport: system optimization and predictive beamforming.}

{\textit{1) System optimization:}} In a notable contribution to RIS-aided ISAC, the work presented in~\cite{2022_Wang} addresses NLoS communications. In the proposed system model, the BS communicates with a target vehicle while concurrently estimating its distance and velocity. In scenarios where the LoS is blocked, the RIS serves the dual purpose of radar sensing and communication. It achieves this by estimating the location of the vehicle and concentrating the signal power in the estimated direction. The method employed for accurate distance and velocity estimation is the maximum likelihood method. The communication performance is evaluated based on data rate, considering the estimation accuracy. Similar problems were also tackled in~\cite{2023_Feng,2023_Zhimin}. Additionally, in~\cite{2023_Li4}, the authors explored an RIS-assisted ISAC framework designed for high-speed railways.

The authors of~\cite{2022_Shao} and~\cite{2022_Shao2} introduced a system model featuring an ISAC BS communicating with multiple vehicles while simultaneously sensing their locations. An innovative aspect of the model involves the deployment of RISs on the rooftops of the vehicles to enhance the sensing performance by adjusting the phases of the echo signals. To address the challenge of joint communications and sensing, the authors devised a grid-based parametric model. They formulated the joint estimation problem as a compressive sensing problem and employed a message-passing algorithm utilizing a progressive approximation method. This approach was shown to be effective in reducing computational complexity. Furthermore, an expectation-maximization algorithm was employed to fine-tune the grid parameters, alleviating potential model mismatch issues.

{\textit{2) Predictive beamforming:}} To leverage the predictability of motion in vehicular networks, the authors of~\cite{2023_Meng,2023_Meng2} and~\cite{2024_Li2} introduced a low-complexity predictive beamforming scheme for a system where a BS communicates with multiple vehicles equipped with STAR-RISs. In this system model, STAR-RISs are mounted on the tops of the vehicles to enhance both the sensing and communication performance for users inside the vehicles. The BS transmits a communication signal, part of which is reflected back to the BS via the STAR-RIS for location sensing, while the other part is transmitted to serve the user inside the vehicle.

The sensing model aims to predict the distance, angle, and velocity of each vehicle, and the BS constructs the channel matrix based on these estimated quantities. Two models are employed for this purpose: the state evolution model and the measurement model. The measurement model is obtained from the collected echo signals, while the state evolution model estimates the distance, angle, and velocity of each vehicle based on geometric relationships~\cite{2020_Fan}.

\begin{figure} 
         \centering
         \includegraphics[width=1\columnwidth]{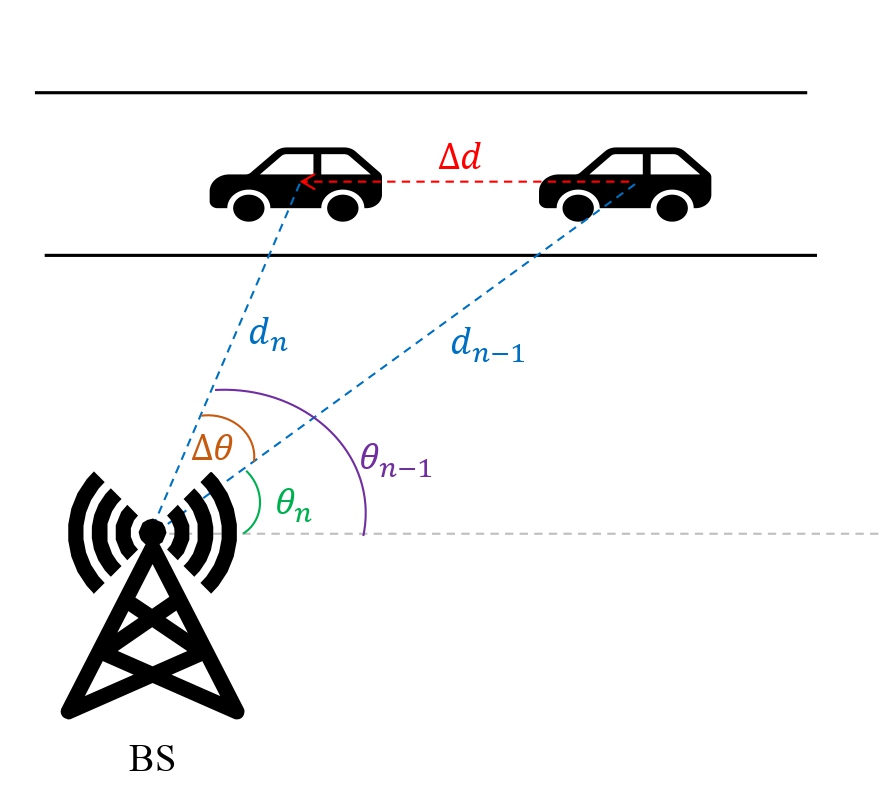}
         \vspace{0.05cm}
        \caption{State evolution model for predictive beamforming.}
        \label{fig:PB}
\end{figure}

To illustrate, consider a BS attempting to align the beam toward a vehicle moving along a straight road, as shown in Fig.~\ref{fig:PB}. In the initial stage, the BS conducts pure radar sensing to determine the distance, angle, and velocity of the vehicle in the first time slot. In the subsequent time slots, the BS, on the basis of employing a number of geometric relationships, estimates the values of these quantities. These relationships are derived in~\cite{2020_Fan}. To derive these relationships, common assumptions and simplification are assumed such as that the vehicle is locally moving at a constant speed, which leads to estimation errors. To correct these errors, a Kalman filter is employed to combine the state evolution model with the echo measurement model and obtain refined estimates. The channel matrix is then constructed, and the resources are allocated by solving optimization problems. While the system can operate by observing the reflected echoes from the vehicles directly, the presence of STAR-RISs at the top of the vehicles enhances sensing performance, leading to better-constructed channels and improved communication performance.

{ \textbf{Lessons learned:} The integration of ISAC within intelligent transport applications highlights its significant potential in enhancing mobility and safety. One key lesson learned is the advantage of leveraging the predictability of vehicle motion, which allows for the development of low-complexity solutions for both communication and sensing. Research showcases two primary categories of studies: system optimization and predictive beamforming. In system optimization, innovative models like the deployment of RIS on vehicles demonstrate enhanced communication and sensing capabilities, addressing challenges like NLoS communications and improving estimation accuracy for distance and velocity. In predictive beamforming, the use of STAR-RISs mounted on vehicles enables more effective location sensing and communication by constructing channel matrices based on estimated vehicle states, showcasing the advantages of a proactive approach to beam alignment. Moreover, the studies emphasize that the combination of RIS and advanced beamforming techniques can substantially improve the performance of communication systems in intelligent transport applications. For instance, the predictive beamforming schemes that utilize geometric relationships and Kalman filtering for estimating vehicle parameters lead to better channel construction and resource allocation. This integration not only enhances the efficiency of communication links but also allows for more accurate sensing, ultimately contributing to the reliability and safety of intelligent transport systems. }

\begin{figure*}
  \centering
  \begin{tabular}{c c}
    \includegraphics[width=0.95\columnwidth]{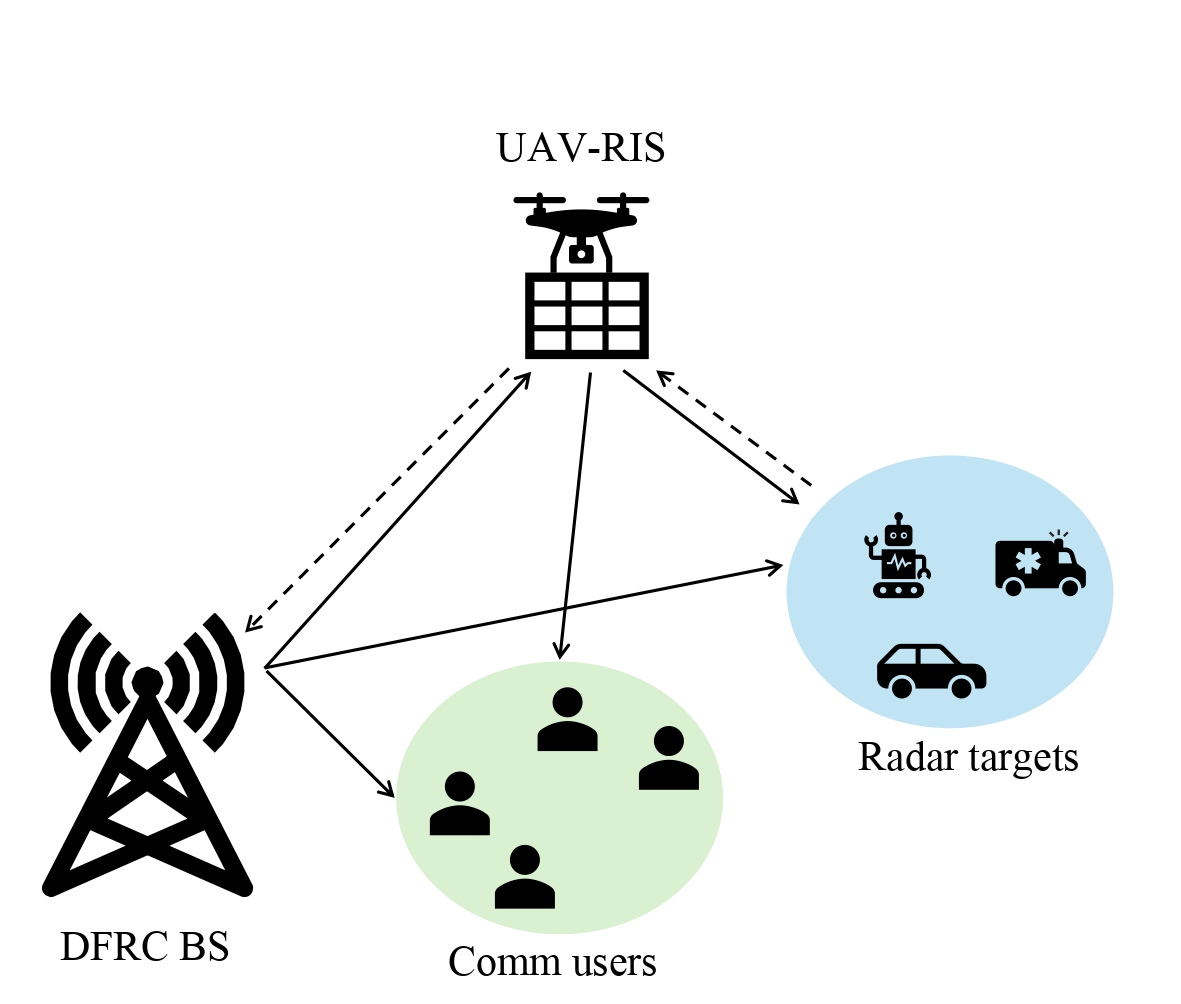} &
      \includegraphics[width=0.95\columnwidth]{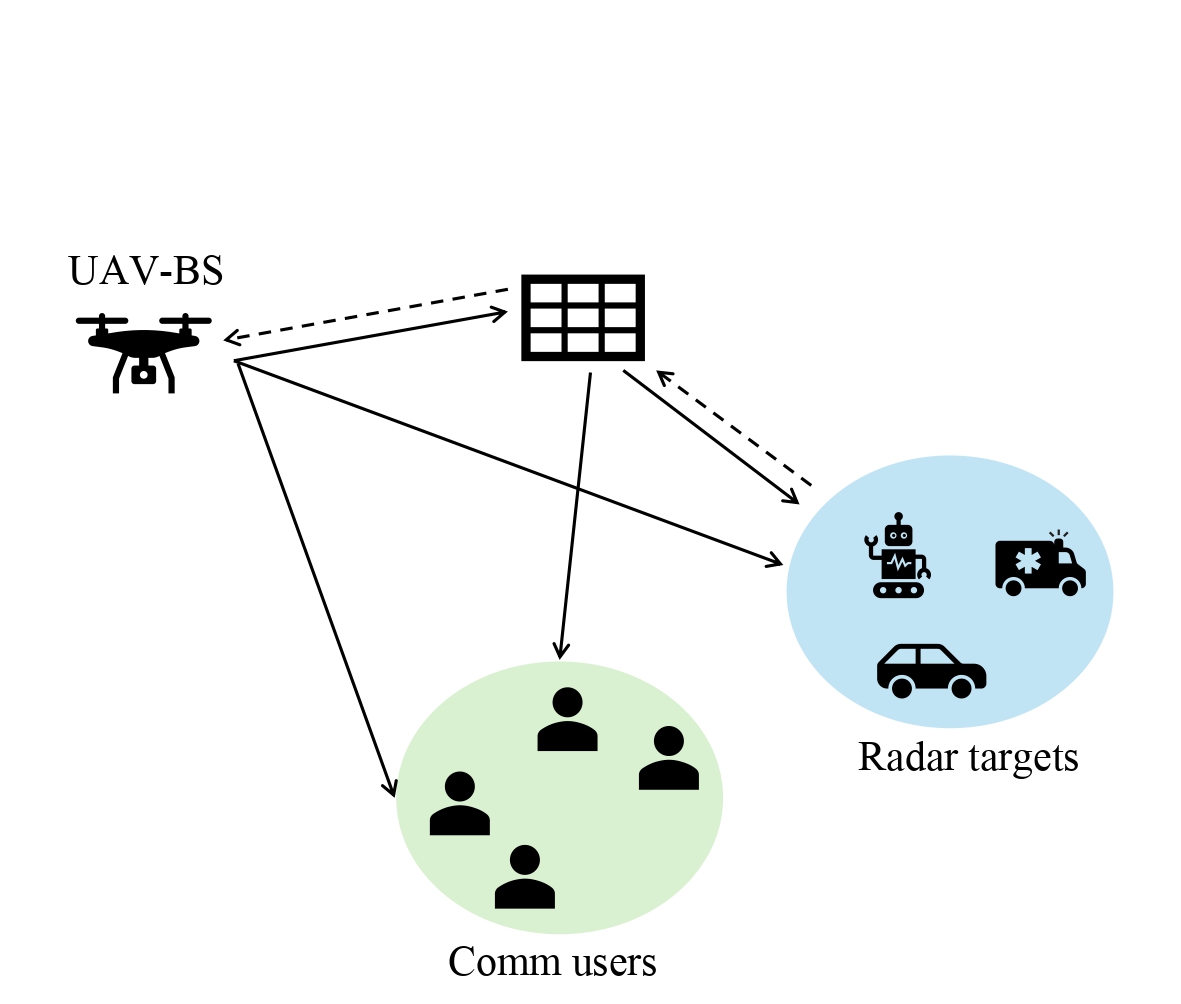}  \\
      \scriptsize (a)    &
      \scriptsize (b)   \\
  \end{tabular}
    \medskip
  \caption{In RIS-assisted DFRC systems, UAVs can be used as an (a) aerial RIS or (b) aerial DFRC BS.}
  \label{fig:UAV}
\end{figure*}

{\textbf{Challenges, opportunities and open research directions:} One significant challenge is the need for robust algorithms capable of handling the dynamic nature of vehicular environments. The reliance on predictive models and motion predictability assumes consistent vehicle behavior, which may not always hold true due to factors such as varying speeds, sudden stops, or erratic movements. Additionally, implementing low-complexity solutions that efficiently manage communication and sensing tasks simultaneously remains difficult, particularly in environments with high user density and potential interference. Moreover, ensuring reliable connectivity in scenarios with NLoS conditions poses a critical hurdle, as traditional communication methods may fail to perform optimally.}

{Despite these challenges, there are numerous opportunities for advancing research in this area. One promising direction is the exploration of advanced machine learning techniques to enhance real-time decision-making and resource allocation in ISAC systems, allowing for adaptive responses to changing vehicular dynamics. Research into the development of hybrid systems that can seamlessly integrate various communication paradigms, such as V2X, and advanced sensing methods could yield significant performance enhancements. Additionally, exploring the use of multi-agent systems for collaborative sensing and communication among vehicles can foster enhanced situational awareness and improved traffic management.}

{
\subsection{UAV}
Utilizing UAVs as part of DFRC systems has proven effective due to their ability to adjust their locations, improving overall system performance. In RIS-assisted DFRC setups, UAVs can be leveraged for two primary purposes, as illustrated in Fig.~\ref{fig:UAV}. The first purpose is to employ the UAV as an aerial RIS, which can further enhance channel controllability. Alternatively, the UAV can function as a DFRC BS, improving signal transmission. While an example of utilizing aerial RIS has been discussed in Section~\ref{sec:jamming} fot anti-jamming, we discuss the second system in this subsection.}

{In~\cite{2024_Zhongqing} an RIS was utilized alongside a UAV BS to enhance DFRC performance. In this setup, the UAV performed dual roles by transmitting signals to multiple users while simultaneously conducting sensing tasks. The objective was to maximize the weighted sum of the average sum rate for the communication users and the sensing SNR for the radar target while managing the inherent tradeoff between these objective functions. This was achieved by jointly optimizing the RIS phase shifts, the UAV trajectory, UAV beamforming, and user scheduling. To address the optimization problem, the authors introduced an iterative AO algorithm to derive an efficient suboptimal solution. Notably, the resulting UAV trajectory design significantly improved the Pareto-optimal front for the sum rate and sensing SNR, thereby enhancing both communication and sensing functionalities in DFRC.}

{\textbf{Lessons learned:} The use of UAVs in DFRC systems has proven beneficial due to their flexibility in adjusting their locations to enhance overall system performance. In RIS-assisted DFRC configurations, UAVs can serve one of two functions: acting as an aerial RIS or functioning as a DFRC BS. When the RIS is integrated with a UAV BS, significant improvements in the Pareto-optimal front can be achieved, thanks to optimal trajectory design of the UAV, which enhances the functionalities of communication and sensing in DFRC systems.}

{\textbf{Challenges, opportunities and open research directions:} One major challenge of utilizing a UAV as part of the RIS-assisted DFRC system is the dynamic nature of UAVs, which can complicate the optimization of communication and sensing functions, requiring advanced algorithms. Additionally, ensuring reliable communication links between UAVs, ground users, BSs and RISs may pose difficulties due to varying environmental conditions and potential interference. However, these challenges also create opportunities for innovation, such as the development of new strategies for cooperative sensing and communication, leveraging UAV mobility to improve coverage and performance in complex environments. Open research directions include exploring advanced machine learning techniques for adaptive resource allocation, enhancing the robustness of UAV-based networks against jamming and other threats, and investigating the integration of energy-efficient approaches to extend UAV operation time. }

\subsection{Sub-Space Rotation}
\begin{figure*}
  \centering
  \begin{tabular}{c c c}
    \includegraphics[width=0.65\columnwidth]{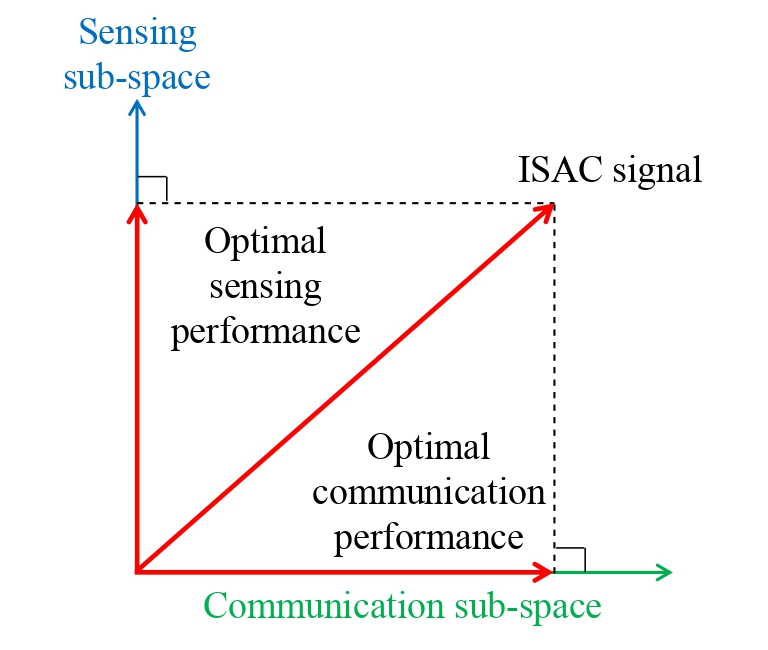} &
      \includegraphics[width=0.65\columnwidth]{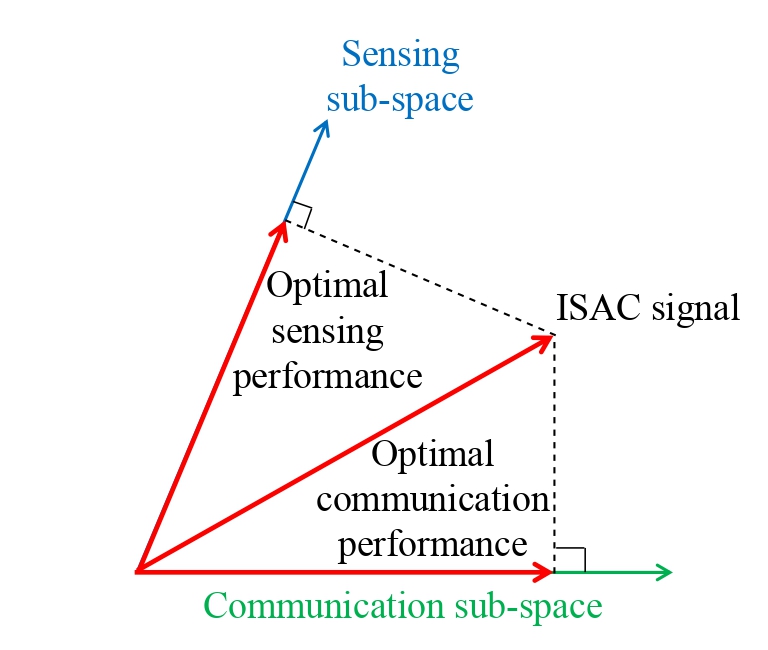} & \includegraphics[width=0.65\columnwidth]{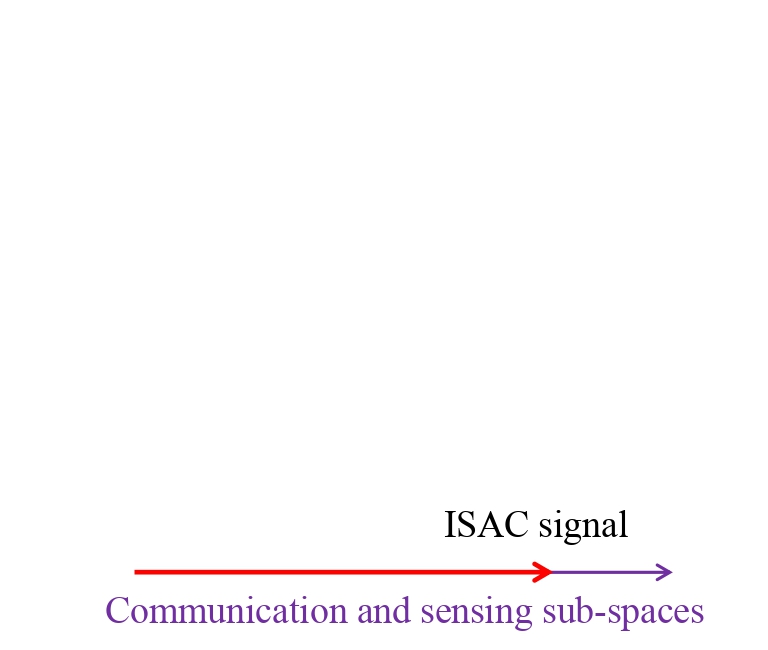} \\
      \scriptsize (a)   &
      \scriptsize (b)  &
      \scriptsize (c) \\
  \end{tabular}
    \medskip
  \caption{Three levels of coupling between the communications and sensing sub-spaces: (a) weak coupling, (b) moderate coupling, and (c) strong coupling.}
  \label{fig:ssr}
\end{figure*}

The authors of~\cite{2023_meng3} presented a unique and interesting approach to exploit the use of RISs in DFRC.

The ISAC signals lie within the space spanned by communication and sensing sub-spaces. The correlation between these sub-spaces can be quantified by their intersection angle. Consequently, the performance of sensing and communication is related to the projection of the ISAC signal onto the respective sub-space. To illustrate these ideas, the authors defined three levels of coupling. In the weakly coupled scenario (illustrated in Fig.~\ref{fig:ssr}(a)), the two sub-spaces are nearly orthogonal, leading to projections close to the minimum. Due to the orthogonality between the communications and sensing sub-spaces, satisfying both requirements simultaneously becomes challenging. For instance, if a communication user and a radar target are situated at opposite locations relative to the BS, attempting to design an optimal beam for one adversely affects the other. In moderately coupled scenarios, as depicted in Fig.~\ref{fig:ssr}(b), the angle between the sensing and communication sub-spaces is less than $90^\circ$, resulting in a better tradeoff between communications and sensing compared to the case of weak coupling. Finally, in the scenario of strong coupling, illustrated in Fig.~\ref{fig:ssr}(c), the two sub-spaces are perfectly aligned, leading to optimal performances for both sensing and communications. An example of this occurs when a communication user is also a radar target; in such a scenario, forming beams towards the direction of the communication user yields optimal communication and sensing performance.

The authors of~\cite{2023_meng3} aimed to leverage the RIS to enhance the coupling between the communication and sensing sub-spaces, thereby reducing the tradeoff between the two functions. To evaluate the effectiveness of this approach, the authors generated Pareto-optimal curves for sensing performance (quantified by the CRB of the AoA estimation) and the communication performance (quantified by the achievable rate). Fig.~3(c) in~\cite{2023_meng3} illustrates the behavior of the improved Pareto-optimal front based on simulations conducted in~\cite{2023_meng3}. These results demonstrate that the sub-space rotation approach can achieve a point in the upper-left corner of the Pareto-optimal front, which is unattainable in the weakly coupled case.

{\textbf{Lessons learned:} Achieving optimal performance for both sensing and communication simultaneously is inherently challenging for the cases where there is weak coupling between communications and sensing  sub-spaces. This limitation emphasizes the need for innovative strategies to manage the tradeoffs associated with conflicting objectives in communication and sensing. Leveraging RISs can significantly improve the coupling between these sub-spaces, leading to enhanced performance for both functions. The ability to reach points on the Pareto-optimal front that were previously unattainable in weakly coupled scenarios underscores the effectiveness of using RISs to maximize the coupling between the communication and sensing sub-spaces.}

{\textbf{Challenges, opportunities and open research directions:} A significant challenge lies in optimizing the tradeoffs between communication and sensing performance, particularly in weakly coupled scenarios where the two functions conflict. This necessitates the development of sophisticated algorithms and frameworks that can intelligently manage these tradeoffs while ensuring robust performance in dynamic environments. Additionally, there is an opportunity to explore novel architectures and configurations of RISs to further enhance the coupling between communication and sensing sub-spaces.  }

\subsection{Near-Field Operation}
Recently, several research works have explored the near-field operation of RIS-assisted DFRC. {While near-field beam-focusing can improve both the communication and sensing performance of ISAC by directing the beams toward physical locations characterized by specific angles and distances, the mathematical formulations of the beam-focusing are more complicated.}

{\textit{1) Performance analysis:} In~\cite{2022_Wang3}, the authors aimed to establish the fundamental limits of RIS-aided near-field localization and explore potential performance improvements in RIS-aided communication. The study focused on the near-field operation of RIS-aided ISAC, emphasizing the careful control of the phase response of an RIS to ensure that the signals scattered by different parts of the surface coherently superimpose at the receivers. The paper also introduced the concept of continuous intelligent surfaces and established the performance limits of RIS-aided communication and RIS-aided localization. Additionally, the study derived near- and far-field ISAC signal models, addressed the optimal design of the phase response for RISs in various scenarios (including those with obstacles and multiple surfaces), and quantified the performance gain in localization and communication facilitated by RISs in complex wireless environments.}

{\textit{2) System optimization:}} In~\cite{2023_Palmucci}, the authors concentrated on optimizing a full-duplex DFRC system, comprising a single communication user and a single radar target. Initially, the authors derived an exact expression for the CRB, considering the cases of both LoS and NLoS propagation paths. Subsequently, the authors formulated a problem to jointly minimize the power of the effective self-interference signal in the uplink case at the radar receiver while simultaneously maximizing the achievable rate for the DL communication user, subject to the exact CRB constraint. The joint optimization problem was transformed into an equivalent MSE minimization problem, consisting of two terms: the self-interference power at the radar and the MSE at the communication user. Simulation results showcased that the proposed scheme significantly outperforms the conventional far-field DFRC scheme without an RIS.


{\textbf{Lessons learned:} Studies have established the fundamental limits of RIS-aided near-field localization and communication, highlighting performance gains from optimized phase control of RIS surfaces, especially in complex environments. Additionally, system optimization efforts have focused on minimizing self-interference and improving tracking accuracy, particularly under position uncertainties, showing significant improvements over conventional far-field DFRC systems. These findings emphasize the importance of careful phase design and optimization for RIS in near-field operation to maximize system performance.}

{\textbf{Challenges, opportunities and open research directions:} A major challenge of near-field operation is the complexity of the mathematical models required for beam-focusing, as the focus shifts from far-field approximations to more intricate near-field formulations. The coherent superposition of signals at receivers, especially in environments with obstacles or multiple surfaces, requires precise control of the RIS phase response, which complicates the system design and optimization. Moreover, the self-interference in full-duplex systems and the tracking loss due to narrow beams, known as the  \textit{deafness problem}, add further computational complexity to optimizing system performance. These challenges are particularly pronounced when dealing with position uncertainties and dynamic environments, making it difficult to maintain robust performance across communication and radar functions.}

{The ability to focus beams on specific angles and distances enables more precise sensing and communication, opening the door for improved localization accuracy and enhanced system performance in complex environments. This creates opportunities for developing continuous intelligent surfaces and novel optimization algorithms that can adapt to both near- and far-field conditions. Open research directions include further exploration of joint optimization strategies, such as minimizing self-interference and tracking errors over dynamic scenarios, and refining beamforming techniques to account for user mobility and trajectory prediction. Additionally, advancing methods to handle position uncertainties and improving the robustness of RIS designs under varying environmental conditions remain key areas for future research.}

\subsection{Extended Reality}

In 6G and beyond networks, the applications of human-machine interactions are expected to flourish, with extended reality (XR) playing a crucial role. This motivation prompted the authors of~\cite{2023_ma} to explore an ISAC framework in the context of XR, with the aid of an RIS. Specifically, a practical positioning approach utilizing the multiple signal classification (MUSIC) algorithm is proposed, enhanced by specially designed RIS configurations. Additionally, the paper addressed the joint optimization of the transmit beamformer and the RIS phase shifts, aiming to maximize channel capacity under CRB constraints. The optimization problem was tackled through AO with gradient projection and manifold optimization. Simulation results were then presented, showcasing the feasibility of the proposed positioning algorithm in XR applications.

{\textbf{Lessons learned:} The integration of RIS into ISAC frameworks holds great potential for XR applications in 6G and beyond networks. Such an integration demonstrates the feasibility of these technologies, suggesting they could be instrumental in realizing efficient human-machine interactions in future networks.}

{\textbf{Challenges, opportunities and open research directions:} Ensuring accurate positioning in dynamic environments particularly for XR, where real-time responsiveness is crucial, is a key challenge. However, the use of RIS-enhanced configurations, combined with algorithms such as MUSIC, offers promising opportunities to improve positioning accuracy and channel capacity in human-machine interactions. Open research directions include refining optimization methods for dynamic XR scenarios, enhancing the robustness of positioning algorithms, and addressing the tradeoffs between complexity and performance in practical deployments.}

{
\subsection{Movable Antennas}
The recently proposed concept of movable antennas refers to a specific type of antenna system designed to adjust its position or orientation to optimize signal reception or transmission~\cite{2024_Zhu1}. Movable antennas play a significant role in channel customization by enabling dynamic adjustments to optimize signal quality based on the specific needs of a communication channel. In systems where channel conditions vary, movable antennas can be aligned or repositioned to reduce interference, enhance signal strength, and maintain a stable connection. By customizing the channel through antenna movement, users can ensure the best possible performance for different applications, such as data transmission, broadcasting, or secure communications.
}

{Motivated by the fact that both RISs and movable antennas share the common goal of customizing the propagation environment, the authors of~\cite{2024_Haisu} investigated utilizing both to improve DFRC performance. In particular, the authors studied an RIS-assisted DFRC system equipped with a two-dimensional (2D) array of movable antennas. The angles of the targets that need to be estimated are assumed to be in the NLoS area of the BS, so the RIS is used to create a virtual LoS link to boost sensing and communication performance. An optimization problem was then formulated to maximize the radar’s minimum beampattern gain while adhering to communication SINR constraints. A method based on SDR, sequential rank-one constraint relaxation, and SCA was then proposed to solve the optimization problem, where the transmit beamforming, RIS phase shifts, and the positions of the movable antennas were designed. The simulation results underscored the benefit of using movable antennas over fixed-position antennas, especially in the high transmit SNR regime, where an improvement of 29\% in the radar’s minimum beampattern gain was achieved.}

{\textbf{Lessons learned:} Movable antennas enable dynamic channel customization by adjusting their position or orientation to optimize signal quality, particularly in environments where channel conditions vary. When paired with RISs, these systems can improve both radar sensing and communication performance in DFRC systems. Simulation results revealed that this approach significantly outperforms traditional fixed-position antennas, especially at high SNR levels, demonstrating the versatility and effectiveness of movable antennas in advanced wireless communication and sensing applications.}

{\textbf{Challenges, opportunities and open research directions:} The integration of movable antennas and RISs in DFRC systems presents several challenges, opportunities, and open research directions. A key challenge lies in the complex optimization of beamforming, RIS phase shifts, and antenna positioning to achieve real-time performance improvements, particularly in dynamic environments with varying channel conditions. This opens up opportunities for developing more efficient algorithms that can handle these highly non-convex problems while ensuring scalability for large-scale networks. Additionally, the potential of movable antennas to reduce interference and enhance multi-user communication opens new avenues for their application in dense urban and vehicular networks. Open research directions include exploring the impact of movable antennas in three-dimensional (3D) array configurations, investigating their energy efficiency, and addressing the practical limitations of hardware implementations, such as the mechanical constraints and power consumption of continuously adjusting antennas.}

{
\section{Machine Learning for RIS-assisted ISAC}  \label{sec:ML_RIS_ISAC}
Machine learning tools have emerged as potential candidates for providing low-complexity solutions for various communication and sensing applications. The application of machine learning techniques in RIS-assisted ISAC has been investigated from several perspectives. This section presents a literature review of machine learning methods for RIS-assisted ISAC, including both RIS-assisted RCC and RIS-assisted DFRC systems. A summary of the works presented in this section is provided in Table~\ref{tab:ML}.
}

\begin{table*}
\footnotesize
\centering
\caption{{References on machine learning for RIS-assisted ISAC.}}
\begin{tabular} {|m{1.5cm} | m{1.3cm}| m{1.8cm} |m{6.3cm}| m{4cm}|} 
 \hline 
 Reference & Year & RCC or DFRC & Goal  & Methods  \\  [0.5ex] 
 \hline 
 \hline 
  \cite{2023_Saikia} & 2023 & RCC  & Resource allocation to maximize the communication mutual information subject to constraints on the radar signal power and the interference energy at the radar receiver   & Meta-reinforcement learning - Markov decision process \\   
   \hline
  \cite{2022_Zhong} & 2022 & DFRC & Joint waveform design to maximize
the SINR for the radar and minimize the communication MUI energy & Decoupled online learning network - Deep neural network \\
 \hline 
\cite{2022_Xiangnan} & 2022 & DFRC & Resource allocation to maximize the communication capacity subject to ergodic constraints & Proximal policy optimization - Reinforcement learning \\
 \hline 
 \cite{2024_Xiuli} & 2024 & DFRC & Resource allocation to minimize the transmit power with constraints on the communication SINR and the sensing CRB & Meta-learning gradient descent - Lagrange multipliers -  Manifold optimization\\
 \hline 
  \cite{2024_Saikia} & 2024 & DFRC & Resource allocation to to minimize the squared position error bound subject to communication QoS constraints &  Hybrid deep reinforcement learning - Markov decision process \\
 \hline 
   \cite{2024_Ye} & 2024 & DFRC & Resource allocation to maximize the spectral efficiency of DFRC &  Unsupervised learning - Image-shaped channel samples algorithm  \\
    \hline 
   \cite{2022_Liu7} & 2022 & DFRC & Channel estimation for DFRC &  Deep neural network \\
 \hline 
      \cite{2023_Liu7} & 2023 & DFRC & Channel estimation for DFRC &   Convolutional neural network \\
 \hline 
    \cite{2023_Liu4} & 2023 & DFRC & Channel estimation for DFRC &  Deep neural network -  Extreme learning machine \\
 \hline 
    \cite{2024_Du1} & 2024 & DFRC & Channel estimation and environment mapping for DFRC &  Complex-valued depth residual convolutional neural network - Parallel factor tensor decomposition \\
 \hline 
 \cite{2023_Zhu4}  & 2023 & DFRC & PLS enhancement for STAR-RIS-assisted ISAC to prevent the unauthorized disclosure of legitimate
user information to a potential eavesdropping sensing target &  Deep reinforcement learning  \\
  \hline 
  \cite{2023_Liu9} & 2023 & DFRC & PLS enhancement for RIS-assisted ISAC via incorporating artificial noise &  Soft actor-critic based deep reinforcement learning - Entropy maximization for policy diversity \\
  \hline 
\end{tabular}
\label{tab:ML}
\end{table*}

\subsection{Interference Management in RCC}
In an attempt to reduce the complexity associated with solving complex optimization problems for resource allocation in RIS-assisted RCC,~\cite{2023_Saikia} investigated the application of deep reinforcement learning to address the resource allocation optimization problem. In their system model, a LoS link was presumed to be available between the radar and the target, and unavailable between the communication BS and the users. Consequently, the BS relied on the RIS to transmit the communication signals. The problem was formulated to maximize the mutual information for the communication system, subject to constraints on the radar signal power and the interference energy at the radar receiver.

Instead of relying on traditional optimization-based methods, the authors opted for a meta-reinforcement learning-based algorithm. In this approach, an agent was guided by a reward-based procedure with the objective of discovering the optimal policy to maximize the cumulative rewards. A Markov decision process was employed to integrate current and past actions and rewards into the policy, facilitating the attainment of an optimal solution.

Through simulations, the authors showcased that the proposed method can obtain the optimal solution six times faster than traditional AO-based solutions. The reinforcement learning based approach also surpassed other learning-based methods, including the twin delayed deep deterministic and deep deterministic policy gradient, in terms of communication mutual information and radar detection probability.

{\textbf{Lessons learned:} The application of machine learning for RIS-assisted RCC is still in at an early stage. Nevertheless, existing works have demonstrated the effectiveness of deep reinforcement learning in addressing complex optimization problems such as resource allocation in RIS-assisted ISAC systems. Meta-reinforcement learning, used in this context, enables the agent to learn how to learn optimal policies across different tasks or scenarios. This meta-learning capability is advantageous in complex environments where conditions and requirements may vary, allowing the system to adapt more effectively to changing conditions. This approach could pave the way for more adaptive and intelligent communication and radar coexistence systems, capable of optimizing performance dynamically in real-time scenarios.}

\textbf{Challenges, opportunities and open research directions:} In reinforcement learning models, the agent relies on the environment to acquire the optimal policy, and the effectiveness of these systems is highly influenced by the Markov decision process utilized. Therefore, obtaining accurate system descriptions is essential. Additionally, the agent may make inaccurate decisions when confronted with situations it has not been trained on. Combining learning-based methods with theoretical ones can mitigate against this limitation, aiming for a tradeoff between system complexity and performance.

{
\subsection{DFRC Waveform Design}
The optimal joint design of DFRC waveforms is challenging due to the absence of metrics that can quantify the optimal balance between communications and sensing. This has led researchers to employ machine learning tools to provide enhanced DFRC waveform designs.}

The study in~\cite{2022_Zhong} presented an innovative approach for waveform design in RIS-assisted DFRC systems. The primary goal of the paper was to simultaneously maximize the SINR for the radar and minimize the communication MUI energy. Addressing the intricacies of the challenging waveform design optimization problem, the authors introduced a decoupled online learning network based algorithm. The proposed framework comprises three key modules: trainable network parameters, loss calculation, and back-propagation. The loss calculation module computes the cost function with fixed network parameters. Following this, a deep learning optimizer is employed to train the network parameters, minimizing the loss function and ensuring convergence. Simulation results provided compelling evidence for the effectiveness of the learning-based method. The proposed approach achieved a 0.69 dB enhancement in radar SINR and a remarkable 138.8\% improvement in the sum rate when compared to theoretical methods.

{\textbf{Lessons learned:} Designing optimal joint waveforms for DFRC systems is complex due to the lack of clear metrics balancing communication and sensing. To address this, machine learning tools such as deep learning have been explored. Research in this area highlights the effectiveness of a decoupled online learning network for waveform optimization, showing that this method can significantly improve radar SINR and communication performance. The use of trainable parameters, loss calculation, and back-propagation ensures effective training, leading to enhanced outcomes over traditional methods.}

{\textbf{Challenges, opportunities and open research directions:} Challenges in joint design for DFRC systems include the difficulty in quantifying the optimal tradeoff between radar sensing and communication, as well as the complexity of waveform design in dynamic environments. However, the integration of machine learning, particularly deep learning, offers significant opportunities by enabling adaptive and data-driven waveform optimization. Open research directions include developing more robust and interpretable machine learning models, exploring real-time adaptive systems, and establishing standardized metrics to evaluate the balance between communication and sensing in DFRC systems.}

{
\subsection{Resource Allocation}
Resource is another key challenge in RIS-assisted DFRC systems due to the diverse system models and problem formulations as explained in Section~\ref{sec:RA_DFRC}. Since several traditional optimization algorithms suffer from high complexity and lack of ability to avoid converging to local solutions, machine learning tools have been proposed in a number of research works as replacements.
}

A proximal policy optimization based beamforming design was proposed in~\cite{2022_Xiangnan} for an RIS-assisted DFRC system in the THz band. The focus was on maximizing the communication capacity subject to ergodic constraints. In tackling this challenging optimization problem, the authors introduced a proximal policy optimization algorithm. Since the literature on this topic lacks in-depth exploration for ISAC systems in the THz band and its relevant considerations, the proposed algorithm is deemed effective for the joint optimization problem adopted in the paper.

{The work~\cite{2024_Xiuli} targeted minimizing transmit power by jointly optimizing both active and passive beamforming at the BS and RIS, respectively, with constraints on the communication SINR and the sensing CRB. Due to the non-convexity of the optimization problem as well as coupling between the optimization variables and the high-dimensional optimization variables, solving the problem using the traditional optimization approaches is prohibitively complex. Drawing on meta-learning techniques, the authors introduced a meta-learning algorithm based on gradient descent to iteratively update the optimization variables and solve the non-convex optimization problem. Further, the constrained optimization problem was transformed into its dual form by employing Lagrange multipliers. Meta-learning neural networks were then used to adjust the optimization variables through local loss functions, with network parameters updated via global loss functions.}

{The authors in~\cite{2024_Saikia} introduced C-RAN to enable collaboration between multiple BSs to enhance the cooperation benefits for both communication and sensing functions. They formulated an optimization problem to minimize the squared position error bound, which reflected system performance by optimizing the transmit beamformer, RIS phase shifts, and subcarrier assignment. Additionally, to adjust the RIS phase shifts, the authors proposed an RIS-assisted cooperative ISAC protocol. This protocol used collected measurements to refine the agent's location and velocity estimates and reconstruct the environmental map with greater accuracy. The large feasible space posed computational challenges, making traditional gradient-based or exhaustive search methods impractical. To address this, a framework based on Markov decision processes was developed and a hybrid deep reinforcement learning algorithm was presented to efficiently navigate the problem.}

{To achieve a low-complexity and high-performance beamforming design, the authors of~\cite{2024_Ye} leveraged image-shaped channel samples and introduced an ISAC beamforming neural network model to maximize the spectral efficiency of DFRC. By using unsupervised learning, the loss function integrated important performance metrics, such as communication and sensing channel correlation and sensing channel gain, without requiring labeled data. Simulation results demonstrated that the proposed method offers 25\% improvement in the sensing SNR compared to the sub-space rotation theoretical algorithm of~\cite{2023_meng3} with 75\% reduction in the running time.}

{\textbf{Lessons learned:} Traditional optimization algorithms often struggle with high-dimensional and non-convex problems, making it difficult to find optimal solutions without getting stuck in local optima. Machine learning techniques have emerged as powerful alternatives. These methods reduce computational complexity and provide more efficient solutions to beamforming and resource allocation problems. For example, proximal policy optimization has been successfully applied to maximize communication capacity under ergodic constraints in the THz band, offering a promising solution where conventional methods fall short. Additionally, meta-learning has proven effective in tackling joint optimization challenges by iteratively updating optimization variables using gradient descent. This approach efficiently addresses the coupling between variables in complex, high-dimensional spaces.} {Furthermore, advanced techniques such as hybrid deep reinforcement learning are crucial for solving large-scale decision-making problems in RIS-assisted ISAC systems. In conjunction with C-RAN, hybrid deep reinforcement learning enhances system performance by enabling collaboration between multiple BSs for both communication and sensing purposes. The use of unsupervised learning in ISAC beamforming models has also demonstrated significant improvements in spectral efficiency and sensing accuracy while reducing the need for labeled data. }

{\textbf{Challenges, opportunities and open research directions:} Ensuring robust performance in varying environmental conditions, such as interference and fading, is a key challenge when employing machine learning for resource allocation in RIS-assisted DFRC systems. Furthermore, the integration of advanced machine learning techniques into these systems requires thorough understanding and refinement to ensure reliability and adaptability, particularly in real-world applications. Despite these challenges, one promising avenue is the exploration of hybrid machine learning approaches that combine reinforcement learning and deep learning for more efficient resource management and optimization strategies. There is also potential for further research into unsupervised learning methods that optimize system performance without the need for extensive labeled datasets. Exploring the design of adaptive algorithms that can dynamically respond to changing environments and user requirements will also be crucial.}

{
\subsection{Channel Estimation}
Channel estimation is another challenging task in ISAC due to the lack of capability of the RIS and the passive targets to send or process pilot signals.
}

 Tthe authors of~\cite{2023_Liu4,2023_Liu7,2022_Liu7} employed a channel estimation framework for RIS-assisted ISAC based on neural networks and extreme learning machines. While the complexity of solving the channel estimation problem by standard theoretical approaches is prohibitively high, machine learning tools can significantly reduce the complexity of the estimation procedure. Two types of input-output pairs are proposed for the extreme learning machines: the first pair is generated using the original received DFRC signals, and the second pair is based on the least-squares estimation results. To enhance the estimation accuracy, data augmentations are leveraged to enrich the training samples. The numerical results provided by the authors showed that the proposed machine learning approach achieves a comparable performance to the least-squares benchmark estimator, with significantly lower computational complexity.
 
{An ISAC algorithm capable of performing simultaneous channel estimation, positioning, and environment mapping was introduced in~\cite{2024_Du1}. The authors employed a complex-valued depth residual convolutional neural network assisted channel estimation algorithm to address the limited robustness of traditional algorithms. In addition to this, the authors exploited the sparsity of the mmWave channel through parallel factor tensor decomposition to extract factor matrices that contained crucial channel parameters, including complex path gain, direction-of-arrival, direction-of-departure, and time delay. Using these parameters, an environment mapping was achieved based on the geometric relationships between the position and channel parameters. Simulation results indicated that the proposed algorithm maintains strong ISAC performance even at low transmit SNRs.}

{\textbf{Lessons learned:}  Employing machine learning techniques, such as neural networks and extreme learning machines, can substantially reduce the complexity associated with traditional estimation methods. The development of input-output pairs for extreme learning machines, including those derived from the original received DFRC signals and least-squares estimation results, enhances the estimation process. Moreover, data augmentation techniques have proven effective in enriching training samples, thereby improving estimation accuracy. Algorithms that leverage advanced models, like complex-valued depth residual convolutional neural networks and tensor decomposition, can effectively extract crucial channel parameters, facilitating simultaneous channel estimation, positioning, and environment mapping while maintaining robust performance, even at low transmit SNRs.}

{\textbf{Challenges, opportunities and open research directions:} Machine learning for channel estimation in ISAC systems faces several challenges. One significant issue is the limited availability of labeled data for training models, primarily due to the inability of RIS and passive targets to send or process pilot signals effectively. Additionally, ensuring model robustness against noise and varying channel conditions is critical, as machine learning models can be sensitive to these factors. Computational complexity remains a concern, as advanced algorithms may still require considerable computational resources for training and inference, which poses a problem for real-time applications where low latency is essential. There is also the risk of overfitting due to limited training data, which can lead to poor generalization in new scenarios. Finally, integrating machine learning based techniques into existing ISAC frameworks presents compatibility challenges that need to be addressed for successful deployment.}

{Despite these challenges, there are numerous opportunities and open research directions in leveraging machine learning for channel estimation in an ISAC system. Hybrid approaches that combine machine learning with traditional estimation techniques can capitalize on the strengths of both methodologies to improve accuracy while maintaining efficiency. Furthermore, the adoption of unsupervised and semi-supervised learning methods could help mitigate the data scarcity issue by effectively utilizing unlabeled data. Advances in edge computing and distributed learning frameworks present promising opportunities for real-time processing of channel estimation algorithms. Research into adaptive learning models that dynamically adjust to changing conditions can enhance estimation accuracy, while cross-domain learning techniques could further improve model performance. Finally, establishing standardized evaluation metrics and encouraging collaboration between theoretical research and practical implementations will help bridge the gap between model development and real-world applications, ultimately refining machine learning approaches in ISAC systems.}

{
\subsection{Physical Layer Security}}
The authors of~\cite{2023_Zhu4} addressed a DFRC problem with an aid of a STAR-RIS. Specifically, the objective of the work was to prevent the unauthorized disclosure of legitimate user information to a potential eavesdropping sensing target. The authors seek to maximize the average security rate of the legitimate user by jointly optimizing the receive filters and transmit beamforming in the BS, along with the transmitting and reflecting coefficients of the STAR-RIS, subject to constraints on the minimum echo SNR of the radar target and the achievable rate of the legitimate user. Due to the complexity of the problem, finding the globally optimal solution is challenging. To address this challenge, two deep reinforcement learning algorithms were presented for the joint design of receive filters, transmit beamforming of BS, and transmitting and reflecting coefficients of the STAR-RIS.

Tackling the same security aspect of RIS-assisted ISAC but taking a different approach, the authors of~\cite{2023_Liu9} explored secure DFRC transmission via incorporating artificial noise, as discussed in Section~\ref{sec:pls}. To maximize the secrecy rate, the authors jointly optimized the transmit beamformer, the artificial noise vector, and the RIS phase shifts. Employing a soft actor-critic based deep reinforcement learning algorithm, policy diversity was enhanced through entropy maximization, guaranteeing convergence to the globally optimal solution.

{\textbf{Lessons learned:} The complexity solving optimization problems concerning PLS while trying to minimize the compromisation in the communication and sensing performance underscores the need for innovative solutions. The application of deep reinforcement learning algorithms has emerged as an effective strategy to address this challenge by enabling adaptive optimization of critical parameters in dynamic environments. Additionally, techniques like soft actor-critic-based deep reinforcement learning, which promotes policy diversity through entropy maximization, can facilitate convergence to globally optimal solutions, demonstrating the significant potential of machine learning approaches to bolster PLS in RIS-assisted ISAC systems.}

{\textbf{Challenges, opportunities and open research directions:} One primary challenge is the complexity of the optimization problems involved, which often require balancing multiple objectives, such as maximizing secrecy rates while ensuring effective communication and sensing capabilities. The high dimensionality of the problem space, combined with the need for real-time processing, can make it difficult for traditional machine learning methods to converge to optimal solutions. Additionally, the quality of training data is crucial; however, in PLS scenarios, labeled datasets that accurately represent eavesdropping scenarios may be limited or difficult to generate. Furthermore, ensuring the robustness of machine learning models against adversarial attacks and changing environmental conditions remains a significant concern, as these factors can adversely affect the reliability of the security enhancements.}

{Despite these challenges, there are numerous opportunities and open research directions for leveraging machine learning in PLS for RIS-assisted ISAC systems. Research can focus on creating novel algorithms that incorporate uncertainty modeling and adversarial training to enhance the robustness of PLS mechanisms against various threats. Exploring unsupervised and semi-supervised learning techniques can also help address the limitations of labeled data by effectively utilizing available unlabeled data for training. Moreover, the integration of multi-agent systems could facilitate collaborative strategies among multiple RIS units to optimize PLS collectively.}

\section{Holographic ISAC}  \label{sec:holo}
Holographic ISAC refers to the use of RHSs for transmission and/or reception in ISAC. While research in this area is still limited, promising results and prototypes have been presented in the literature. In this section, we first discuss the beamforming design of RHS-assisted RCC and RHS-assisted DFRC, followed by a summary of the implementation efforts of RHS-assisted DFRC along with experimental results. We conclude the section by presenting current works on SIM-assisted ISAC. A summary of the paper diussed in this section is presented in Table~\ref{tab:holo}.

\begin{table*}
\footnotesize
\centering
\caption{{References on holographic ISAC.}}
\begin{tabular} {|m{1.5cm} | m{1.3cm}| m{1.8cm} |m{6.3cm}| m{4cm}|} 
 \hline 
 Reference & Year & RCC or DFRC & Goal  & Methods  \\  [0.5ex] 
 \hline 
 \hline 
  \cite{2024_Du} & 2024 & RCC  & Beamforming design to maximize the communication weighted sum bits per unit bandwidth subject to a constraint on the radar SINR   & AO - Fractional programming  \\   
   \hline
     \cite{2022_Haobo} & 2022 & DFRC  & Beamforming design to minimize the beampattern mismatch error while setting a threshold on the communication QoS and prototyping   & AO - ZF - SDR  \\   
       \hline
     \cite{2024_Zhuoyang} & 2024 & DFRC  & Beamforming design to maximize the communication and radar SINRs with RHSs employed for both signal transmission and echo reception   & AO - Rayleigh quotient-based method  \\  
   \hline
        \cite{2024_Adhikary} & 2024 & DFRC  & Beamforming design in cell-free network to maximize the sensing SINR and system
energy efficiency   & Variational autoencoder based scheme - Transformer-based algorithm  \\  
   \hline
\cite{2024_Kexin} & 2024 & DFRC  & Beamforming design for BD-RHS-assisted ISAC to maximize the sum rate for multiple communication users and minimize the largest eigenvalue of the CRB matrix for multiple targets   & AO - Symmetric unitary projection - SCA \\  
   \hline
   \cite{2024_Ziwei}  & 2024 & DFRC  & Beamforming design for RHS-assisted ISAC with RSMA transmission to maximize the radar permcne metrics for detection, tracking and localization targets while adhering to a communication sum rate constraint & AO - SCA -  Difference-of-convex programming  \\  
    \hline
    \cite{2024_Ziwei1}  & 2024 & DFRC  & Beamforming design for RHS-assisted ISAC with RSMA transmission to maximize the secrecy sum rate subject to sesning CRB constraint & AO - SCA -  SOCP  \\ 
    \hline
    \cite{2023_Wei3}  & 2023 & DFRC  & Beamforming design for RIS-assisted holographic ISAC to maximize the radar SINR subject to communication SNR constraints & AO - Maximin procedure - difference of convex programming \\ 
   \hline
      \cite{2024_Wei1}  & 2024 & DFRC  & Beamforming design for RIS-assisted holographic ISAC to maximize the minimum radar SINR subject to communication SNR constraints & AO - Maximin procedure - difference of convex programming \\ 
   \hline
       \cite{2024_Ziqing} & 2024 & DFRC  & Beamforming design to minimize the CRB of the AoA while maintaining a threshold on the communication SINRs and prototyping   & AO - SDR  \\   
       \hline
\end{tabular}
\label{tab:holo}
\end{table*}

{
\subsection{Holographic Beamforming for RCC}
In~\cite{2024_Du}, the authors explored incorporating RHS into the communication transmitter of an RCC system to facilitate holographic beamforming. Specifically, the system model consists of a time division multiplexing UL multi-user communication system enhanced by RHS backscatter, operating alongside radar sensing. In this RCC system, the radar transmits a probing signal to track a distant target, while multiple communication users attempt to capture the electromagnetic waves emitted by the radar. Each user employs an RHS to transmit its own data to the BS. The authors formulated an optimization problem aimed at maximizing the weighted sum spectral efficiency, subject to a radar QoS constraint. The solution to this problem is obtained by employing AO and fractional programming. The inclusion of RHSs on the user side showed significant improvement compared to methods like ZF, with the weighted sum spectral efficiency increasing from 7 bits/ssec/Hz to 35 bits/sec/Hz (a 5-fold increase) when RHSs consisting of 60 elements were used. This improvement was even greater (nearly 8-fold) when a larger RHS with 100 elements was employed.
}

{\textbf{Lessons learned:} Incorporating RHS into communication transmitters can significantly enhance system performance, particularly in multi-user UL scenarios coexisting with radar sensing. The use of RHS for holographic beamforming allows users to effectively capture radar-emitted signals and transmit their own data to the BS via backscatter, showcasing the potential of RHS to optimize both communication and sensing tasks. The significant gains in the weightedsum spectral efficiency, particularly when using larger RHS arrays, highlight the scalability and effectiveness of RHS in improving system throughput, especially when compared to traditional techniques like ZF. }

{\textbf{Challenges, opportunities and open research directions:} One major challenge lies in the complexity of optimizing joint radar and communication performance, particularly in scenarios where CSI is imperfect or unavailable, which may limit the effectiveness of beamforming and resource allocation strategies. Additionally, the scalability of RHS, especially as the number of users and the size of the RHS arrays increase, poses challenges in terms of computational complexity and system overhead. However, these challenges also open up opportunities for the development of more efficient optimization algorithms and machine learning-based solutions to manage the CSI and improve the adaptability of beamforming strategies. Future research could explore dynamic and real-time optimization approaches that account for mobility, varying radar targets, and environmental conditions. Moreover, there is potential to investigate hybrid active-passive systems that combine the strengths of both active antennas and RHS to further enhance performance in joint communication and sensing applications.}

{
\subsection{Holographic Beamforming for DFRC}
Holographic beamforming for DFRC has gained increasing attention lately due to its ability to generate more directed beams towards communication users and radar targets, thereby enhancing ISAC performance. The research in this area can be categorized into five main categories.}

\begin{figure*}
  \centering
  \begin{tabular}{c c}
    \includegraphics[width=0.95\columnwidth]{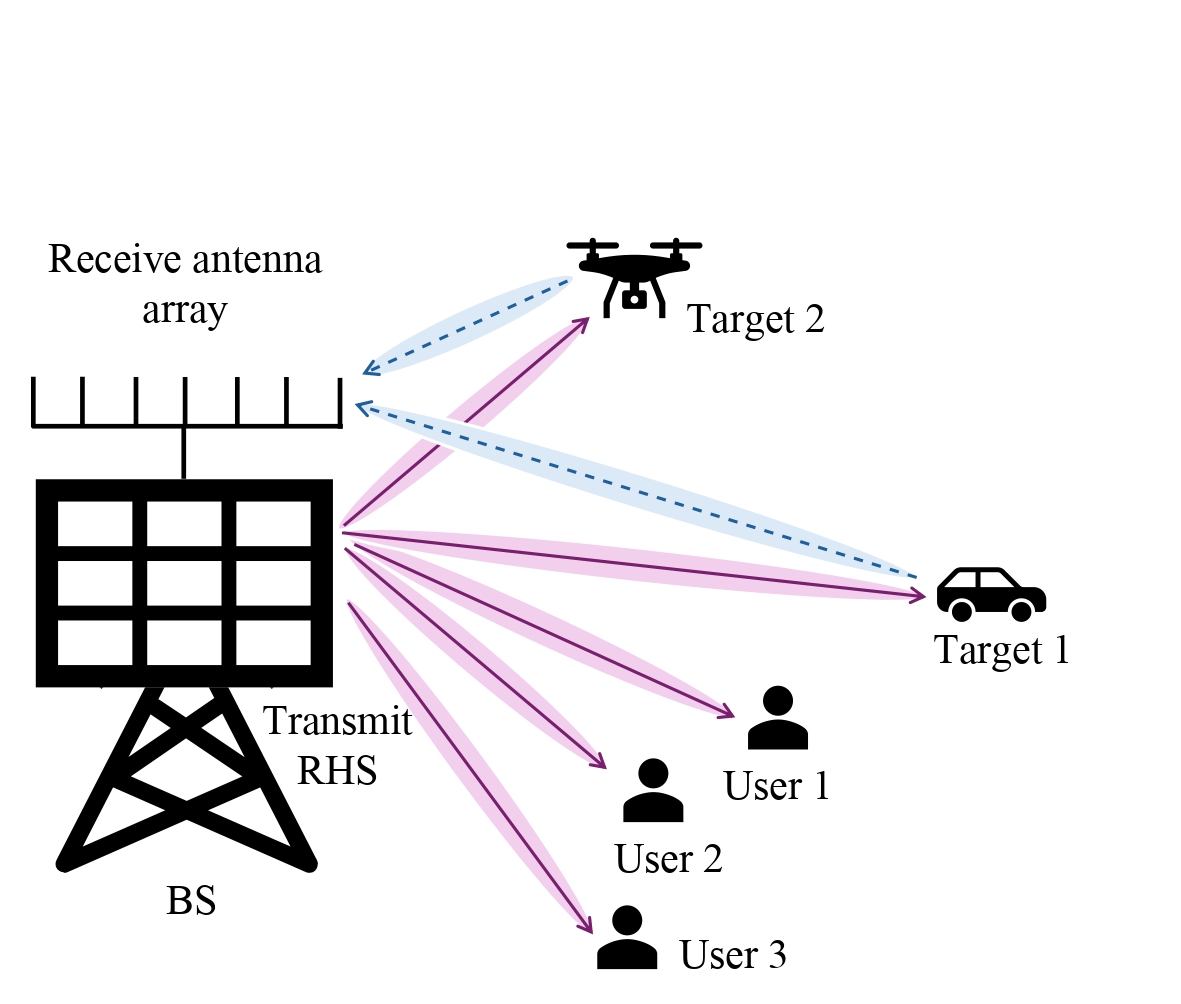} &
      \includegraphics[width=0.95\columnwidth]{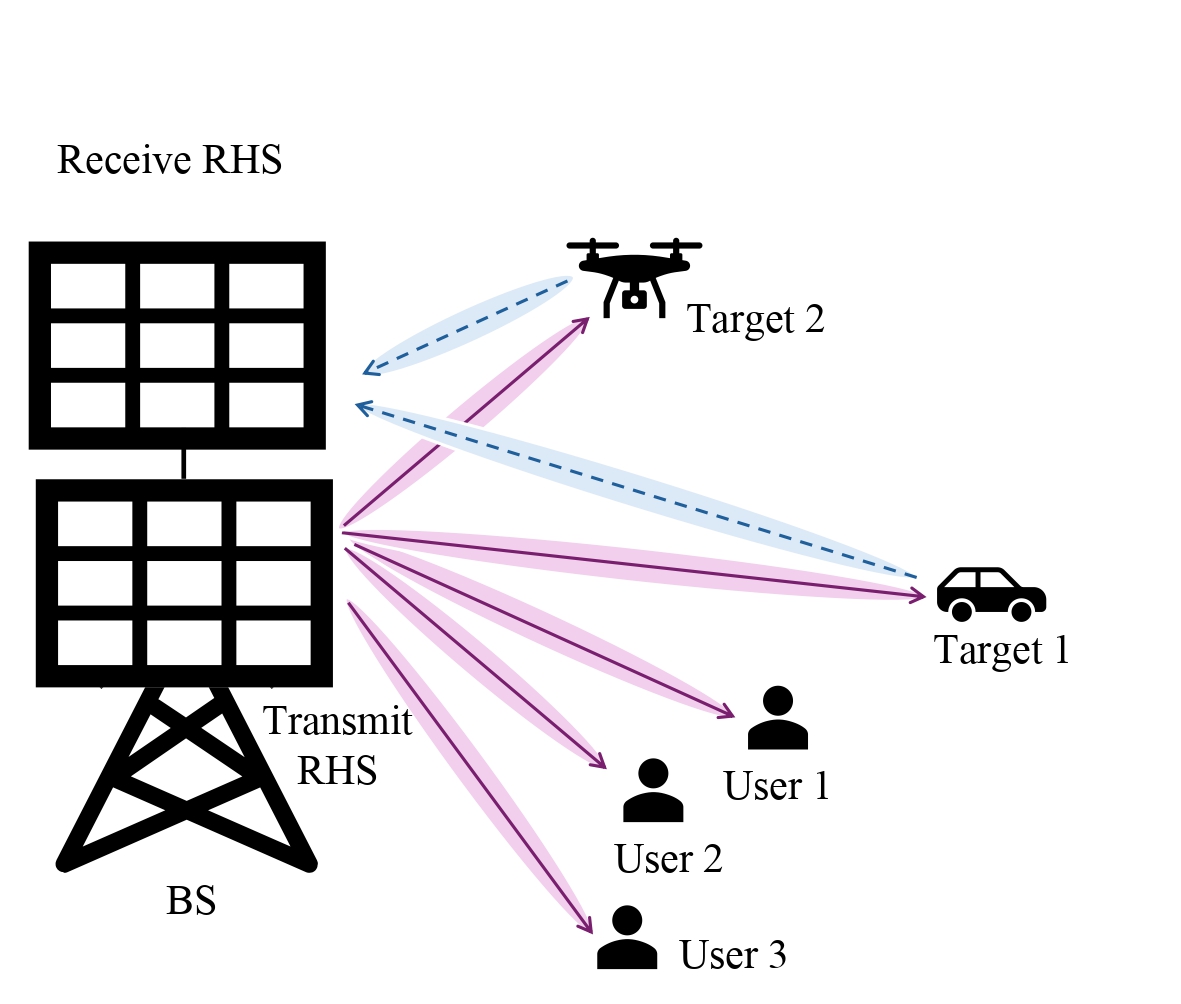}  \\
      \scriptsize (a)    &
      \scriptsize (b)   \\
  \end{tabular}
    \medskip
  \caption{System models for holographic DFRC where (a) an RHS is employed as a transmitter and an antenna array is employed as a reciver and (b) an RHS is employed as a transmitter and another RHS is employed as a receiver.}
  \label{fig:RHS_DFRC_RA}
\end{figure*}

{\textit{1) Resource allocation:} One of the early works on holographic beamforming for DFRC was presented in~\cite{2022_Haobo}, where the authors proposed an ISAC transmission scheme for a BS serving multiple DL communication users and simultaneously detecting multiple radar targets. Unlike RIS-assisted ISAC systems, the BS and the RHS are co-located within the same unit to allow holographic beamforming. As shown in Fig.~\ref{fig:RHS_DFRC_RA}~(a), two types of antennas/surfaces are employed: RHSs for signal transmission and a regular antenna array for receiving the targets' echo signals. The BS and the RHS employ digital and analog beamforming, respectively, to transmit the ISAC signals. This allows the system to sense many targets simultaneously and provide different data streams to the DL users.}

{Following the digital signal processing at the BS, the signals are sent to the RF chains which in turn modulate them and then deliver them to the feed of the RHS. At the RHS, the modulated signals are converted to radiation patterns through the waveguide and metamaterial elements for analog processing, where an analog beamformer is used to determine the radiation amplitudes of the RHS elements.}

{In the reception phase, the communication users receive the ISAC signals broadcast by the RHS and decode their data streams. At the same time, the BS receives the echo signals reflected by the targets via the antenna array and processes them. In this process, different holographic patterns are produced, each corresponding to a single feed. The analog beamformer is then designed by superposing the output holographic patterns in a weighted manner, a technique known as holographic-pattern-division multiple access (HDMA)~\cite{2022_Deng}. The beamformer is them optimized by choosing the weights of the holographic patterns to minimize the beampattern mismatch error while setting a threshold on the communication QoS.}

{In order to tackle the complexity of the optimization problem, the authors proposed an AO framework where they alternate between optimizing the digital and analog beamformers. The digital beamforming sub-optimization problem is tackled using the ZF approach to mitigate inter-user interference for all communication users. To optimize the weights of the analog beamformer, the sub-optimization problem is reformulated as a quadratic program, and solved using the SDR technique. This optimization problem was applied and tested with a real-world prototype, this generated promising results as will be discussed in Section~\ref{sec:proto_RHS_DFRC}.}

{While~\cite{2022_Haobo} employed an antenna array for echo signal reception, a system model with both transmitting and receiving continuous-aperture RHSs was investigated in~\cite{2024_Zhuoyang}. To facilitate a design with continuous pattern, the authors proposed a continuous-to-discrete transformation of the RHS pattern using the Fourier transform, translating the continuous beamforming design into a discrete one. The problem was then formulated by creating a joint objective function aiming to balance the performance of multi-target sensing with the requirements of multi-user communication. To tackle the non-convex problem with coupled variables, the authors proposed an AO algorithm, where the transmit beamforming design is solved by breaking it down into a series of feasibility sub-problems, and the receive beamforming is handled using a Rayleigh quotient based method. The proposed RHS-based ISAC system model showed a significant improvement in the Pareto-optimal front of the sensing SINR versus the communication SINR when the authors compared RHSs containing 81 elements to discrete antenna arrays with the same aperture length and antenna spacing of half a wavelength. For instance, the authors demonstrated a 44\% improvement in the sensing SINR when an RHS is utilized instead of a discrete antenna array at a communication SINR requirement of 0 dB. The improvement in the Pareto-optimal front becomes more pronounced at higher communication SINR requirements. For example, a 210\% improvement in the sensing SINR is observed at a communication SINR requirement of 25 dB. This highlights the advantage of RHSs in improving the tradeoffs between communications and sensing, thereby enhancing ISAC performance.}

{\textit{2) Cell-free holographic DFRC:} Holographic beamforming improves cell-free MIMO by generating highly directed and adaptive beams, enhancing the precision and efficiency of sensing and communication across a wide area. This leads to better signal quality, reduced interference, and more efficient resource utilization, ultimately improving overall system performance in terms of connectivity, energy efficiency, and user experience.}

{Building on this concept, the authors of~\cite{2024_Adhikary} proposed a holographic ISAC framework for cell-free networks. Their method utilized an artificial intelligence based technique for beamforming design. An optimization problem was formulated to maximize the sensing SINR and system energy efficiency. The proposed algorithm decomposed the optimization problem into two sub-problems: a sensing sub-problem and a power allocation sub-problem. A variational autoencoder based scheme was employed to solve the sensing sub-problem, followed by a transformer-based mechanism to allocate power to users based on the sensing information. Simulation results demonstrated that this method outperformed other learning-based techniques, such as long short-term memory (LSTM) and gated recurrent unit algorithms, achieving energy efficiency improvements between 15\% and 17\%. }

{\textit{3) Beyond-diagonal RHSs (BD-RHSs):} The study in~\cite{2024_Kexin} investigated the use of BD-RHS as part of the DFRC base station in the mmWave band. The deployment of BD-RHS at the BS adds additional design flexibility provided by the fully connected structure of the RHS matrix. The authors proposed an efficient two-stage algorithm for beamforming design at the transmitter, aiming to jointly maximize the sum rate for multiple communication users while minimizing the largest eigenvalue of the CRB matrix for multiple sensing targets. Numerical results indicated that the transmitter-side BD-RIS-aided mmWave ISAC outperformed conventional diagonal-RHS-aided systems in both communication and sensing performance.}

{4) RSMA: The interplay between RSMA and RHS-assisted DFRC has been explored in several studies. In~\cite{2024_Ziwei}, the authors exploited RSMA to address interference management between communication and sensing in holographic ISAC. In the system model, three types of targets (which are also users) exist: communication/detection users, communication/tracking users, and communication/localization users. The authors introduced sensing QoS performance metrics for each of these user types and formulated an optimization problem with the sensing QoS criterion as the objective function while adhering to communication QoS constraints for all users. To tackle this challenging optimization problem, an AO procedure was utilized to divide the optimization problem into three sub-problems, one for each optimization variable. The SCA algorithm was applied to address the QoS communication constraints, and a difference-of-convex programming method was used to resolve the rank-one nonconvex constraints.}

{In~\cite{2024_Ziwei1}, a time-division sensing-communication mechanism was designed for holographic DFRC. To address the issue of interference management and enhance system PLS, the authors implemented a RSMA transmission scheme, where the common stream was designed to include both a useful signal and an artificial noise. This approach also took into account imperfect CSI and modeled the eavesdropper's channel in a detailed manner, providing an upper bound on the CSI error.}

{The authors derived the secrecy outage probability and constructed an optimization problem with the secrecy sum rate as the objective function. The goal was to optimize the common and private stream beamformers along with the timeslot duration allocated for communication and sensing. To tackle the optimization problem, an AO-based approach was proposed using a second-order cone programming (SOCP) algorithm. Furthermore, the S-procedure, Bernstein’s inequality, and SCA methods were employed to address the objective function and non-convex constraints.}

{\textit{5) RIS-assisted holographic DFRC:} Leveraging the advantages of both RHSs and RISs, the authors of~\cite{2023_Wei3} and~\cite{2024_Wei1} explored the use of both for DFRC. They considered a wideband DFRC system with an RHS at the transmitter and an RIS reflector. The authors jointly designed the receive filter, along with the digital, holographic, and passive beamformers, to maximize the worst-case radar SINR while ensuring a communication SINR threshold across all users. To solve the resulting optimization problem, they developed an AO method involving CM and difference of convex constraints. Numerical results showed that adding an RIS to the RHS-assisted DFRC improved the radar SINR by 20\% for low communication QoS requirements and by 300\% for high ones.}

{\textbf{Lessons learned:} Holographic beamforming has shown significant potential in enhancing the performance of DFRC systems through more directed and adaptive beams. This beamforming technique allows simultaneous communication with multiple users and detection of radar targets, creating a more efficient use of resources compared to systems with conventional arrays. Recent research in the area can be categorized into five key topics. First, resource allocation for RHS-assisted DFRC with RHSs can be used for signal transmission only or employed for signal transmission and echo signal reception. Second, cell-free DFRC systems can benefit from holographic beamforming to mitigate interference between different targets and users, which leads to better communication and sensing performance.} {Third, BD-RHSs can introduce additional design flexibility, leading to better performance in both communication and sensing by employing innovative beamforming algorithms. Fourth, the interplay between RSMA and holographic beamforming for DFRC has been investigated to manage interference between communication and sensing, with optimization frameworks enhancing both the system’s power allocation and secrecy. Lastly, RIS-assisted holographic DFRC combines RHSs and RISs to maximize ISAC performance, showing substantial improvements in sensing and communication performance through joint beamforming design and optimization techniques. These five aspects collectively push the boundaries of ISAC system performance.}

{\textbf{Challenges, opportunities and open research directions:} Employing RHSs for DFRC systems presents several challenges. First, hardware complexity is a major issue, as integrating both communication and sensing functionalities into a single platform requires advanced beamforming components that are both cost-effective and scalable. Additionally, real-time beamforming optimization is challenging, as it requires highly precise control over the phase and amplitude of each surface element to ensure simultaneous communication and radar sensing performance. Interference management between communication and radar signals, particularly in dynamic environments, also remains a significant hurdle, as the two functions must coexist without degrading each other's performance.}

{Despite these challenges, RHSs present significant opportunities for enhancing DFRC systems. The ability to create highly adaptive and directed beams enables improved accuracy in both communication and radar operations, especially in environments where signal interference or multi-path effects are common. RHSs also allow for fine-grained control over the beam shape, which can be leveraged for multi-target tracking and multi-user communication, offering significant improvements in system efficiency. Furthermore, the integration of machine learning algorithms with RHS-based DFRC systems opens up possibilities for real-time optimization of beamforming patterns, which could dramatically improve energy efficiency, signal quality, and user experience in next-generation networks.}

{Several open research directions exist for the future development of RHS-based DFRC systems. One key area is the design of efficient optimization algorithms that can handle the complex, non-convex beamforming problems associated with RHSs, such as joint communication and radar signal design. Another research direction is exploring robust beamforming solutions that can account for real-world uncertainties, such as CSI and RHS imperfections. Additionally, further work is needed to develop scalable and cost-effective hardware solutions, as well as explore the use of quantum computing or other advanced technologies to push the limits of beamforming accuracy and system performance.}

\subsection{RHS-Assisted ISAC Prototype} \label{sec:proto_RHS_DFRC}
A prototype of an RHS-Assisted ISAC system was presented by the authors of~\cite{2023_Haobo}. In particular, the authors designed a one-dimensional (1D) RHS with a length, width, and thickness of \unit[15]{$\text{cm}$}, \unit[3]{$\text{cm}$}, and \unit[0.17]{$\text{cm}$}, respectively. The substrate is made up of four layers and serves as a waveguide. Each of the sixteen feed lines is linked to a metamaterial element. The other end of the feed lines is connected to a field programmable gate array (FPGA) output pin, which supplies a bias voltage to the component.

The ISAC transceiver module, which serves as an ISAC BS here for both transmitting and receiving ISAC signals. The host computer employs an FPGA to control the radiation amplitudes of the RHS elements. Signal frequency conversion between baseband and passband is facilitated by a frequency converter that performs both down-conversion and up-conversion. For echo reception, a standard horn antenna (LB-75-20-C-SF) with a frequency range of \unit[10-15]{GHz} is utilized.

The user module acquires the communication stream by receiving and decoding the ISAC signal transmitted by the ISAC transceiver module. The receive antenna captures the ISAC signal, which is then sent to the frequency converter for down-conversion to the equivalent baseband signal.

The target module simulates radar targets by generating controlled radar echo signals, as demonstrated in~\cite{2021_Ma}. Activation of the target module occurs when the receiving antenna captures the ISAC signal broadcasted by the RHS. To emulate targets at different ranges, variations in the time delays have been implemented.

The experimental results show that that the RHS exhibits slightly higher gains towards the directions of interest compared to the phased array. However, the power consumption of the phased array is 30 times greater than that of the RHS. This suggests that the RHS is capable of supporting ISAC with similar performance and significantly lower power consumption compared to the phased array.

The experimental results also  indicate that the estimated range closely aligns with the actual range, affirming the viability of sensing through holographic ISAC. Furthermore, the data rate between the BS and the user is measured at \unit[5]{Mbit/sec}, underscoring that communication between the BS and the user remains supported even as the BS conducts radar sensing concurrently.

\begin{figure} 
         \centering
    \includegraphics[width=\columnwidth]{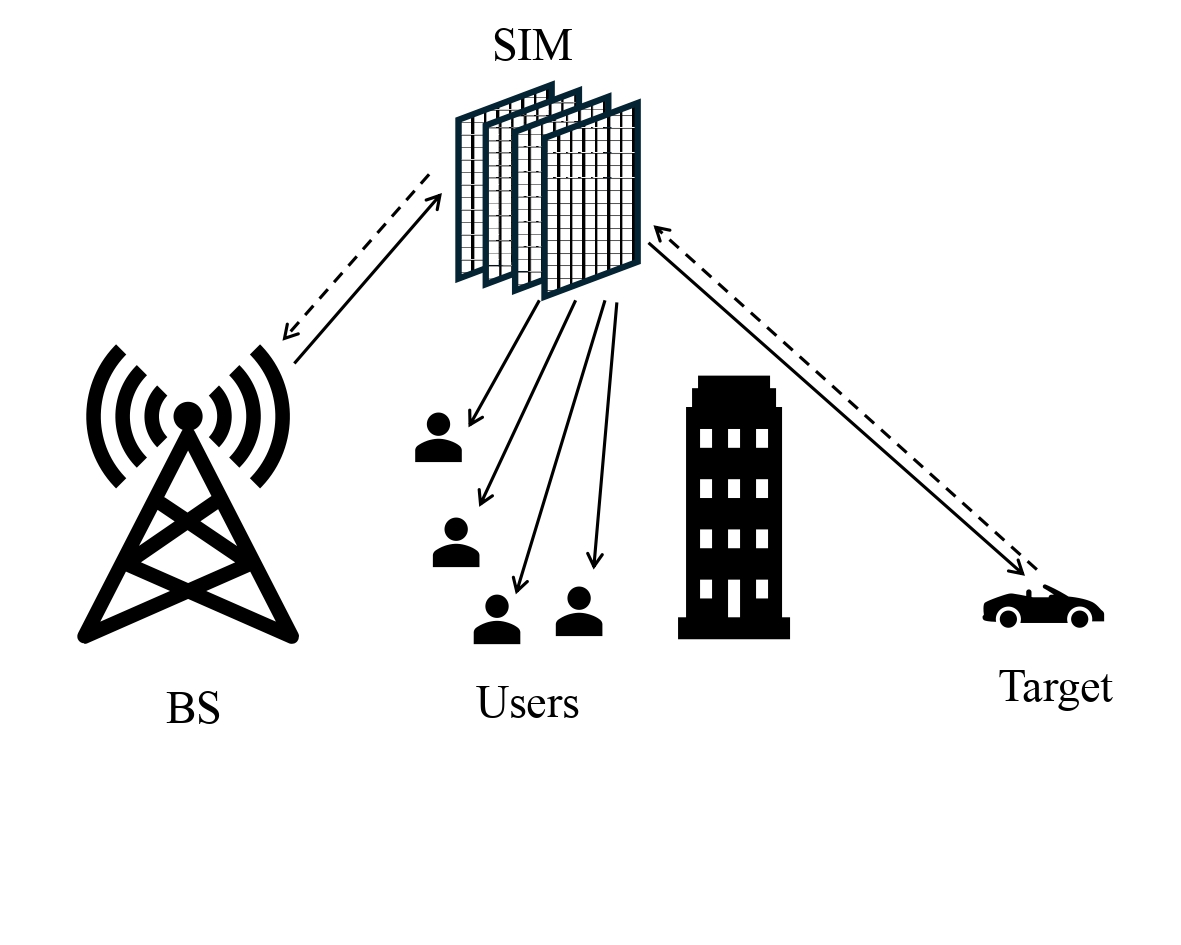}
        \caption{SIM-assisted DFRC system model presented in~\cite{2024_Ziqing}.}
        \label{fig:holo_SIM}
\end{figure}

{
\subsection{Holographic Beamforming via SIM}
SIMs consist of multiple layers of RHSs; therefore, they are more capable of generating sharp holographic beams. Moreover, when used as an entity between the transmitter and the receiver, SIMs are better able to generate directed beams toward the targets and users, as shown in~\cite{2024_Ziqing}. The authors of this work considered a system model consisting of a BS, an extended target, and multiple communication users. Holographic beamforming is performed using a SIM placed between the transmitter and the receiver, as illustrated in Fig.~\ref{fig:holo_SIM}.
}

{An optimization problem was formulated in~\cite{2024_Ziqing} to minimize the CRB of the AoA estimation for extended target by jointly optimizing the communication and sensing beamforming matrices, as well as the end-to-end transmission matrix of the SIM. This optimization must be performed while ensuring the minimum SINR constraints for each of the communication users. To tackle this non-convex optimization problem, the authors proposed an AO-based approach that utilizes the SDR algorithm. Notably, the authors observed almost linear decrease in the CRB as the number of layers increases. For instance, a difference of \unit[8]{dB} is observed between a SIM with one layer (i.e., a standard RIS) and a SIM with seven layers when the communication SINR threshold is set to \unit[0]{dB}. Furthermore, SIMs with a higher number of layers are more resistant to performance degradation when the communication SINR requirement increases, compared to those with fewer layers. This is because more layers allow for greater system design flexibility through more degrees of freedom.}

{\textbf{Lessons learned:} Research works highlight the advantage of utilizing SIMs consisting of multiple layers to generate sharp holographic beams to improve ISAC performance. The findings indicated a direct correlation between the number of layers in the SIM and the CRB, with notable differences observed between a single-layer SIM and a seven-layer SIM. Additionally, SIMs with more layers demonstrated enhanced resilience to performance degradation under increasing communication SINR demands, attributed to their increased design flexibility and degrees of freedom.}

{\textbf{Challenges, opportunities and open research directions:} One significant challenge of employing SIMs for ISAC is the complexity of designing and optimizing the multilayer structures, which requires advanced algorithms capable of handling non-convex optimization problems. Additionally, channel estimation is more challenging in the case of SIMs due to the presence of multiple layers, necessitating the estimation of a multitude of channels. This can significantly increase overhead and degrade system performance. While this issue can be resolved to some extent by estimating average channels instead of instantaneous ones, it leads to the presence of imperfect CSI that needs to be accounted for in the beamforming design step.}

{Despite these challenges, there are ample opportunities and open research directions in this domain. For instance, the development of adaptive algorithms that dynamically optimize metasurface configurations in real time could enhance the efficiency and effectiveness of ISAC systems. Additionally, investigating machine learning techniques to predict and adapt to changing environments could revolutionize how ISAC systems operate, allowing for smarter resource allocation and user prioritization. Finally, interdisciplinary research that combines insights from communication theory, sensor networks, and materials science can drive fundamental innovations in stacked intelligent metasurfaces, fostering the development of next-generation ISAC applications in various fields, including autonomous driving and smart cities.}

{
\subsection{RHS-Assisted ISAC Prototype}
The authors of the work discussed in the previous subsection also developed a prototype for the system model shown in Fig.~\ref{fig:holo_SIM}. In their experimental setup, the authors fabricated the SIM, in which each layer consists of a radiating layer, a receiving layer, and a ground plane. Moreover, a printed circuit board was utilized to represent the layout of a uniform planar array made up of $16 \times 16$ unit cells, with a total area of of \unit[961]{$\text{cm}^2$}. In the experimental setup, a universal software radio peripheral serves as the signal source, transmitting a signal at a frequency of \unit[5.8]{GHz} with a power budget of \unit[10]{dBm}. Since the system configuration utilizes a single antenna, no beamforming scheme is applied at the transmitter. Also, the transmission coefficient matrix for each scenario is obtained through an exhaustive search during communication or sensing experiments. 
}

{Three configurations are considered in the experiments: 1-layer, 2-layer, and 3-layer SIMs, along with five different coordinates for the transmitters. The measurements revealed an average error of the AoA estimation of $3.6^\circ$ for a single-layer SIM (i.e., an RIS). This error is reduced to $1.5^\circ$ (58\% lower error) for a 2-layer SIM and $1.4^\circ$ (61\% lower error) for a 3-layer SIM.}

\section{Conclusion}  \label{sec:conc}
This survey provided a comprehensive review of the existing literature on ISAC assisted by metasurfaces, discussing the associated challenges and opportunities. {The article started with a brief background on ISAC and metasurfaces. Then, it summarized existing research on RIS-assisted ISAC, where metasurfaces were employed as separate entities between the transmitters and receivers considering two integration levels in ISAC: RCC and DFRC. Finally, the article summarized the state of the art in holographic ISAC, which encompasses the utilization of metasurfaces at both the transmitter and receiver sides. Within each category, we presented lessons learned as well as prominent opportunities, providing a holistic perspective on the evolving landscape of RIS-assisted ISAC.}




%





\ifCLASSOPTIONcaptionsoff
  \newpage
\fi





\bibliographystyle{IEEEtran}
\bibliography{IEEEabrv,Bibliography}
%


\end{document}